\documentclass[a4paper,11pt]{article}
\pdfoutput=1
\usepackage{jheppub}

\usepackage[utf8]{inputenc}

\allowdisplaybreaks

\usepackage{amssymb}
\usepackage{amsmath}
\usepackage{mathrsfs}
\usepackage{tikz}
\usepackage{braket}
\usepackage{color}
\usepackage{xcolor}
\usepackage{graphicx}
\usepackage{slashed}
\usepackage{bbold}
\usepackage{breqn}

\allowdisplaybreaks

\newcommand{\e}{\epsilon}

\newcommand{\zero}{{(0)}}
\newcommand{\one}{{(1)}}
\newcommand{\two}{{(2)}}
\newcommand{\three}{{(3)}}
\newcommand{\als}{\alpha_s}
\newcommand{\alsmu}{\alpha_s(\mu)}

\newcommand{\Ord}{\mathcal{O}}

\newcommand{\nn}{\nonumber}

\newcommand{\df}{d}

\newcommand{\Gcusp}{\Gamma^{\text{cusp}}}
\newcommand{\Lp}{L_\perp}

\newcommand{\rd}{\mathrm{d}}
\newcommand{\ry}{\mathrm{y}}
\newcommand{\rx}{\mathrm{x}}

\newcommand{\rT}{\mathcal{T}}

\newcommand{\cL}{\mathcal{L}}
\newcommand{\cO}{\mathcal{O}}
\newcommand{\cC}{\mathcal{C}}

\newcommand{\cP}{{\mathcal P}}

\newcommand{\hU}{\widehat{U}}
\newcommand{\hDe}{\mathbf{\Delta}}

\definecolor{darkred}{rgb}{0.7,0.0,0.0}
\definecolor{darkblue}{rgb}{0.0,0.0,0.9}
\definecolor{darkgreen}{rgb}{0.0,0.5,0.0}
\definecolor{brown}{rgb}{0.0,0.0,0.0}
\newcommand{\red}{\color{darkred}}
\newcommand{\blue}{\color{darkblue}}
\newcommand{\green}{\color{darkgreen}}
\newcommand{\orange}{\color{orange}}

\newcommand{\cyan}{\green}

\definecolor{bblue}{rgb}{0.0,0.0,1}

\definecolor{bred}{rgb}{0.7,0.0,0.0}

\def\bT{{\bf{T}}}
\newcommand{\id}{\mathbf{1}}
\newcommand{\intlim}[3]{\int_{#1}^{#2}\! \rd #3 \,}
\newcommand{\cusp}{\mathrm{cusp}}

\def\cC{\mathcal{C}}

\def\cI{\mathcal{I}}

\def\cL{\mathcal{L}}

\def\cN{\mathcal{N}}
\def\cO{\mathcal{O}}

\def\cP{\mathcal{P}}

\DeclareMathOperator*{\SumInt}{%
\mathchoice%
  {\ooalign{$\displaystyle\sum$\cr\hidewidth$\displaystyle\int$\hidewidth\cr}}
  {\ooalign{\raisebox{.14\height}{\scalebox{.7}{$\textstyle\sum$}}\cr\hidewidth$\textstyle\int$\hidewidth\cr}}
  {\ooalign{\raisebox{.2\height}{\scalebox{.6}{$\scriptstyle\sum$}}\cr$\scriptstyle\int$\cr}}
  {\ooalign{\raisebox{.2\height}{\scalebox{.6}{$\scriptstyle\sum$}}\cr$\scriptstyle\int$\cr}}
}

\preprint{MPP-2025-159}

\title{The N$^3$LO Twist-2 Matching of Helicity TMDs and SIDIS $q_\ast$ Spectrum}

\author[a]{Yu Jiao Zhu}
\emailAdd{yzhu@mpp.mpg.de}

\affiliation[a]{Max-Planck-Institut f\"{u}r Physik, Werner-Heisenberg-Institut, Boltzmannstr. 8, 85748 Garching, Germany}

\abstract{We compute the  twist-2 matching of transverse momentum dependent (TMD) helicity parton
distribution and fragmentation functions  at next-to-next-to-next-to-leading order (N$^3$LO) in QCD.
This calculation entails the complete set of  next-to-next-to-leading order (NNLO) 
Dokshitzer-Gribov-Lipatov-Altarelli-Parisi (DGLAP) splitting functions govering the evolution  of helicity-dependent  parton distribution functions (PDFs) and fragmentation functions (FFs).
Within TMD factorization framework, we quantify the  impact of  radiative corrections by completing the  next-to-next-to-next-to-leading logarithmic (N$^3$LL) prediction for lepton-hadron  transverse momentum imbalance in semi-inclusive deep inelastic scattering (SIDIS).  
Our results provide the most precise theoretical input for probing the helicity structure and 
confined motion  of quarks and gluons at future electron-ion collider (EIC).
}

\begin{document}

\maketitle

\section{Introduction}
\label{sec:introduction}
Transverse-Momentum-Dependent (TMD) Parton Distribution Functions (PDFs)
parameterize the  confined motion of parton constituents inside a nucleon target.
TMD Fragmentation Functions (FFs) describe the  hadronization process of parton decay  into  color-singlet hadrons.
On top of that, 
TMDs also carry novel information about the nucleon’s internal  flavor and spin structures,
small-x TMDs encodes Regge asymptotics of QCD governed by
Balitsky-Fadin-Kuraev-Lipatov (BFKL) evolution~\cite{Balitsky:1978ic,Kuraev:1977fs} and beyond.
TMDs are key objects in the next QCD frontier, 
and are  essential for understanding the glue that binds us all~\cite{Accardi:2012qut}.

A longstanding objective in QCD is to uncover the fundamental mechanisms of color confinement.
On the other hand, QCD is known for the  converse of confinement-- the celebrated  
property of asymptotic freedom~\cite{Gross:1973id,Politzer:1973fx}.
Thanks to the hierarchy between short- and long-distance scales, the dynamics at different regimes can be systematically disentangled within the framework of effective field theory. Schematically, the cross section admits a factorized structure
\begin{align}
\sigma (Q,\Lambda_{\text{QCD}})= 
\cC(Q,\mu)\otimes \langle \cO \rangle (\mu,\Lambda_{\text{QCD}}) + \cO\left(\frac{\Lambda^2_{\text{QCD}}}{Q^2}\right)\,,
\end{align}
where $\mu$ is the factorization scale. Factorization based on  Collins-Soper-Sterman (CSS) formalism~\cite{Collins:1988ig}
has been proven for Drell-Yan process and  single inclusive hadron production~\cite{Bodwin:1984hc,
Collins:1984kg,
Collins:1988ig,
Collins:1989gx,
Nayak:2005rt}.

The principle of factorization underlies all theoretical predictions of QCD.
To the UV, the renormalization of $\alpha_s$ is independent of the external momenta flowing into the Green functions,
so that UV physics is completely factorized from collider measurements performed on asymptotic final states.
To the IR, long-distance physics is parametrized while  short-distance dynamics are computed
from Feynman amplitudes, again thanks to the asymptotic freedom. 
UV  and IR  cutoﬀ dependence drop out in the \emph{relations} among physical quantities.

In the context of TMDs, 
factorization provides the gateway that bridges partonic high-energy interactions to what is observed at the detectors:
the confined motion can never been probed directly,
it manifests through correlations with the  distribution of final-state gauge bosons, leptons, and hadrons,
owing to the unification 
of  Standard Model  gauge group  
${\text{SU}}(3)\times {\text{SU}}(2)\times {\text{U}}(1)_{\text{Y}}~({\text{SU}}(2)\times {\text{U}}(1)_{\text{Y}}\to\text{U}(1)_{\text{EM}}$. Indeed, 
TMDs are indispensable non-perturbative input for a variety of benchmark observables in the Standard Model,
including 
Drell-Yan process~\cite{
Dokshitzer:1978yd,
Parisi:1979se,
Collins:1984kg,
Arnold:1990yk,
Ladinsky:1993zn,
Bozzi:2010xn,
Becher:2010tm,
Becher:2011xn,
Bertone:2019nxa},
semi-inclusive deep-inelastic scattering (SIDIS)~\cite{Ji:2004wu,
Ji:2004xq,
Liu:2018trl,
Fang:2024auf,
Gao:2022bzi,
Ebert:2021jhy,Fang:2024auf},
electron-positron annihilation to hadrons and jets~\cite{Collins:1981uk,Collins:1981va,Neill:2016vbi,Gutierrez-Reyes:2018qez,Gutierrez-Reyes:2019vbx,Gutierrez-Reyes:2019msa},
Higgs boson production~\cite{Berger:2002ut,Bozzi:2005wk,Gao:2005iu,Echevarria:2015uaa,Neill:2015roa,Bizon:2017rah,Chen:2018pzu,Bizon:2018foh},
top quark pair production~\cite{Zhu:2012ts,Li:2013mia,Catani:2014qha,Catani:2018mei}, 
 $J/\psi$ production~\cite{Echevarria:2024idp, Copeland:2023wbu, Kishore:2018ugo,
 Maxia:2024cjh,Banu:2024ywv,Kishore:2024bdd,Chakrabarti:2022rjr,Boer:2023zit,Bor:2022fga,Bacchetta:2018ivt,Boer:2016bfj},
as well as Energy-Energy (EEC) or Charge-Charge Correlator~\cite{Moult:2018jzp,Gao:2019ojf,Gao:2024wcg,Kang:2024otf,Kang:2023big,Liu:2024kqt,Monni:2025zyv}
at lepton and hadron colliders.

Helicity TMDs in particular probe spin-momentum correlations and are central to understanding the decomposition of nucleon spin. SIDIS provides direct access to such distributions through transverse momentum and spin asymmetry measurements.
In this paper, we consider lepton-jet  transverse  momentum imbalance in lepton-hadron colliders 
with polarized high-energy lepton beams scattering off longitudinally polarized nucleon targets:
\begin{align}
 \ell(p_\ell,\lambda_\ell) + N(P_N, S_{\parallel}) \to \ell'(p_{\ell'},\lambda_{\ell'}) + h(P_h) + X(P_X)
\,.\end{align}
We derive the TMD factorization formula for this observable within the framework of Soft-Collinear Effective Theory (SCET)~\cite{Bauer:2000ew,Bauer:2000yr,Bauer:2001ct,Bauer:2001yt}, which enables a systematic separation of perturbative and nonperturbative contributions in the small transverse momentum regime. Using  our new analytic expressions, we obtain a high-precision prediction for the $q_\ast$ spectrum, which provides a complementary probe to the TMD dynamics, alongside 
the conventional $q_T$-differential cross section in SIDIS.

This work is motivated by the broader program of global analyses of TMD distributions. Over the past decade, global fits of unpolarized and polarized TMDs have been performed using SIDIS, Drell-Yan, and $e^+e^-$ annihilation data within the TMD factorization framework~\cite{Sun:2014dqm,Bertone:2019nxa,Scimemi:2019cmh,Moos:2023yfa,Moos:2025sal,Almasy:2011eq,Bacchetta:2017gcc,Bacchetta:2019sam,Bacchetta:2022awv,Bacchetta:2024qre,Bacchetta:2025ara}. These fits have reached  approximately N$^4$LL accuracy in the unpolarized sector, while efforts to include spin-dependent TMDs--such as helicity and transversity distributions--are currently underway~\cite{Yang:2024drd,Bacchetta:2024yzl}. Despite this progress, theoretical control over helicity-sensitive observables still lags behind that of their unpolarized counterparts. In particular, the full next-to-next-to-next-to-leading order (N$^3$LO) matching for helicity TMDs has so far been unavailable. This limits the precision of phenomenological fits, especially in light of the high-statistics spin-polarized SIDIS measurements anticipated from the Electron-Ion Collider (EIC).

The study of helicity TMDs is inherently linked to the evolution of helicity PDFs and FFs, 
which is governed by polarized DGLAP~\cite{Altarelli:1977zs,Gribov:1972ri,Dokshitzer:1977sg} splitting functions.
 Unlike in the unpolarized case, where splitting functions are known up to NNLO~\cite{Moch:2004pa,Vogt:2004mw,Mitov:2006ic,Moch:2007tx,Chen:2020uvt} and partially known to N$^3$LO~\cite{Moch:2017uml,Gehrmann:2023ksf,Gehrmann:2023cqm,Gehrmann:2023iah,Falcioni:2023luc,Falcioni:2023vqq,Falcioni:2023tzp,Moch:2023tdj,Basdew-Sharma:2022vya,Falcioni:2024xyt}, their polarized counterparts have long remained incomplete beyond NLO~\cite{Mertig:1995ny,Vogelsang:1995vh,Vogelsang:1996im,Zijlstra:1993sh,Behring:2019tus,Stratmann:1996hn,Rijken:1997rg}. In recent years, significant effort has been devoted to computing the NNLO helicity splitting functions~\cite{Moch:2014sna,Moch:2015usa,Blumlein:2021enk,Blumlein:2021ryt,Blumlein:2022gpp}, driven by the need  for high-precision QCD evolution in global analyses and improved theoretical control over spin asymmetries in polarized deep-inelastic scattering. 
These splitting functions are also essential for implementing helicity-sensitive parton evolution in Monte Carlo simulations of polarized scattering processes.
Moreover, they play a key role in the ongoing effort to resolve the longstanding proton spin puzzle--the question of how the proton’s spin is distributed among its constituent quarks and gluons. Helicity PDFs provide a direct link between measurable observables and the partonic spin content of the nucleon, and their precise extraction requires accurate knowledge of both their functional forms and scale evolution. In this context, higher-order polarized splitting functions and matching coefficients are indispensable for reducing theoretical uncertainties in spin decomposition.
In particular, recent progress in understanding helicity evolution at small~$x$ has revealed potentially non-negligible contributions to the proton spin from the low-$x$ region~\cite{Kovchegov:2015pbl,Kovchegov:2016weo,Kovchegov:2017jxc,Kovchegov:2017lsr}.
 Beyond their phenomenological relevance, higher-order splitting functions also offer a complementary theoretical motivation:  they provide the database for investigating  crossing symmetry and beyond: the generalization of Gribov-Lipatov reciprocity beyond  amplitude-level crossing~\cite{Chen:2020uvt}.

In this paper, we take a significant step toward completing the perturbative framework for helicity-dependent observables in QCD. We compute the twist-2 matching coefficients for helicity TMD PDFs and FFs at N$^3$LO accuracy. As an outcome, we derive  the full set of NNLO polarized DGLAP splitting functions in analytic form. These results establish a new level of precision for helicity evolution and TMD matching, placing helicity observables on equal theoretical footing with their unpolarized counterparts. We further illustrate the numerical impact of higher-order corrections by performing a N$^3$LL-resummed calculation of the transverse momentum spectrum in SIDIS. Together, these developments provide a solid foundation for future global analyses of spin-dependent data and for interpreting precision measurements at the EIC.

\section{The TMD Factorization of $q_\ast$ Spectrum}
We define leptonic momentum transfer as  $q=p_l-p_{\ell'}$,
with virtuality $Q^2=-q^2$.  
We also introduce the standard set of kinematic invariants
\begin{align}
x=\frac{Q^2}{2 P_N\cdot q}\,,\quad y=\frac{P_N\cdot q}{P_N\cdot p_\ell}\,,\quad z=\frac{P_N\cdot P_h}{P_N\cdot q}\,.
\end{align}
In conventional SIDIS analyses,
one measures  the  transverse-momentum spectrum  $q_T$  of the virtual photon in hadron-hadron frame,
or  equivalently the $P^T_h$   of the  identified hadron in photon-hadron frame
where $ \vec P^\perp_h=-z \vec q_\perp$. 
Instead of the conventional $q_T$  spectrum, 
we propose to measure the  angular correlations in the transverse plane, defined as follows
\begin{align}
\label{eq: fullth-lepton-jet}
    \frac{\rd \sigma^h}{\rd \cos\phi}=&
     \int
    \rd \sigma_{\ell +N\to\ell'+h+X}\times\delta(\cos\phi_{\ell' h}-\cos \phi)\,,
  \end{align}
where  $ \phi_{\ell' h}$ is the lepton-hadron azimuthal angle in the \emph{lepton-target} frame.
The observable select events with a fixed azimuthal angle $\cos\phi_{\ell' h}=\cos \phi$ .
An equivalent but  Lorentz-invariant  way of parametrization is to select events with fixed momentum projection $q_\perp \cdot v$,  
where $v$ is the normal 4-vector  to the event plane
\begin{align}
\label{eq:v-lepron-jet}
v_\mu \equiv  \frac{2 i \epsilon_{ \mu \nu \rho \sigma} p^\nu_\ell  P^\rho_h P^\sigma_N  }{\sqrt{s_{\ell h}s_{\ell N} s_{h N}}}\,,\quad v\cdot v=-\vec v\cdot \vec v=-1\,.
\end{align}
We introduce  the notation of transverse momentum imbalance 
between the scattered lepton $\ell'$ and the identified hadron $h$
\begin{align}
\label{eq:qy-def}
q_\ry\equiv\vec q_\perp\cdot \vec v = - q_\perp \cdot v\,,
\end{align}
where $q_\ry$ quantifies the deviation of the scattered lepton  $\ell'$  from the event plane,
its relation to the azimuthal angle is 
\begin{align}
   \sin \left(\pi-\phi_{\ell' h}\right)=\frac{q_\ry}{q_T}\,.
\end{align}
In this way, we are considering the  $q_\ast$ spectrum ($q_\ast=|q_\ry|$) in the lepton-target frame as a variant of the conventional $q_T$ spectrum in the photon-target frame. Note however, in addition to the laboratory frame construction,
$\epsilon(q, p_\ell, P_h, P_N)$ can also be  interpreted as momentum shift of the lepton in Breit frame or momentum shift of the hadron in Trento frame~\cite{Bacchetta:2004jz}.

The  $q_\ast$ spectrum with $q_\ast\ll Q$ delivers sensitivity to the TMD dynamics, the partonic picture can be imaged as follows:
if only born-level contribution $N+\gamma^\ast \to h$ were considered,  then the hadron, the Proton, 
and the  leptons are coplanar, i.e., the volume form exactly vanishes $\epsilon(p_{\ell'}, p_\ell,P_h, P_N)\equiv 0$.
In general, events with multiple jets $N+\gamma^\ast \to h +X$ contribute, but in the back-to-back limit $\epsilon( p_{\ell'}, p_\ell,P_h, P_N)\to 0$,
all the final-state radiations in $X$ must be collimated, with momenta collinear either to the target $N$ or the detected hadron $h$.
The collinear splitting carries a non-perturbative transverse scale of  $K_T$ -- the intrinsic scale of TMD motion. 
As a result, the momentum of the outgoing lepton $\ell'$ should balance that transverse momentum flow out of the event plane,
leaving small deviations of  $q_\ry \sim K_T$ in the $v$-direction.
Such an observable is known as transverse momentum imbalance between the scattering-off lepton $\ell'$ and the detected hadron $h$,
whose differential cross section is given by
\begin{align}
\label{eq:dcrosec}
\frac{\rd \sigma_{\ell+N\to \ell'+h+X}}{\rd\delta \vec p_J^\perp \rd x y\rd y  z\rd z   }=\frac{2 \pi \alpha^2}{4Q^4} L_{\mu \nu}(P_\ell,P_{\ell'}) W^{\mu \nu}(q, P_N, P_h)\,.
\end{align}
where $\delta \vec p_J^\perp=\vec P_h^\perp/z -\vec q_\perp$ is the difference  in  
transverse momentum
between the jet 
and the photon, $q_\ry=\delta \vec p_J^\perp\cdot v$ in the lepton-hadron frame.
 Moreover, the hadronic/lepton tensors  are 
\begin{align}
W^{\mu\nu}(q,P_N,P_h)=&\prod_X \frac{\rd^3 P_X}{(2\pi)^3 2 E_X} \delta^{(4)}(q+P_N-P_h-P_X) 
\langle N| J^{\dagger \mu}|h,X\rangle  \langle h,X | J^\nu | N \rangle \,.
\nn\\
L^{\mu\nu}=&2 \delta_{\lambda_\ell \lambda_{\ell'}}\left[ \left( p_\ell^\mu p_{\ell'}^\nu+p_\ell^\nu p_{\ell'}^\mu -\frac{Q^2}{2} g^{\mu\nu}\right)+i \lambda_l \epsilon^{\mu\nu \ell \ell'} \right]\,.
\end{align}
To separate perturbative and non-perturbative dynamics associated with the scale hierarchy $q_\ast\ll Q$,
we will derive a factorization formulae from  perspectives of  Soft-Collinear Effective Theory (SCET)~\cite{Bauer:2000ew,Bauer:2000yr,Bauer:2001ct,Bauer:2001yt}.
The active modes in SCET are collinear partons moving in well-separated light-like directions and soft gluons mediating long-range interactions among the jets.
After  BPS field redefinition~\cite{Bauer:2001yt}, soft and collinear sectors decouple in the Lagrangian.
Physically, this reflects that soft gluons cannot resolve the internal structure of a jet, but are only sensitive to its total quantum numbers and direction. 
In SCET, multi-direction collinear jets interact through a complete set of hard operators $\cO_i$. The corresponding Wilson coefficients $\cC_i$ encode the short-distance dynamics, with the latter modded out from the low-energy theory. The hard operators are constructed such that their ultraviolet divergences reproduce the infrared divergences of multi-leg amplitudes. In renormalized perturbation theory, the SCET Lagrangian takes the form 
\begin{align}
\label{eq:scet-lag}
\cL_{\text{SCET}}=\sum_{ {\blue j}=1} ^N \cL_{\blue j} + \cL_{\red \text{soft}}+\sum_i Z_i \cC_i \cO_i\,.
\end{align}
The Lagrangian is soft zero-bin subtracted, which is equivalent to dividing the collinear fields by the vacuum expectation values of zero-bin Wilson lines~\cite{Feige:2014wja,Lee:2006nr}.
At amplitude level when $|X\rangle\simeq|X_{\red \text{soft}} \rangle\prod^N_{\blue j}  |X_{\blue j} \rangle$, 
the amplitude is mediated through effective vertices of $\cO_i$ and is factorized~\cite{Feige:2014wja}, 
e.g., for the  leading jet production in SIDIS
\begin{align}
\label{eq:fac-qbqbg}
\langle X| {\orange\gamma^\ast}; {\blue{N}}\rangle \simeq 
 \sum_{\vec \lambda}
 |\cC_{\vec \lambda}\rangle 
  \langle X_{\red \text{soft}}  X_{\blue n \cyan \bar n}; {\cyan  h}| \cO^{\vec \lambda}_{\text{scet}} |{\blue{N}}\rangle\,, 
  \quad
  \vec \lambda \in \{
{\blue n^\pm }
{ \green{ \bar n^\pm} }
\}\,,
\end{align}
each hard operator $\cO^{\vec \lambda}_{\text{scet}}$ is a local product of helicity fields with definite little group scalings~\cite{Moult:2015aoa}.
The hard amplitude $\mathcal{C}(Q^2,\mu)$   disentangles from  the  infrared (IR) dynamics associated with the hadronic state ${\blue N}$ ,
and can be computed at partonic level
\begin{align}
\label{eq:scet-qqb}
\langle {\green{q}}|J^\mu| {\blue{q}}\rangle = 
\,\mathcal{C}(Q^2,\mu)
\langle {\green{q}}|\bar\chi_{\green{{\bar n}}}|0\rangle
\gamma^\mu
\langle 0| \chi_{\blue{n}}| {\blue{q}}\rangle\,.
\end{align}
In this case, the leading-power collinear expansion is exact ($=$ instead of $\simeq$), 
as each  collinear trajectory contains  just one particle--taking the collinear limit simply does nothing.
Furthermore,  since pure virtual-loop corrections in SCET
are scaleless, they vanish identically, consequently
 \begin{align}
\label{eq:pole-transmu}
  \langle {\green  q}| \cO^{\vec \lambda}_{\text{scet}} |{\blue q} \rangle
 =  \langle {\green  q}| \cO^{\vec \lambda}_{\text{scet}}| {\blue q}\rangle_{\text{tree}}\left(1+\delta_{Z_{\vec \lambda}} (\epsilon_{\text{IR}})\right)\,,
\end{align}
so that the IR poles of a  helicity amplitude is reproduced
 by UV renormalization counter-term diagrams provided by $\delta  {Z_{\vec \lambda}}= {Z_{\vec \lambda}}-1$~\cite{Moult:2015aoa}.
This is as expected since the IR divergences in the theory above a matching scale must match with the UV divergences of the low-energy effective theory below that matching scale.
The discussion above can be summarized as in the following factorization formulae~\cite{Feige:2014wja,Bauer:2000ew,Bauer:2000yr,Bauer:2001ct,Bauer:2001yt}
\begin{align}
\langle {\green{h}};    {X_{{\blue{N}}\green{h}}};X_{\red{ \text {s}}}|J^\mu|{\blue{N}}\rangle\simeq &
\mathcal{C}(Q^2,\mu)
 \langle X_{\red{ \text {s}}}| \rT \left[
  Y^\dagger_{\green{\bar n}}Y_{{\blue{n}}}
  \right]
  |0\rangle
\langle {\green{h}}; {X_{\green{h}}}|\bar\chi_{\green{{\bar n}}}|0\rangle
\gamma^\mu
\langle {X_{\blue{N}}}| \chi_{\blue{n}}| {\blue{N}}\rangle\,.
\end{align}
 Once the hard and soft-collinear degrees of freedom are disentangled,  the next step is to enumerate  all allowed hard events,
and within each hard label, zoom in to the soft-collinear subprocesses and count over  particles  in a proper way:
the classes of momenta entering into the soft-collinear phase space 
differ only by their typical size of energy or rapidity, as a result, the phase space should be integrated over with a rapidity cutoff $\nu$,
here we employ exponential cutoff  scheme~\cite{Li:2016axz}.
Additionally,  a zero-bin subtraction is performed to remove double-counting between collinear and soft sectors,
yielding genuine collinear beam functions~\cite{Manohar:2006nz}. Consequently, we have hard-collinear-soft factorization   
\begin{align}
\label{eq:hadronic-fac}
&W^{\mu\nu}
\simeq 
\prod \rd X_{\blue{\text{c}}\red{ \text {s}}}
n\cdot \bar n
\delta(\bar n \cdot p-x \bar n \cdot P_N)
\delta( n \cdot p_J-  n \cdot P_h/z)
\delta^{(2)}(\delta \vec p_J^\perp+\vec P^\perp_{X_{\blue{\text{c}}\red{ \text {s}}}})
\sum_f  |\mathcal C_f (Q^2,\mu)|^2
\nn\\
\times&
\langle 0|
\overline\rT \left[ Y^\dagger_{{\blue{n}}}
Y_{\green{\bar n}}
\right]  |X_{\red{ \text {s}}}\rangle
\langle X_{\red{ \text {s}}}|
\rT [ Y^\dagger_{\green{\bar n}}Y_{{\blue{n}}}]
|0\rangle 
\mathrm{Tr} [
\langle {\blue N};S_{\parallel}
|
\chi_{\blue{n}}|
{X_{\blue{N}}}
\rangle 
\langle {X_{\blue{N}}}|
\bar\chi_{\blue{n}}
|{\blue
N}; S_{\parallel}
\rangle
\gamma^\nu
\langle 0|
\chi_{{\green{\bar n}}}
|{\green{h}}, X_{\green{h}}\rangle
\langle{\green{h}}, X_{\green{h}}|
\bar\chi_{{\green{\bar n}}}|0\rangle
\gamma^\mu
]
\nn\\
=&
2 n\cdot \bar n \, z
\sum_f  H_f(Q^2,\mu)
\int \frac{\rd b}{2\pi}
e^{i b \delta p_J^\ry} 
 \mathcal{S}_{n\bar n}(b,\mu)
 \mathrm{Tr} [
 \mathscr{B}_{f/N}(S_{\parallel},
 x,b,E_n,\mu)
\gamma^\mu
\mathscr{D}_{h/f}(z,b,
E_{\bar n},\mu)
\gamma^\nu
]\,,
\end{align}
the factor $2$ in the front originates from spin average for the operator definition of TMD FFs. Furthermore, in the last step of eq.~(\ref{eq:hadronic-fac}),
the $\rx$-direction integral is performed, 
so that the impact parameter $b_\perp=b \, v$
is aligned along the $\ry$-direction.
By applying  SCET Fierz identity 
\begin{align}
1 \otimes 1=\frac{1}{2} \biggl[ \frac{\slashed{\bar n}}{2} \otimes \frac{\slashed n}{2}-\frac{\slashed{\bar n} \gamma_5}{2} \otimes \frac{\slashed n \gamma_5}{2}-\frac{\slashed{\bar n} \gamma_\perp^\mu}{2} \otimes
 \frac{\slashed n \gamma_{\perp\mu}}{2} \biggr]\,,
\end{align}
to the  TMD factorization formula in eq.~(\ref{eq:hadronic-fac}), the spin-correlation decomposes into a spin-sum  and a spin-asymmetry contribution
\begin{align}
\label{eq:SIDIS-fac}
x y z \frac{\rd \sigma_{\ell+N\to \ell'+h+X}}{2\rd  q_\ast \rd x\rd y\rd z}
\simeq&\, \sum_f   \int \frac{\rd  b}{2\pi} e^{i  b q_\ry}\,
\left[H_f(Q^2,\mu) \mathcal{S}_{n\bar n}\left(b_\perp\,,\mu\,,\nu\right)\right]
z\mathcal{F}_{h/f}\left(z\,,\frac{b_\perp}{z}\,,E_{\bar n}, \mu,\nu\right)
\nn\\
\times
&
\left(
\sigma^{\text{U}}_0\times x\mathcal{B}_{f/N} (x\,,b_\perp\,,E_n, \mu,\nu) 
+\lambda_\ell S_{\parallel}\times\sigma^{\text{L}}_0\times x \Delta \mathcal{B}_{f/N} (x\,,b_\perp\,,E_n, \mu,\nu)
\right)\,,
\end{align}
where  $\alpha=e_R^2/(4 \pi)$ and
$H_f(Q^2,\mu)=|\cC_f(Q,\mu)\rangle\langle \cC_f(Q,\mu)|$ is the  square of the hard matching coefficient~\cite{Becher:2006mr,Baikov:2009bg}.
The born-level unpolarized cross-section  $\sigma^{\text{U}}_0$ and its polarized counterpart $\sigma^{\text{L}}_0$, are given by
\begin{align}
\sigma^{\text{U}}_0=2\pi \alpha^2 \frac{1+(1-y)^2}{Q^2}\,,\quad \sigma^{\text{L}}_0=2\pi \alpha^2 \frac{1-(1-y)^2}{Q^2} \,.
\end{align}
The denominator $2\rd  q_\ast $ takes into account the reflection symmetry of the events with respect to the event plane.
Again, we stress that our factorization formula consists of genuine collinear beam functions ~\cite{Collins:1982wa,Collins:2011zzd,Echevarria:2012js,Ji:2004wu}
\begin{align}
    \mathcal{B}_{q/N}& (x\,,b_\perp\,,E_n, \mu,\nu)
    \equiv
    \lim_{\nu\to\infty}
    Z^q_B(b_\perp\,,E_n, \mu,\nu)\frac{\mathcal{B}^{0}_{q/N} (x\,,b_\perp\,,E_n, \mu,\nu)}{\mathcal{S}^{0}_{n\bar n}(b_\perp\,,\mu,\nu)}\,,
\end{align}
where the superscript $0$ denotes  unsubtracted collinear beam functions.
The beam or soft function is defined with a rapidity regulator
~\cite{Collins:2011zzd,Becher:2010tm,Echevarria:2011epo,Echevarria:2012js,
Chiu:2012ir,Li:2016axz,Echevarria:2015usa,Echevarria:2015byo,Echevarria:2016scs,Becher:2011dz,Ebert:2018gsn}.
In this work, we adopt exponential regulator~\cite{Li:2016axz} for both soft~\cite{Li:2016ctv} and collinear sectors~\cite{Luo:2019bmw,Luo:2019hmp,Luo:2019szz,Luo:2020epw}, as 
it explicitly preserves non-abelian exponentiation~\cite{Gatheral:1983cz,Frenkel:1984pz}, 
and the zero-bin soft function within exponential regularization scheme is identical to the TMD soft function.

The TMD beam functions and fragmentation functions are  the conventional ones but are parameterized one-dimensionally along  the $v$-direction since $\frac{b_\perp}{b}=v$.
Note that  for the beam functions or TMD PDFs,
the incoming beam provides a canonical reference frame for transverse momentum.
In the case of TMD FFs, we have used parton frame TMD fragmentation functions $\mathcal{F}_{h/f}$,
where the parton initialing the decay processes has zero transverse momentum. It is related to the hadron-frame TMD fragmentation functions by
\begin{align}
\label{eq: parton-hadron-frame}
\mathcal{F}_{h/f}\left(z\,,\frac{b_\perp}{z}\,,E, \mu,\nu\right)=z^2 \mathcal{D}_{h/f}\left(z\,,b_\perp\,,E, \mu,\nu\right)\,,
\end{align}
the later is defined by enforcing that the detected hadron has zero transverse momentum~\cite{Collins:2011zzd,Luo:2019hmp,Luo:2019bmw,Luo:2020epw}.
The TMD soft correlation function $\mathcal{S}_{n\bar n}$ is defined as a vaccum expectation value of the time-ordered Wilson lines
\begin{align}
  \label{eq:soft}
\qquad \mathbf{S}_{12\dots m}(b_\perp) = \langle 0 |T[\boldsymbol{O}_{n_1\dots n_m }(0)] \overline{T}[\boldsymbol{O}_{n_1\dots n_m }^\dagger (b_\perp)] | 0 \rangle\,,
\end{align}
where $\boldsymbol{O}_{n_1\dots n_m}(x) = \prod_i^m \boldsymbol{Y}_{n_i} (x)$, 
with $\boldsymbol{Y}_{n_i}(x) = \exp[ i \int \rd s\, n_i \cdot A_s(s n_i + x) \mathbf{T}_i]$ a semi-infinite light-like soft Wilson line, 
and $n_i= p_i/p_i^0$ the light-like direction of the incoming or outgoing parton with $b_\perp \cdot n_i=0$.
Each Wilson line is aligned  with  the  classic trajectories  of the hard particles,
from the origin where hard scattering take place to the distant future where final-state particles are detected.
The  Wilson line can couple to infinitely many soft gluons through eikonal interactions.
Thanks to the non-abelian exponentiation theorem~\cite{Gatheral:1983cz,Frenkel:1984pz},
 the TMD soft  correlation function is an exponential of web diagrams with a dipole color structure, provided that 
  no more than three  Wilson lines are considered~\cite{Catani:1999ss,Catani:2000pi,Gao:2023ivm}
 \begin{align}
  \mathbf{S}_{12\dots m}\equiv\exp\Bigl[
  -\sum_{i<j\in \{12\dots m\}} \mathbf{T}_i \cdot \mathbf{T}_j\, \mathbf{s}_{ij}
   \Bigr]\,,\quad m\leq 3
\end{align}   
where $\mathbf{s}_{ij}$ is the universal soft exponent, it is related to the conventional back-to-back TMD soft function~\cite{Li:2016ctv} through a boost
\begin{align}\label{eq:S_dipole}
 \mathbf{s}_{ij} =\mathbf{s}_{\perp}\,\Bigl(L_b,L_\nu+\ln\frac{n_i\cdot n_j}{2}\Bigr)
  \,.
\end{align}
For SIDIS soft function, the dipole color structure leads to naive Casimir scaling up to N$3$LO.
Indeed, by  color conservation
\begin{align}
 \mathcal{S}^i_{n\bar n}\left(b_\perp\,,\mu\,,\nu\right)
  =
  \exp\Bigl[
  C_i \,\mathbf{s}_{n \bar n}\left(b_\perp\,,\mu\,,\nu\right)
  +\mathcal{O}(\alpha^4_s)
    \Bigr]\,,
    \end{align}
  $C_i=C_F$ for fundamental representation, $C_i=C_A$ for adjoint representation.
We have assumed here that the TMD soft function $ \mathcal{S}^i_{n\bar n}$ takes the same expression no matter $n$ and $\bar n$ being incoming or outgoing. 
In \cite{Collins:2004nx}, it was argued with contour deformation that the TMD soft function takes the same universal expression for Drell-Yan, 
$e^+e^-$ and semi-inclusive deep elastic scattering.
\section{Renormalization Group Evolution and $q_\ast$ Spectrum with RG Improved Perturbation Theory}
\begin{figure}
    \centering
    \includegraphics[width=0.45 \textwidth]{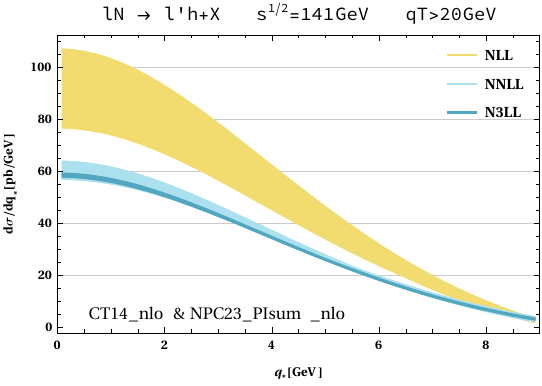}
    \includegraphics[width=0.45 \textwidth]{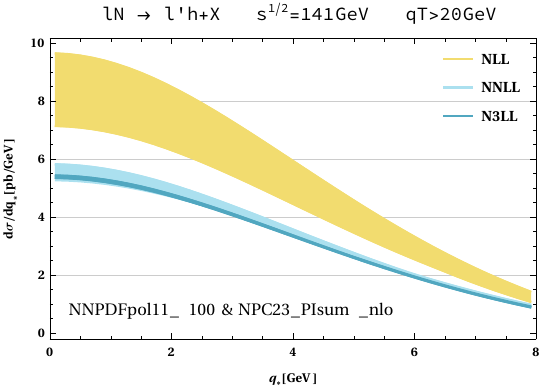}
    \caption{Resummed $q_\ast$ distributions at EIC energies.}
    \label{fig:qy-eic}
\end{figure}

\begin{figure}
    \centering
    \includegraphics[width=0.45 \textwidth]{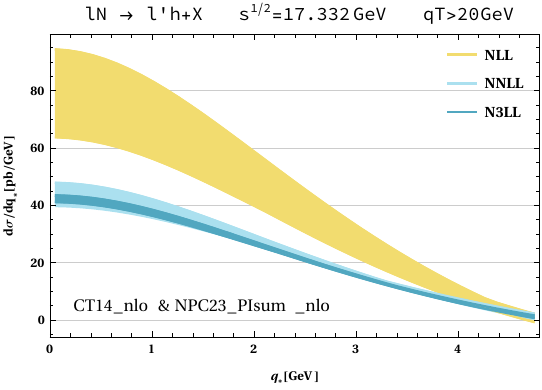}
    \includegraphics[width=0.45 \textwidth]{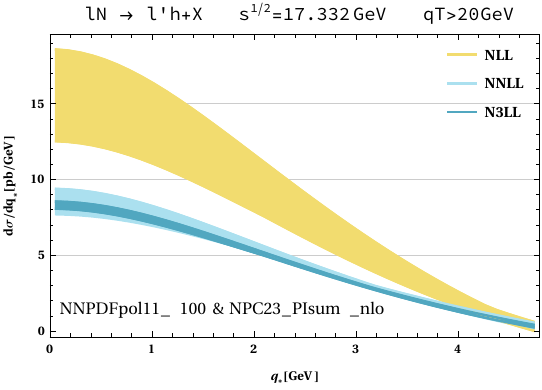}
    \caption{Resummed $q_\ast$ distributions at COMPASS energies.}
    \label{fig:qy-compass}
\end{figure}

We first introduce two conventional integrals in the context of Sudakov problems~\cite{Becher:2006mr}
\begin{align} \label{eq:Kw_def}
K_\Gamma (\mu_0, \mu)
=- \intlim{\alpha_s(\mu_0)}{\alpha_s(\mu)}{\alpha_s} \frac{\Gamma(\alpha_s)}{\beta(\alpha_s)}
   \intlim{\alpha_s(\mu_0)}{\alpha_s}{\alpha_s'} \frac{1}{\beta(\alpha_s')}
\,,
\quad
A_\gamma(\mu_0, \mu)
= -\intlim{\alpha_s(\mu_0)}{\alpha_s(\mu)}{\alpha_s} \frac{\gamma(\alpha_s)}{\beta(\alpha_s)}
\,.\end{align}
From RG consistence,
it follows  that the $\mu$-evolution is linear in $\ln \mu$ to all  orders in $\alpha_s$, for  all the relevant building blocks~\ref{sec:RG}.
The hard Wilson coefficients obey the following  renormalization group equation 
\begin{align}
\label{eq:hardwil-evo}
\frac{\rd |\cC(\mu)\rangle}{\rd \ln \mu} \equiv&\, \widehat\Gamma_H(\mu)  |\cC(\mu)\rangle\,,
\quad
\widehat\Gamma_H(\mu)=\, \gamma_\cusp (\alpha_s) \hDe_H(\mu) +\widehat \gamma_H(\alpha_s)\,,
\end{align}
where the non-cusp piece $\widehat \gamma_H(\alpha_s)$ is proportional to the identity operator
\begin{align}\label{eq:non_cusp_simplify}
\widehat\gamma_H(\alpha_s)=\sum_{i=q,g}\frac{\gamma^H_i}{2}\id \,.
\end{align}
 $ \hDe_H(\mu)$ is a process-dependent color matrix that does not depend on $\alpha_s$
\begin{align}
\hDe_H(\mu)= - \sum_{i<j} {\bf T}_i \cdot {\bf T}_j  \ln\frac{\sigma_{ij}\, s_{ij}- i0 }{\mu^2}\,,
 \end{align}
$\sigma_{ij}=-1$ if both $i$ and $j$ are incoming or outgoing, and $1$ otherwise.
By  color identity 
\begin{equation}
-2\sum_{i<j}\bT_i\cdot\bT_j=\sum_i \bT_i^2 
\end{equation}
we find $ \hDe_H(\mu)- \hDe_H(\mu_0)$ is linear in $\ln\mu$ with trivialized color factors
\begin{align}
 \hDe_H(\mu)- \hDe_H(\mu_0)
 =\id (n_g C_A + n_q C_F) \ln \left( \frac{\mu_0}{\mu} \right)\,.
\end{align}
Consequently, the solution for the evolution  in eq.~(\ref{eq:hardwil-evo}) is
\begin{align}
\hU_H(\mu_0,\mu)
=&
 e^{(n_g C_A + n_q C_F) K_\cusp(\mu_0,\mu)}
\times
 \cP \exp\Bigg[-A_\cusp(\mu_0, \mu)\,
\hDe_H(\mu_0) -  A_H (\mu_0,\mu)\Bigg]
\,.\end{align}
The $\mu$-evolution of the TMD soft  function is~\cite{Kidonakis:1998bk,Kidonakis:1998nf,Aybat:2006mz,Aybat:2006wq}
\begin{align}
  \label{eq:softRG}
  \frac{\rd\mathbf{S}_{1\dots m}}{\rd\ln \mu} = \mathbf{\Gamma}_S^\dagger \cdot \mathbf{S}_{1 \dots m} + \mathbf{S}_{1 \dots m} \cdot \mathbf{\Gamma}_S  \,,
\end{align}
 the anomalous dimension $\mathbf{\Gamma}_S$ takes the form 
\begin{align}
\mathbf{\Gamma}_S= \gamma_\cusp (\alpha_s) \hDe_S(\mu,\nu) +\widehat\gamma_S(\alpha_s)\,,
\end{align}
where $ \hDe_S(\nu;\mu)$ is a process-dependent color matrix that does not depend on $\alpha_s$
\begin{align}
\hDe_S(\nu;\mu)=  \sum_{i<j} {\bf T}_i \cdot {\bf T}_j  \ln\left(\frac{\sigma_{ij}\,  n_i\cdot n_j- i0 }{  2}\right)\left(\frac{\nu^2}{\mu^2}\right)\,,
 \end{align}
and the soft anomalous $\widehat\gamma_S(\alpha_s)$  takes the form
\begin{align}
  \label{eq:Gammas}
 \widehat\gamma_S(\alpha_s)= - \sum_i \frac{1}{2} \gamma^S_i\mathbf{1}\,.
\end{align}
The $\mu$-evolution for the TMD soft correlation functions is solved by
\begin{align}
\hU_S(\nu;\mu_0,\mu)
= &
e^{-(n_g C_A + n_q C_F) K_\cusp(\mu_0,\mu)- A_S (\mu_0,\mu)}
\times
 \cP \exp\Bigg[-A_\cusp(\mu_0, \mu)\,
\hDe_S(\mu_0) \Bigg]
\,.
\end{align}
 The  rapidity  renormalization group (RRGs)~\cite{Chiu:2011qc,Chiu:2012ir} evolution is
\begin{align}
  \label{eq:softnuRG}
  \frac{\rd \mathbf{S}_{1\dots m}}{\rd \ln \nu} =
\frac{1}{2} \left( \mathbf{\Gamma}_R^\dagger(\mu, \mu_b) \cdot \mathbf{S}_{1\dots m} + \mathbf{S}_{1\dots m} \cdot \mathbf{\Gamma}_R(\mu, \mu_b) \right) \,,
\end{align}
where  $\mathbf{\Gamma}_R(\mu,\mu_b)$  denotes the Collins-Soper (CS) kernel~\cite{Collins:1981uk,Collins:1987pm,Collins:2011zzd}
\begin{align}
\label{eq:nuAD}
\mathbf{\Gamma}_R(\mu, \mu_b) = &\, \sum_i   \left(
-\int^{\mu^2}_{\mu_b^2} \frac{d\bar{\mu}^2}{\bar{\mu}^2} \gamma^i_{\rm cusp} [\alpha_s(\bar{\mu})] + \gamma^i_R[\alpha_s(\mu_b)] \right)\,,
\end{align}
its $\mu$-evolution is controlled by  
 the cusp anomalous dimension $\gamma^i_{\rm cusp}$ which originates from the UV renormalization of the cusped Wilson loops~\cite{Brandt:1981kf,Brandt:1982gz}. On the other hand,  the rapidity anomalous dimension $\gamma^i_R(\mu_b)$ has both  perturbative part as $b_T \to 0$ and  genuine non-perturbative part as $b_T \to \infty$. 
 Tremendous progress has been made in the nonperturbative study of the Collins–Soper kernel~\cite{Kang:2024dja,LatticePartonLPC:2022eev,Shanahan:2020zxr,Shanahan:2021tst,Avkhadiev:2024mgd,Schlemmer:2021aij,Li:2021wvl,LatticeParton:2020uhz}, and perturbative accuracy has now reached N{}$^4$LL~\cite{Li:2016ctv,Moult:2022xzt,Duhr:2022yyp}. Up to N{}$^3$LL, naive Casimir scaling holds and one may write, e.g.,
 \begin{align}
\gamma_i^S=C_i \gamma^S,\quad \gamma^i_R=C_i \gamma_R\,.
\end{align}
To  N${}^4$LL strict Casimir scaling is violated due to the presence of the quartic Casimirs, requiring a 
generalized  scaling scheme~\cite{Boels:2017ftb,
Boels:2017fng,Boels:2017skl,Lee:2019zop,Lee:2021uqq,Lee:2022nhh,Das:2019btv,Das:2020adl,Moch:2018wjh,Henn:2019rmi,Grozin:2017css,
vonManteuffel:2020vjv,Agarwal:2021zft,Moult:2022xzt,Duhr:2022yyp}. 
The rapidity RG for TMD soft function is solved by
\begin{align}
&U_R(\nu_S,\nu;\mu_S)=
\exp\left[
\int_{ \nu_S}^\nu \mathbf{\Gamma}_R(\mu_S,\mu_b) \rd \ln \bar\nu
\right]
=
\left(\frac{\nu^2}{\nu_S^2}\right)
^
{
(n_g C_A + n_q C_F)
 \left(
  -A_\cusp(\mu_S, \mu_b)+ \frac{1}{2}\gamma_R(\mu_b)
  \right)
  }\,.
\end{align}
The combined solution of $\mu$-RG and $\nu$-RG leads to a two-dimensional kernel, which simultaneously resums large logarithms in factorization and rapidity scales.
A convenient scale evolution scheme is to evolve the hard and soft sectors from their natural scales to the common factorization scale $(\mu\,,\nu)$
which we set to be the natural scales of the TMD beam functions,
this leads to the following evolution kernels for perturbative  resummation 
\begin{align}
\label{eq: evolve-3 jets}
&\hU_H\hU^\dagger_H(\mu_H,\mu)
\hU_S\hU^\dagger_S(\nu;\mu_S,\mu)
U_R(\nu_S,\nu;\mu_S)
=
\exp
\bigg[
-2A_H(\mu_H,\mu)
-
2A_S(\mu_S,\mu)
\nn\\
+&
(n_g C_A + n_q C_F)
\left(2 K_\cusp(\mu_H,\mu_S)+
\left(\frac{1}{2}\gamma_R(\mu_b)
-A_\cusp (\mu_S,\mu_b)
\right) \ln \frac{\nu^2}{\nu_S^2}
\right)-A_\cusp(\mu_H,\mu_S)
\nn\\
\times&
\left(-2\sum_{i<j} \bT_i\cdot \bT_j \ln \frac{s_{ij}}{\mu_H^2}\right)
-
A_\cusp(\mu_S,\mu)
\left(-2\sum_{i<j} \bT_i\cdot \bT_j \left(\ln \frac{s_{ij}}{\nu^2}-\ln \frac{n_i\cdot n_j}{2} \right)
\right)
\bigg]\,.
\end{align}
 Eq.~(\ref{eq: evolve-3 jets}) provides a general evolution function  for processes admitting a color-dipole structure,
and is expected to hold for reactions with no more than 3-jets.
RG invariance of the cross section is guaranteed by linear relations among the anomalous dimensions.
Indeed, 
with a bit color algebra  we find
\begin{align}
2 \hDe_H(\mu)
+2\hDe_S(\nu;\mu)
=2\sum_i \gamma^i_{\textrm{cusp}}\ln{\frac{2E_i}{\nu}}\,.
    \end{align}
This is consistent with the RG equations of 
various TMD beam/fragmentation functions $\mathcal{G}_i\,,i\in\{q\,,g\}$ 
\begin{align}
\label{eq:rg-beam}
    \frac{\rd}{\rd\ln\mu}\ln \mathcal{G}_i= -\gamma^i_{\rm cusp}\ln \frac{4 E_i^2}{\nu^2}+\gamma^B_{i} \,,
\end{align}
provided that $\gamma_i^H+\gamma_i^{B}-\gamma_i^{S}=0$ holds to all orders.
The beam/fragmentation functions  are regulator dependent, regulator independent  physical TMDs
are obtained by dressing them   with the cloud of soft gluons  
\begin{align}
f^q_1 \left(x\,,b_\perp\,,\xi^n, \mu\right)=&\,\mathcal{B}_{q/N} (x\,,b_\perp\,,E_n, \mu,\nu)\sqrt { \mathcal{S}_{n\bar n}(b_\perp, \mu,\nu)}\,,
\nn\\
D^q_1\left(z\,,\frac{b_\perp}{z}\,,\xi^{\bar n}, \mu\right)=& \,\mathcal{F}_{h/q}\left(z\,,\frac{b_\perp}{z}\,,E_{\bar n}, \mu,\nu\right) \sqrt{\mathcal{S}_{n\bar n}(b_\perp, \mu,\nu)}\,.
\end{align}
They obey  Collins-Soper equations~\cite{Collins:1981uk,Collins:1987pm,Collins:2011zzd}
\begin{align}
\label{eq:rg-CS}
    \frac{\rd}{\rd\ln \sqrt{ \xi^i}}\ln \mathcal{\widetilde{G}}_i (x_i, b_\perp, \xi^i,\mu)
    \equiv 
    &
   \, \Gamma_R^i(\mu,\mu_b) 
   =-2 A^i_{\rm cusp}(\mu,\mu_b)+\gamma^i_R(\mu_b)\,,
    \nn\\
  \frac{\rd}{\rd\ln\mu }\ln \mathcal{\widetilde{G}}_i (x_i, b_\perp, \xi^i,\mu)
  \equiv 
  &
\,  \Gamma_\mu^i(\mu, \xi^i)
    =\, - \gamma^i_{\rm cusp}(\alpha_s(\mu))\ln\frac{\xi^i}{\mu^2} -\gamma^i_{H} \,,
\end{align}
where $\mu_b=b_0/b, b_0=2e^{-\gamma_E}$,  $\xi^i$ are the Collins-Soper scales~\cite{Collins:1981uk,Collins:1987pm,Collins:2011zzd}.
These equations are consistent with SCET RRGs~\cite{Chiu:2011qc,Chiu:2012ir}   since the rapidity and hard scale in eq.~(\ref{eq:rg-beam}) are  paired.
Solution to the two-dimensional Collins-Soper equations are
\begin{align}
\label{eq:rg-CS-solution}
\frac{\mathcal{\widetilde{G}}_i (x_i, b_\perp, \xi^i,\mu)}{\mathcal{\widetilde{G}}_i (x^i, b_\perp, \xi_0^i,\mu_0)}
=&
\exp\bigg[
\int_{ \mu_0}^{\mu}
 \Gamma_{\bar\mu}^i(\bar\mu, \xi^i)
  \rd \ln \bar \mu
  + 
\int_{ \xi^i_0}^{\xi^i} {\Gamma}_R(\mu_0,\mu_b) \rd \ln \sqrt{\bar {\xi^i}}
\bigg]
\nn\\
=&
e^{-2 K^i_{\rm cusp}(\mu_0,\mu)+A^i_H(\mu_0,\mu)}
\times
\left(
\frac{\xi^i}{\mu^2_0}
\right)^
{
A^i_{\rm cusp}(\mu_0,\mu)
}
\times
\left(
\frac{\xi^i}{\xi^i_0}
\right)^
{
-A^i_{\rm cusp}(\mu_0,\mu_b)
+\frac{1}{2}
\gamma_R^i(\mu_b)
}
\,.
\end{align}
Putting everything together, we obtain a TMD factorization theorem for $q_\ast$ spectrum, expressed directly in terms of physical TMDs
\begin{align}
\label{eq:SIDIS-fac-CS}
x y z \frac{\rd \sigma_{\ell+N\to \ell'+h+X}}{2\rd  q_\ast \rd x\rd y\rd z}
\simeq&\, \sum_q   \int \frac{\rd  b}{2\pi} e^{i  b q_\ry}\,
H_q(Q^2,\mu)
z D_1^{q}\left(z, \frac{b_\perp}{z},\xi^{\bar n}_0,\mu_0\right)
\nn\\
\times
&
\left(
\sigma^{\text{U}}_0\times x f_1^q(x,b_\perp,\xi^{n}_0,\mu_0)
+\lambda_\ell S_{\parallel}\times\sigma^{\text{L}}_0\times x g_1^q(x,b_\perp,\xi^{n}_0,\mu_0)
\right)
\nn\\
\times&
\,\prod_ie^{-2 K^i_{\rm cusp}(\mu_0,\mu)+A^i_H(\mu_0,\mu)}
\times
\left(
\frac{\xi^i}{\mu^2_0}
\right)^
{
A^i_{\rm cusp}(\mu_0,\mu)
}
\times
\left(
\frac{\xi^i}{\xi^i_0}
\right)^
{
-A^i_{\rm cusp}(\mu_0,\mu_b)
+\frac{1}{2}
\gamma_R^i(\mu_b)
}
\,,
\end{align}
where the Collins-Soper scales are
\begin{align}
\label{eq:CS-scale}
\xi^n = Q^2 \frac{x}{x+y-x y}\,,\quad \xi^{\bar n} = Q^2 \frac{x+y-x y}{x}\,.
\end{align}
The factorization framework of eq.~(\ref{eq:SIDIS-fac-CS}) was  pioneered by Collins, Soper, and Sterman (CSS)~\cite{Collins:1981uk,Collins:1981uw,Collins:1984kg} and 
expanded in~\cite{Collins:2011zzd,Ji:2004wu,deFlorian:2001zd,Catani:2000vq,Catani:2010pd,Aybat:2011zv,Ji:2005nu},
and  later reformulated in the framework of Soft 
Collinear Effective Theory~\cite{Becher:2010tm,Becher:2011xn,Becher:2012yn,
Echevarria:2011epo,Echevarria:2012js,Echevarria:2014rua,Ebert:2019tvc,Collins:2017oxh}.
The N$^3$LL  predictions for eq.~(\ref{eq:SIDIS-fac-CS}) is shown in fig.~(\ref{fig:qy-eic},\ref{fig:qy-compass}) for both unpolarized and polarized distributions, where
we incorporate the non-perturbative model given in Ref~\cite{Sun:2014dqm}
\begin{figure}
    \centering
   \includegraphics[width=0.45 \textwidth]{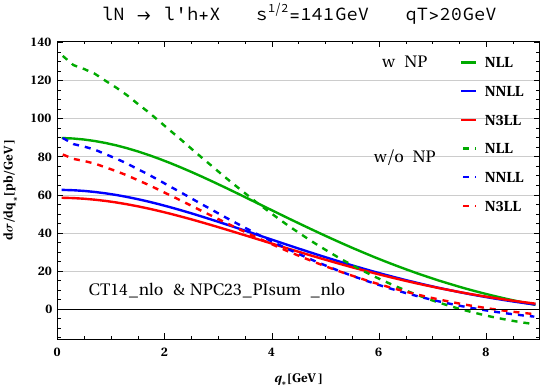}
    \includegraphics[width=0.45 \textwidth]{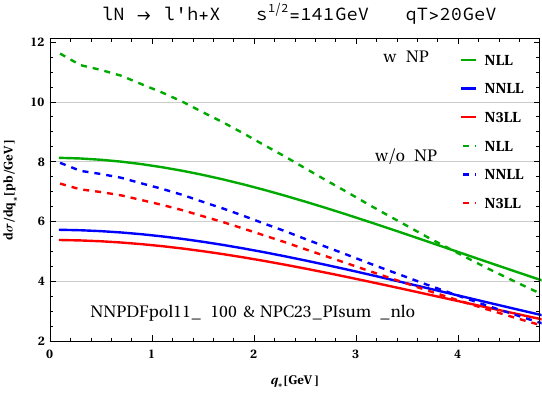}
    \caption{Impact of non-perturbative model function  of BLNY18~\cite{Sun:2014dqm} for pion $q_\ast$-distribution. 
    Here, all the results are implemented with $b_\ast$ prescription, for otherwise the integration diverges.
    }
    \label{fig:qy-np}
\end{figure}
\begin{align}
\label{eq:syy2}
S_{\text{NP}}^{\text{SIDIS}}=
\left({g_1 \over 2 }+ {g_h \over z_h^2}
\right) b^2
+ g_2 \ln\left(\frac{b}{b_\ast}\right)\ln\left(\frac{Q}{Q_0}\right)
\,,
\nn\\
\end{align}
and  accordingly,  the  $b^\ast$ prescription~\cite{Collins:2014jpa} to  regulate Landau poles
\begin{align}
\label{eq:bstar}
b\to b^\ast=\frac{b}{\sqrt{1+b^2/b^2_{\text{max}}}}\,,\quad b_{\text{max}}=1.5\text{GeV}^{-1}\,.
\end{align}
 The impact of the non-perturbative model is shown in fig.~(\ref{fig:qy-np}). 
\section{Twist-2 Matching of  helicity TMDs}
The factorization formulae in eq.~(\ref{eq:SIDIS-fac-CS}) does not require a strict scale hierarchy between $q_\ast$ and $\Lambda_{\text{QCD}}$.
However when $Q\gg q_\ast\gg\Lambda_{\text{QCD}}$,
we will have two equivalent constructions of the effective field theory that has 
 dynamic degrees of freedom carrying  momenta of order $\Lambda_{\text{QCD}}$.
To obtain the EFTs associated with the scale hierarchy  $Q\gg q_\ast\gg\Lambda_{\text{QCD}}$ to the leading power,
one may either integrate out momenta of  off-shellness $\simeq q_\ast\gg \Lambda_{\text{QCD}}$ from the TMDs,
or instead integrate out the modes above $\Lambda_{\text{QCD}}$ once for all.
This corresponds to twist-2 matching onto the PDFs/FFs, 
with DGLAPs resumming   large logarithms associated with $\Lambda_{\text{QCD}}$.
Owing to the consistency  between TMD factorization and the standard collinear factorization,
TMDs  admit  matching onto conventional PDFs/FFs,
we derive their N$^3$LO matching coefficients in this section, which provide the boundary conditions for the numerical results presented in Figs.~(\ref{fig:qy-eic},\ref{fig:qy-compass}).
\subsection{Operator definitions for  the TMD helicity  PDFs and FFs}
\label{sec:defin-helicity}
The TMD helicity parton distribution  functions (TMD helicity PDFs) can be defined in terms of SCET~\cite{Bauer:2000ew,Bauer:2000yr,Bauer:2001yt,Bauer:2002nz,Beneke:2002ph} collinear fields
\begin{align}
  \label{eq:PDFdef}
   \mathcal{B}_{q/N}^{\rm bare}(x,b_\perp,\lambda_N) = &
\int \frac{db^-}{2\pi} \, e^{-i x b^- P^{+}} 
 \langle N(P),\lambda_N | \bar{\chi}_n(0,b^-,b_\perp) \frac{\slashed{\bar{n}}}{2}\gamma_5 \chi_n(0) | N(P) ,\lambda_N\rangle \, ,
  \nn\\
  {\cal B}_{g/N}^{{\rm bare}}(x,b_\perp,\lambda_N) =&
   - x P_+ \int \frac{d b^-}{2 \pi} e^{- i x b^- P^+} \langle N (P),\lambda_N | \widetilde{\cal A}_{n \perp}^{a,\mu} (0, b^-, b_\perp) {\cal A}_{n \perp}^{a,\mu}(0) | N(P),\lambda_N \rangle \,,
\end{align}
where $N(P)$ is a hadron state with momentum $P^\mu = (\bar{n} \cdot P) n^\mu/2 = P^+ n^\mu/2$, and $n^\mu = (1, 0, 0, 1)$ and $\bar{n}^\mu = (1, 0, 0, -1)$.
 $\chi_n = W_n^\dagger \xi_n$ is the gauge invariant collinear quark field~\cite{Bauer:2001ct} in SCET, 
constructed from collinear quark field $\xi_n$ and path-ordered collinear Wilson line $W_n(x) = {\cal P} \exp \left(i g \int_{-\infty}^0 ds\, \bar{n} \cdot A_n (x + \bar{n} s) \right)$.
$A_{n\perp}^{a,\mu}$ is the gauge invariant collinear gluon field with color index $a$ and Lorentz index $\mu$, and the dual collinear gluon field $\widetilde A_{n\perp}^{a,\mu}$ is defined in terms of the transverse Levi-Civita tensor
\begin{align}
\widetilde A_{n\perp}^{a,\mu}\equiv i \e_{\perp}^{\mu \nu} A_{n\perp,\nu}^{a}\,,
\quad \text{and}\quad \e_{\perp}^{\mu \nu}=\frac{P_\sigma \bar n_\rho}{P \cdot \bar n} \e^{\mu \nu \sigma \rho}\,.
\end{align} 
The  helicity distribution function depends on the target helicity $\lambda_N=\pm$, 
however, by rotation invariance and parity symmetry, the dependence is trivially linear in the target helicity,
and we define $ \Delta\mathcal{B}_{i/N}$  through 
\begin{align}
\label{eq:helicity-fac}
\lambda_N \Delta\mathcal{B}_{i/N}^{\rm bare}(x,b_\perp) \equiv&\,\mathcal{B}_{i/N}^{\rm bare}(x,b_\perp,\lambda_N) \,.
\end{align}
For sufficiently small $b_\perp$, the TMD PDFs in Eq.~\eqref{eq:PDFdef} admit operator product expansion onto the  collinear helicity PDFs,
\begin{align}
  \label{eq:PDFOPE}
   \Delta \mathcal{B}_{i/N}^{\rm bare}(x,b_\perp) =& \sum_i  \int_x^1 \frac{d\xi}{\xi} \, \Delta\mathcal{I}_{ij}^{\rm bare}(\xi,b_\perp) \,\Delta \phi_{j/N}^{\rm bare}(x/\xi) + \mathcal{O}(\Lambda_{\text{QCD}} b_\perp) \, ,
  \end{align}
where the summation is over all parton flavors $i$. 
The perturbative matching coefficients $\Delta\mathcal{I}_{ij}^{\rm bare}(\xi,b_\perp)$ in Eq.~\eqref{eq:PDFOPE} are independent of the  hadronic state $N$, 
as a result, one can replace the hadron $N$ with a partonic state $j$ and  compute the matching coefficients within perturbation theory. 
By inserting a complete set of $n$-collinear state $\mathbb{1} = \SumInt_{X_n}  \! d {\rm PS}_{X_n}      | X_n \rangle \langle X_n |$ into the operator definition,
the bare  TMD PDFs can be computed from splitting amplitudes integrated over collinear phase space 
\begin{align}
\label{eq:amp-id1}
\Delta{\cal B}_{ij}^{{\rm bare}} (x, b_\perp)= \lim_{\tau\to0}&\SumInt_{X_n}  \! d {\rm PS}_{X_n} 
e^{- i K_\perp \!\cdot b_\perp} 
e^{ -b_0 \tau \frac{P \cdot K}{P^+}}
\delta(K^+ - (1 - x) P^+) 
   \Delta{\rm \bold P}^{\sigma \rho}_{i \leftarrow j}
   \bar\Gamma_{\sigma \rho}\,,
\end{align}
where $K^\mu$ is the total momentum of $|X_n \rangle$, and $d\text{PS}_{X_n}$ is the collinear phase space measure,
and $\Delta{\rm \bold P}^{\sigma \rho}_{i \leftarrow j}$ is the space-like spin correlator in the target helicity space
\begin{align}
\label{eq:sp-id}
 \Delta{\rm \bold P}^{\sigma \tau}_{q \leftarrow j}
  \equiv
  {\rm \bold{Sp}}_{X_n q_{l}^* \leftarrow j_\sigma}^{*} 
  \Gamma^{l s}
  {\rm \bold{Sp}}_{X_n q_{s}^* \leftarrow j_\rho} ^{}
\equiv
 \langle j_\sigma | \bar{\chi}_n| X_n \rangle   \frac{\slashed{\bar{n}}}{2}\gamma_5  \langle X_n |  \chi_n| j_\rho\rangle
\nn\\
  \Delta{\rm \bold P}^{\sigma \tau}_{g \leftarrow j}
  \equiv
  {\rm \bold{Sp}}_{X_n g_{\mu}^* \leftarrow j_\sigma}^{*} 
  \Gamma^{\mu\nu}
  {\rm \bold{Sp}}_{X_n g_{\nu}^* \leftarrow j_\rho} ^{}
\equiv
-x P_+
 \langle j_\sigma |  {\cal A}^{n \perp}_{a,\mu}| X_n \rangle  i\e^{\mu \nu}_{\perp}    \langle X_n |        {\cal A}^{n \perp}_{a,\nu}| j_\rho\rangle
 \,.
\end{align}
The spin density matrices for the partons in Eq.~(\ref{eq:sp-id}) are 
\begin{align}
\label{eq:defin-spin-proj}
  \Gamma^{l s}=\left[\frac{\slashed{\bar{n}}}{2}\gamma_5 \right]^{ls}\,,
  \quad
\text{and}
  \quad
    \Gamma^{\mu\nu}= i\e^{\mu \nu}_{\perp}\,.
\end{align}
According to eq.~(\ref{eq:helicity-fac}),  the target  spin projectors $\bar\Gamma$ should be given by 
 \begin{align}
 \label{eq:defin-spin-proj2}
 \bar \Gamma^{l s}=\cN_{\text{sch.}} \left[\frac{\slashed{P}}{2}\gamma_5 \right]^{ls}\,,
  \quad
\text{and}
  \quad
  \bar  \Gamma^{\mu\nu}=\cN_{\text{sch.}} \frac{i\e^{\mu \nu}_{\perp}}{2}\,,
\end{align}
 where the factor $ \cN_{\text{sch.}}$ depends on the definition of $\gamma_5$ in dimensional regularization.
Furthermore,  the integral in Eq.~(\ref{eq:amp-id1}) requires a rapidity cutoff to be well defined. In this work, 
we implement such a cutoff via the exponential rapidity regulator $e^{ -b_0 \tau \frac{P \cdot K}{P^+}}$~\cite{Li:2016axz,Luo:2019hmp},
which effectively  restricts the total energies of the collinear radiations by a rapidity scale  $\nu=1/\tau$. 
The TMD FFs are defined as crossings of TMD beam functions,  
and according to parton-hardron frame relation  given in eq.~(\ref{eq: parton-hadron-frame})~\cite{Collins:2011zzd,Luo:2019hmp,Luo:2019bmw}, 
we have,  in dimensional regularization
\begin{align}
  \label{eq:FF_hadron_Frame}
\Delta{\cal F}_{N/q}^{\rm bare} (z, b_\perp/z)  =& z^{1 - 2 \e} \SumInt_{X_n}    \int \frac{db^-}{2\pi}   e^{iP^+  b^-  / z }  
  \langle 0 | 
 \bar \chi_{n}(0,b^-,b_\perp) |  N(P),X \rangle \frac{\slashed{\bar{n}}}{2}\gamma_5 \langle  N(P),X | \chi_{n}(0) | 0 \rangle  \,,
 \nn\\
 \Delta {\cal F}_{N/g}^{{\rm bare}} (z, b_\perp/z) = &
- \frac{P_+}{z^{2\e}} \SumInt_{X_n}  \int \frac{db^-}{2 \pi} 
e^{i P^+  b^- /z} \langle 0 |
\widetilde{\cal A}_{n\perp}^{a,\mu}(0, b^-, b_\perp) | N(P), X \rangle
\langle N(P), X | {\cal A}_{n\perp}^{a,\mu}(0) | 0 \rangle \,,
\end{align}
where $P^\mu = (\bar n \cdot P) n^\mu/2 =  P^+ n^\mu/2$ is the momentum of the final state detected hadron. 
 Similar to Eq.(\ref{eq:amp-id1}), we can compute the partonic  TMD helicity FFs  by  
 \begin{align}
\label{eq:amp-id2}
\Delta{\cal F}_{ij}^{{\rm bare}} (z, b_\perp/z)=z^{1 - 2 \e} \lim_{\tau\to0}&\SumInt_{X_n}  \! d {\rm PS}_{X_n} 
e^{- i K_\perp \!\cdot b_\perp} 
e^{ -b_0 \tau \frac{P \cdot K}{P^+}}
\delta\left(K^+ - \left(\frac{1}{z}-1\right) P^+\right) 
   \Delta{\rm \bold P}^{T,\sigma \rho}_{i \leftarrow j}
   \bar\Gamma_{\sigma \rho}\,,
\end{align}
 where $ \Delta{\rm \bold P}^{T,\sigma \rho}_{i \leftarrow j}$ is the square of the time-like splitting amplitude,
 which can be obtained from the space-like ones in Eq.~(\ref{eq:sp-id})  by analytical continuation.
The spin projectors $\Gamma$ in Eq.~(\ref{eq:FF_hadron_Frame})  and  the dual spin projectors $\bar \Gamma$  in Eq.~(\ref{eq:amp-id2})
 are formally identical to those of the space-like ones in Eq.~(\ref{eq:defin-spin-proj}), but  $P$ is now    the momentum of the detected hardron.
\subsection{Collinear mass factorization and renormalization group equations}
\label{sec:massFac-RG}
The bare TMD helicity PDFs or FFs have divergencies both of UV  and IR origin,
the UV renormalization, the  zero-bin subtraction and mass-factorization  are technically identical to those of the unpolarized case~\cite{Luo:2020epw},
and are summarized in the following collinear mass factorization formula
\begin{align}
  \label{eq:mass-fac-form}
\frac{1}{Z_i^B}  \frac{\Delta{\cal B }_{ij}^{{\rm bare}}(x,b_\perp)}{\mathcal{S}_{0 \rm b} } = & \sum_k \Delta\mathcal{I}_{i k}(x,b_\perp,\mu,\nu) \otimes \Delta\phi_{kj}(x,\mu)  \,,
\nn\\
\frac{1}{Z_i^B}  \frac{\Delta{\cal F}_{ij}^{{\rm bare}}(z,b_\perp/z)}{\mathcal{S}_{0 \rm b} } = & \sum_k  \Delta d_{ik}(z,\mu)\otimes\Delta \mathcal{C}_{kj} (z, b_\perp/z,\mu,\nu)  \,,
\end{align}
where  $\mathcal{S}_{0 \rm b}(\alpha_s)$ is the bare zero-bin soft function which is the same as the TMD soft function~\cite{Li:2016ctv}.
 $Z_i^B$ (see in Sec.~\ref{sec:RC}) are the multiplicative operator renormalization constants for  TMD  PDFs and FFs.
 $\Delta\mathcal{I}_{ij}$ (and $\Delta\mathcal{C}_{ij}$) are the finite coefficient functions.
$\Delta\phi_{ki}$ (and $\Delta d_{ik}$) are the  partonic helicity PDFs (and FFs), 
they evolve with the helicity splitting functions~\cite{Mertig:1995ny,Vogelsang:1995vh,Vogelsang:1996im,Zijlstra:1993sh,Moch:2014sna,Moch:2015usa,Behring:2019tus,Stratmann:1996hn,Rijken:1997rg}.
Up to $\alpha_s^3$, we have
 \begin{align}
\Delta\phi_{ij}(x, \alpha_s) 
=& \delta_{ij} \delta(1-x) - \frac{\alpha_s}{4 \pi} \frac{{\Delta P}^\zero_{ij}(x)}{\epsilon} 
\nn\\
+&  \left(\frac{\alpha_s}{4 \pi}\right)^2 \bigg[ \frac{1}{2 \epsilon^2} \biggl( \sum_k {\Delta P}^\zero_{ik} \otimes {\Delta P}^\zero_{k j}(x) + \beta_0 {\Delta P}^\zero_{ij} (x)  \biggl) - \frac{1}{2\epsilon} {\Delta P}^\one_{ij}(x)  \bigg]  
\nn \\
+&  \left(\frac{\alpha_s}{4 \pi}\right)^3 \bigg[  \frac{-1}{6 \epsilon^3} \biggl(  \sum_{m \,, k }{\Delta P}^{\zero}_{im} \otimes {\Delta P}^{\zero}_{mk} \otimes {\Delta P}^{\zero}_{k j}(x)  + 3 \beta_0 \sum_k {\Delta P}^{\zero}_{ik} \otimes {\Delta P}^{\zero}_{kj}(x)
\nn \\ 
 +&2 \beta_0^2 {\Delta P}^{\zero}_{ij}(x) \biggl)  
+ \frac{1}{6 \epsilon^2} \biggl( \sum_k {\Delta P}^{\zero}_{ik} \otimes {\Delta P}^{\one}_{kj}(x) 
+ 2 \sum_k {\Delta P}^{\one}_{ik} \otimes {\Delta P}^{\zero}_{kj} (x) 
\nn\\
 +& 2  \beta_0 {\Delta P}^{\one}_{ij}(x) +  2  \beta_1 {\Delta P}^{\zero}_{ij}(x) \biggl)
 -\frac{1}{3 \epsilon} {\Delta P}^\two_{ij}(x) \bigg]
 +\Ord(\alpha_s^3)  \,,
\end{align}
 and for  the FFs
\begin{align}
\Delta d_{ij}(z, \alpha_s) 
=& \delta_{ij} \delta(1-z) - \frac{\alpha_s}{4 \pi} \frac{{\Delta P}^{T\zero}_{ij}(z)}{\epsilon} 
\nn\\
+&  \left(\frac{\alpha_s}{4 \pi}\right)^2 \bigg[ \frac{1}{2 \epsilon^2} \biggl( \sum_k {\Delta P}^{T\zero}_{ik} \otimes {\Delta P}^{T\zero}_{k j}(z) + \beta_0 {\Delta P}^{T\zero}_{ij} (z)  \biggl) - \frac{1}{2\epsilon} {\Delta P}^{T\one}_{ij}(z)  \bigg]  
\nn \\
+&  \left(\frac{\alpha_s}{4 \pi}\right)^3 \bigg[  \frac{-1}{6 \epsilon^3} \biggl(  \sum_{m \,, k }{\Delta P}^{T\zero}_{im} \otimes {\Delta P}^{T\zero}_{mk} \otimes {\Delta P}^{T\zero}_{k j}(z)  + 3 \beta_0 \sum_k {\Delta P}^{T\zero}_{ik} \otimes {\Delta P}^{T\zero}_{kj}(z) 
\nn \\ 
+&2 \beta_0^2 {\Delta P}^{T\zero}_{ij}(z) \biggl)  
+ \frac{1}{6 \epsilon^2} \biggl( 2\sum_k {\Delta P}^{T\zero}_{ik} \otimes {\Delta P}^{T\one}_{kj}(z) 
+  \sum_k {\Delta P}^{T\one}_{ik} \otimes {\Delta P}^{T\zero}_{kj} (z) 
\nn\\
 +& 2  \beta_0 {\Delta P}^{T\one}_{ij}(z) 
+  2  \beta_1 {\Delta P}^{T\zero}_{ij}(z) \biggl)
 -\frac{1}{3 \epsilon} {\Delta P}^{T\two}_{ij}(z) \bigg]
 +\Ord(\alpha_s^3)  \,,
\end{align}
where ${\Delta P}_{ij}^{(n)}$ is the $\text{N}^n$LO space-like helicity splitting function,
which is presently  known to NNLO~\cite{Mertig:1995ny,Vogelsang:1995vh,Vogelsang:1996im,Zijlstra:1993sh,Moch:2014sna,Moch:2015usa,Behring:2019tus}.
 ${\Delta P}_{ij}^{T (n)}$ is the $\text{N}^n$LO time-like helicity splitting function,
which is computed at NLO in the work of Refs~\cite{Stratmann:1996hn,Rijken:1997rg}.

The factorized finite coefficient functions obey the following $\mu$-RG equations
\begin{align}
\frac{\df}{\df \ln\mu} \Delta \cI_{ji}^{ }(x,b_\perp,\mu,\nu) = 2 \bigg[ \Gcusp_j(\alsmu) \ln\frac{\nu}{x P_{+}} +& \gamma^B_j(\alsmu) \bigg] \Delta \cI_{ji}^{}(x,b_\perp,\mu,\nu)
\nn\\
- &2 \sum_k \Delta\cI_{jk}^{}(x,b_\perp,\mu,\nu) \otimes {\Delta P}_{ki}(x,\alsmu) \,,
\end{align}
\begin{align}
\frac{\df}{\df \ln\mu}\Delta \cC^{ }_{ij}(z,b_\perp/z,\mu,\nu) = 2 \bigg[ \Gcusp_j(\alsmu) \ln\frac{z\nu}{ P_{+}} +& \gamma^B_j(\alsmu) \bigg] \Delta\cC^{}_{ij}(z,b_\perp/z,\mu, \nu)
\nn\\
-& 2 \sum_k {\Delta P}^T_{ik}(z,\alsmu) \otimes \Delta\cC^{}_{kj}(z,b_\perp/z,\mu,\nu) \, ,
\label{eq:Imu}
\end{align}
and the rapidity evolution equations~\cite{Chiu:2011qc,Chiu:2012ir}
\begin{align}
\frac{\df}{\df\ln\nu}\Delta \cI_{ji}^{}(x,b_\perp,\mu,\nu) =& -2 \left[ \int_{\mu}^{b_0/b_T} \frac{\df\bar{\mu}}{\bar{\mu}} \Gcusp_j(\alpha_s(\bar{\mu})) + \gamma^R_j(\als(b_0/b_T)) \right] \Delta\cI_{ji}^{}(x,b_\perp,\mu,\nu) \, ,
\nn\\
\frac{\df}{\df\ln\nu}\Delta \cC^{}_{ij}(z,b_\perp/z,\mu,\nu) =& -2 \left[ \int_{\mu}^{b_0/b_T} \frac{\df\bar{\mu}}{\bar{\mu}} \Gcusp_j(\alpha_s(\bar{\mu})) + \gamma^R_j(\als(b_0/b_T)) \right] \Delta\cC^{}_{ij}(z,b_\perp/z,\mu, \nu) \,.
\label{eq:Inu}
\end{align}
We present the coefficient function oder-by-order  in terms of  strong coupling $\alpha_s(\mu)/(4 \pi)$,
up to $\Ord(\alpha_s^3)$, the solution to above evolution equations reads
\begin{align}
\label{eq:RGs-1}
\Delta\cI^\zero_{ji}(x,b_\perp,&\,\mu,\nu) = \delta_{ji} \delta(1-x) \, ,\nn
\\
\Delta\cI^\one_{ji}(x,b_\perp,&\,\mu,\nu) = \left( - \frac{\Gcusp_0}{2} \Lp L_Q + \gamma_0^B \Lp + \gamma_0^R L_Q \right) \delta_{ji} \delta(1-x) - {\Delta P}_{ji}^\zero(x) \Lp +\Delta I_{ji}^\one(x) \, , \nn
\\
\Delta\cI^\two_{ji}(x,b_\perp,&\,\mu,\nu) =  \bigg[ \frac{1}{8} \left( -\Gcusp_0 L_Q + 2\gamma^B_0 \right) \left( -\Gcusp_0 L_Q + 2\gamma^B_0 + 2\beta_0 \right) \Lp^2
\nn\\
+& \left(  (-\Gcusp_0 L_Q + 2\gamma^B_0 + 2\beta_0) \frac{\gamma_0^R}{2} L_Q 
-\frac{\Gcusp_1}{2} L_Q + \gamma^B_1  \right) \Lp 
\nn\\
+& \frac{(\gamma_0^R)^2}{2} L_Q^2 + \gamma_1^R L_Q \bigg] \, \delta_{ji} \delta(1-x) 
+ \bigg( \frac{1}{2} \sum_l {\Delta P}^\zero_{jl} \otimes {\Delta P}^\zero_{li}(x) \nn
\\
+& \frac{{\Delta P}^\zero_{ji}(x)}{2} (\Gcusp_0 L_Q - 2\gamma_0^B - \beta_0) \bigg) \Lp^2 + \bigg[ -{\Delta P}^\one_{ji}(x) - {\Delta P}^\zero_{ji}(x) \gamma_0^R L_Q 
\nn\\
- &\sum_l  \Delta I^\one_{jl} \otimes {\Delta P}^\zero_{li}(x) \nn
+ \left( -\frac{\Gcusp_0}{2} L_Q + \gamma_0^B + \beta_0 \right) \Delta I^\one_{ji}(x) \bigg] \Lp 
\nn\\
+& \gamma_0^R L_Q \Delta I^\one_{ji}(x) + \Delta I^\two_{ji}(x) \, ,\nn
\\
\Delta\cI^\three_{ji}(x,b_\perp,&\,\mu,\nu) = \Lp^3\bigg[
\left(\frac{1}{2}\beta_0+\frac{1}{4}(2 \gamma^B_0-\Gcusp_0 L_Q) \right) \sum_{l}{\Delta P}^\zero_{jl} \otimes {\Delta P}^\zero_{li}(x) 
\nn\\
-&\frac{1}{6} \sum_{l\,k } {\Delta P}^\zero_{jl} \otimes {\Delta P}^\zero_{lk} \otimes {\Delta P}^\zero_{ki}(x) 
+\delta_{ji} \delta(1-x) \left( \frac{1}{6}\beta_0^2(2 \gamma^B_0-\Gcusp_0 L_Q)\right.
\nn\\
+&\left.\frac{1}{8} \beta_0(2 \gamma^B_0-\Gcusp_0 L_Q)^2+\frac{1}{48} (2 \gamma^B_0-\Gcusp_0 L_Q)^3 \right) \nn
\\
+&{\Delta P}^\zero_{ji}  \left( -\frac{1}{2} \beta_0 (2 \gamma^B_0-\Gcusp_0 L_Q)-\frac{1}{3}\beta_0^2-\frac{1}{8} (2 \gamma^B_0-\Gcusp_0 L_Q)^2 \right) \nn
\bigg]
\\
+&\Lp^2\bigg[ 
\left(-\frac{3}{2} \beta_0-\frac{1}{2}(2 \gamma^B_0-\Gcusp_0 L_Q) \right)\sum_l \Delta I^\one_{jl} \otimes {\Delta P}^\zero_{li} (x) 
\nn\\
+&\frac{1}{2} \sum_{l\,k } \Delta I^\one_{jl} \otimes {\Delta P}^\zero_{lk} \otimes {\Delta P}^\zero_{ki}(x)
+\frac{1}{2} \sum_l {\Delta P}^\zero_{jl} \otimes {\Delta P}^\one_{li}(x)
\nn\\
+&{\Delta P}^\zero_{ji}(x) \left(-\frac{1}{2}\beta_1-\frac{1}{2}(2 \gamma^B_1-\Gcusp_1 L_Q) \right)
+\delta_{ji} \delta(1-x)  \left ( \frac{1}{4} \beta_1(2 \gamma^B_0-\Gcusp_0 L_Q)\right.
\nn\\
+&\left.\frac{1}{2}\beta_0(2 \gamma^B_1-\Gcusp_1 L_Q)+\frac{1}{4}(2 \gamma^B_0-\Gcusp_0 L_Q)(2 \gamma^B_1-\Gcusp_1 L_Q)
 \right)  \nn
\\
+&\Delta I^\one_{ji}(x) \left( \frac{3}{4} \beta_0(2 \gamma^B_0-\Gcusp_0 L_Q)+\beta_0^2+\frac{1}{8}(2 \gamma^B_0-\Gcusp_0 L_Q)^2  \right)
\nn\\
+&{\Delta P}^\one_{ji}(x) \left(-\beta_0-\frac{1}{2}(2 \gamma^B_0-\Gcusp_0 L_Q) \right)
+\frac{1}{2}\sum_l {\Delta P}^\one_{jl} \otimes {\Delta P}^\zero_{li}(x) 
\bigg] \nn
\\
+&\Lp \bigg[
-\sum_{l }\Delta I^\one_{jl} \otimes {\Delta P}^\one_{li} (x)-\sum_{l }\Delta I^\two_{jl} \otimes {\Delta P}^\zero_{li} (x) -{\Delta P}^\zero_{ji}(x) \gamma^R_1 L_Q
-{\Delta P}^\two_{ji}(x) 
\nn\\
+&\delta_{ji} \delta(1-x)  \biggl( 2 \beta_0  \gamma_1^R L_Q 
 +\frac{1}{2}\gamma_1^R (2 \gamma^B_0-\Gcusp_0 L_Q)L_Q+\frac{1}{2} (2\gamma_2^B-\Gcusp_2 L_Q) \biggl )
 \nn\\
 +&\Delta I^\one_{ji} (x)\left( \beta_1+\frac{1}{2}(2 \gamma^B_1-\Gcusp_1 L_Q) \right) \nn
+\Delta I^\two_{ji} (x)\left(2 \beta_0+\frac{1}{2}(2 \gamma^B_0-\Gcusp_0 L_Q) \right)\bigg]
\nn\\
+&\delta_{ji} \delta(1-x) \gamma^R_2 L_Q+\Delta I^\one_{ji}(x) \gamma^R_1 L_Q+ \Delta I^\three_{ji} (x)\,,
\end{align}
where $\Delta I^{(n)}_{ji} (z)$ are the scale-independent coefficient functions.
 In $\Delta\cI_{ji}^{\three}$ we have used $\gamma_0^R = 0$ to simplify the expression. 
 The scale logarithms are defined by
 \begin{align}
\label{eq:LdefinitionS}
 L_\perp = \ln \frac{b_T^2 \mu^2}{b_0^2} , \quad  L_Q  = 2 \ln \frac{x \,  P_+}{\nu}, \quad L_\nu = \ln \frac{\nu^2}{\mu^2} \,,\quad b_0 =2  e^{- \gamma_E}\,.
\end{align}
Similarly, the RG solution to the fragmentation coefficient functions are
\begin{align}
\label{eq:RGs-2}
\Delta\cC^\zero_{ji}(z,b_\perp/z,&\,\mu,\nu) =  \delta_{ji} \delta(1-z) \, ,\nn
\\
\Delta\cC^\one_{ji}(z,b_\perp/z,&\,\mu,\nu) = \left( - \frac{\Gcusp_0}{2} \Lp L_Q + \gamma_0^B \Lp + \gamma_0^R L_Q \right) \delta_{ji} \delta(1-z) - {\Delta P}_{ji}^{T\zero}(z) \Lp +\Delta C_{ji}^\one(z) \, , \nn
\\
\Delta\cC^\two_{ji}(z,b_\perp/z,&\,\mu,\nu) =  \bigg[ \frac{1}{8} \left( -\Gcusp_0 L_Q + 2\gamma^B_0 \right) \left( -\Gcusp_0 L_Q + 2\gamma^B_0 + 2\beta_0 \right) \Lp^2
\nn\\
+& \left(  (-\Gcusp_0 L_Q + 2\gamma^B_0 + 2\beta_0) \frac{\gamma_0^R}{2} L_Q 
-\frac{\Gcusp_1}{2} L_Q + \gamma^B_1  \right) \Lp 
\nn\\
+& \frac{(\gamma_0^R)^2}{2} L_Q^2 + \gamma_1^R L_Q \bigg] \, \delta_{ji} \delta(1-z) 
+ \bigg( \frac{1}{2} \sum_l {\Delta P}^{T\zero}_{jl} \otimes {\Delta P}^{T\zero}_{li}(z) \nn
\\
+& \frac{{\Delta P}^{T\zero}_{ji}(z)}{2} (\Gcusp_0 L_Q - 2\gamma_0^B - \beta_0) \bigg) \Lp^2 + \bigg[ -{\Delta P}^{T\one}_{ji}(z) - {\Delta P}^{T\zero}_{ji}(z) \gamma_0^R L_Q 
\nn\\
- &\sum_l {\Delta P}^{T\zero}_{jl} \otimes \Delta C^\one_{li}(z) \nn
+ \left( -\frac{\Gcusp_0}{2} L_Q + \gamma_0^B + \beta_0 \right)\Delta C^\one_{ji}(z) \bigg] \Lp
\nn\\
 +& \gamma_0^R L_Q \Delta C^\one_{ji}(z) + \Delta C^\two_{ji}(z) \, ,\nn
\\
\Delta\cC^\three_{ji}(z,b_\perp/z,&\,\mu,\nu) = \Lp^3\bigg[
\left(\frac{1}{2}\beta_0+\frac{1}{4}(2 \gamma^B_0-\Gcusp_0 L_Q) \right) \sum_{l}{\Delta P}^{T\zero}_{jl} \otimes {\Delta P}^{T\zero}_{li}(z) 
\nn\\
-&\frac{1}{6} \sum_{l\,k } {\Delta P}^{T\zero}_{jl} \otimes {\Delta P}^{T\zero}_{lk} \otimes {\Delta P}^{T\zero}_{ki}(z) 
+\delta_{ji} \delta(1-z) \left( \frac{1}{6}\beta_0^2(2 \gamma^B_0-\Gcusp_0 L_Q)\right.
\nn\\
+&\left.\frac{1}{8} \beta_0(2 \gamma^B_0-\Gcusp_0 L_Q)^2+\frac{1}{48} (2 \gamma^B_0-\Gcusp_0 L_Q)^3 \right) \nn
\\
+&{\Delta P}^{T\zero}_{ji}  \left( -\frac{1}{2} \beta_0 (2 \gamma^B_0-\Gcusp_0 L_Q)-\frac{1}{3}\beta_0^2-\frac{1}{8} (2 \gamma^B_0-\Gcusp_0 L_Q)^2 \right) \nn
\bigg]
\\
+&\Lp^2\bigg[ 
\left(-\frac{3}{2} \beta_0-\frac{1}{2}(2 \gamma^B_0-\Gcusp_0 L_Q) \right)\sum_l {\Delta P}^{T\zero}_{jl} \otimes \Delta C^\one_{li} (z) 
\nn\\
+&\frac{1}{2} \sum_{l\,k } {\Delta P}^{T\zero}_{jl} \otimes {\Delta P}^{T\zero}_{lk} \otimes\Delta  C^\one_{ki}(z)
+\frac{1}{2} \sum_l {\Delta P}^{T\zero}_{jl} \otimes {\Delta P}^{T\one}_{li}(z)
\nn\\
+&{\Delta P}^{T\zero}_{ji}(z) \left(-\frac{1}{2}\beta_1-\frac{1}{2}(2 \gamma^B_1-\Gcusp_1 L_Q) \right)
+\delta_{ji} \delta(1-z)  \left ( \frac{1}{4} \beta_1(2 \gamma^B_0-\Gcusp_0 L_Q)\right.
\nn\\
+&\left.\frac{1}{2}\beta_0(2 \gamma^B_1-\Gcusp_1 L_Q)+\frac{1}{4}(2 \gamma^B_0-\Gcusp_0 L_Q)(2 \gamma^B_1-\Gcusp_1 L_Q)
 \right)  \nn
\\
+&\Delta C^\one_{ji}(z) \left( \frac{3}{4} \beta_0(2 \gamma^B_0-\Gcusp_0 L_Q)+\beta_0^2+\frac{1}{8}(2 \gamma^B_0-\Gcusp_0 L_Q)^2  \right)
\nn\\
+&{\Delta P}^{T\one}_{ji}(z) \left(-\beta_0-\frac{1}{2}(2 \gamma^B_0-\Gcusp_0 L_Q) \right)
+\frac{1}{2}\sum_l {\Delta P}^{T\one}_{jl} \otimes {\Delta P}^{T\zero}_{li}(z) 
\bigg] \nn
\\
+&\Lp \bigg[
-\sum_{l } {\Delta P}^{T\one}_{jl} \otimes\Delta C^\one_{li} (z)-\sum_{l } {\Delta P}^{T\zero}_{jl} \otimes \Delta C^\two_{li} (z) -{\Delta P}^{T\zero}_{ji}(z) \gamma^R_1 L_Q
\nn\\
-&{\Delta P}^{T\two}_{ji}(z) 
+\delta_{ji} \delta(1-z)  \biggl( 2 \beta_0  \gamma_1^R L_Q 
 +\frac{1}{2}\gamma_1^R (2 \gamma^B_0-\Gcusp_0 L_Q)L_Q+\frac{1}{2} (2\gamma_2^B-\Gcusp_2 L_Q) \biggl )
 \nn\\
 +&\Delta C^\one_{ji} (z)\left( \beta_1+\frac{1}{2}(2 \gamma^B_1-\Gcusp_1 L_Q) \right) \nn
+\Delta C^\two_{ji} (z)\left(2 \beta_0+\frac{1}{2}(2 \gamma^B_0-\Gcusp_0 L_Q) \right)\bigg]
\nn\\
+&\delta_{ji} \delta(1-z) \gamma^R_2 L_Q+\Delta C^\one_{ji}(z) \gamma^R_1 L_Q+\Delta C^\three_{ji} (z)\,.
\end{align}
The anomalous dimensions appeared above are identical to those in the space-like ones and we have  suppressed their dependence on the exact flavor.
The logarithms appeared in the fragmentation coefficient functions are defined as
\begin{align}
\label{eq:LdefinitionT}
 L_\perp = \ln \frac{b_T^2 \mu^2}{b_0^2} , \quad L_Q = 2 \ln \frac{ P_+}{ z \, \nu}, \quad L_\nu = \ln \frac{\nu^2}{\mu^2} \,,\quad b_0 =2  e^{- \gamma_E}\,,
\end{align}
which differ from thsoe in Eq.~\eqref{eq:LdefinitionS} only in $L_Q$.
Both space-like and time-like coefficient functions depend on the rapidity regulator being used. 
The physical and rapidity-regulator-independent coefficient functions can be obtained by 
multiplying the genuine collinear coefficient functions with the squared root of the TMD soft functions ${\cal S}(b_\perp, \mu, \nu)$~\cite{Luo:2019hmp,Luo:2019bmw}
\begin{align}
  \label{eq:TMD- PDF-FF}
 \Delta \mathscr{I}_{ij}(x, b_\perp, \mu) =&\Delta {\cal I}_{ij}(x, b_\perp, \mu, \nu) \sqrt{{\cal S}_i(b_\perp, \mu, \nu)} \,,
\nn\\
\Delta \mathscr{C}_{ij}(z, b_\perp/z,  \mu) = &\Delta{\cal C}_{ij}(z, b_\perp/z,  \mu, \nu) \sqrt{{\cal S}_j(b_\perp, \mu, \nu)} \,.
\end{align}
\subsection{$\gamma_5$ in dimensional regularization and scheme transformations}
\label{sec:g5}
The computation of polarized TMDs requires careful treatment of the genuinely four-dimensional objects $\gamma_5$ and the Levi-Civita tensor $\epsilon^{\mu\nu\rho\sigma}$, whose definitions must be consistently extended to $D = 4 - 2\epsilon$ dimensions.
However, the anti-commutativity of $\gamma_5$ and the cyclicity of the Dirac trace cannot be simultaneously preserved in dimensional regularization. 
One  approach is to retain the four-dimensional Dirac algebra of $\gamma_5$ by evaluating the trace from a prescribed `reading point'~\cite{Korner:1991sx,Kreimer:1993bh}, 
thereby avoiding ambiguities associated with non-cyclic traces. 
This method is employed in Refs.~\cite{Matiounine:1998re,Rijken:1997rg} to obtain the polarized DIS coefficient functions
and DGLAP splitting functions in $\overline{\text{MS}}$ scheme. 
 The “reading point” prescription is  re-examined in the course of computing helicity TMDs, 
 where we reproduce the two-loop scheme transformation factor $z_{\text{ps}}^{(2)}$, introduced below.
Another approach is to abandon the anti-commutativity of $\gamma_5$, and instead read in $\gamma_5$ from the effective vertex. 
This leads to the use of the \texttt{HVBM} scheme~\cite{tHooft:1972tcz,Breitenlohner:1977hr}, or the closely related \texttt{Larin} prescription~\cite{Larin:1991tj,Zijlstra:1992kj,Larin:1993tq}, for the consistent treatment of $\gamma_5$ in dimensional regularization.
In   \texttt{HVBM} scheme, one proceeds as~\cite{Ravindran:2003gi} 
\begin{enumerate}
\item [1.] Evaluate integrals first 
\item [a)] Use multilinear property $\textrm{Tr}[\slashed{ l_1} \slashed{l_2}\dots \gamma_5 ]= l_1^{\mu_1} l_2^{\mu_2}\times\dots \textrm{Tr}[\gamma^{\mu_1} \gamma^{\mu_2}\dots \gamma_5 ]$
\item [b)] Evaluate  tensor-like Feynman integrals and phase space integrals in D-dimensions 
\item [2.] Begin explicit definition of $\gamma_5$
\item [c)] Replace the $\gamma_5$-matrix by
\begin{align}
\gamma_\mu\gamma_5=\frac{i}{6}\e_{\mu\rho\sigma\tau}\gamma^{\rho}\gamma^{\sigma}\gamma^{\tau} \quad \text{or} \quad \gamma_5=\frac{i}{24}\e_{\mu\rho\sigma\tau}\gamma^\mu\gamma^{\rho}\gamma^{\sigma}\gamma^{\tau}
\nn
\end{align}
\item [d)] Compute trace of Dirac matrix  in D dimensions 
\item [e)] Contract the Levi-Civita tensors in four dimensions 
\end{enumerate}
In $\texttt{Larin}^{+}$ scheme~\cite{Gutierrez-Reyes:2017glx}, one replaces the $\gamma_5$-matrix by
\begin{align}
\gamma^{+}\gamma_5 \to \frac{i \e_{\perp }^{\alpha\beta}}{2}\gamma_{\alpha}\gamma_{\beta}\,,
\end{align}
and supplement it by the D-dimensional relation
\begin{align}
\e_{\perp }^{\alpha_1\beta_1}\e_{\perp }^{\alpha_2\beta_2}=
-g_{\perp }^{\alpha_1\alpha_2}g_{\perp }^{\beta_1 \beta_2}+g_{\perp }^{\alpha_1\beta_2}g_{\perp }^{\beta_1\alpha_2}\,.
\end{align}
Accordingly, the normalization factors presented in Eq.~(\ref{eq:defin-spin-proj2}) are
\begin{eqnarray}
\cN_{\text{sch.}}=
\left\{\begin{array}{cc}
1              &\texttt{HVBM}\,,
\nn\\
(1-\epsilon)^{-1}(1-2\epsilon)^{-1}& \texttt{Larin}^{+}\,.
\end{array}
\right.
\end{eqnarray}
We have verified that the \texttt{HVBM} 
and $\texttt{Larin}^{+}$  prescriptions yield identical results for helicity TMDs  at $\text{N}^3$LO.
Both  \texttt{HVBM} and $\texttt{Larin}^{+}$ schemes break the anticommutativity of the $\gamma_5$ matrix, 
leading to a renormalization of the non-singlet baryon   current   despite its conservation (number of quarks
minus antiquarks of each flavor)~\cite{Larin:1991tj}, 
and causing a violation of the Adler–Bardeen theorem for the non-renormalization of the axial anomaly beyond one-loop~\cite{Adler:1969er}. 
To restore anticommutativity and the associated Ward identities, it is necessary to introduce additional evanescent counterterms to correct the axial current prior to collinear mass factorization. Alternatively, after mass factorization, the finite coefficient functions and the corresponding collinear PDFs (or FFs) are subjected to a factorization-scheme transformation to obtain the correct expression in $\overline{\text{MS}}$
\begin{align}
\Delta{\cal B}^{\text{sub.}} \equiv\frac{1}{Z^B}  \frac{\Delta{\cal B }^{{\rm bare}}}{\mathcal{S}_{0 \rm b} } = &  \Delta\mathcal{I}_L \otimes \Delta\phi_L
=  \left(\Delta\mathcal{I}_L\otimes Z_i^{-1} \right)\otimes \left(Z_i\otimes\Delta\phi_L \right) 
=   \Delta\mathcal{I} \otimes \Delta\phi\,,
\nn\\
\Delta{\cal D}^{{\text{sub.} }}\equiv\frac{1}{Z^B}  \frac{\Delta{\cal D}^{{\rm bare}}}{\mathcal{S}_{0 \rm b} } = & \Delta d_L\otimes\Delta \mathcal{C}_L 
=\left(\Delta d_L \otimes Z_i^T\right) \otimes\left((Z^{T}_i)^{-1}\otimes\Delta \mathcal{C}_L\right)=\Delta d\otimes\Delta \mathcal{C}\,,
\end{align}
where above we introduce the UV and soft zero-bin subtracted functions $\Delta{\cal B}^{\text{sub.}}$ and $\Delta{\cal D}^{{\text{sub.} }}$, and suppress all the function arguments and matrix indices.
The transformation factors are defined for each of the flavor group indices (see the flavor group decomposition in apendix~\ref{sec:favor-decom}).
For flavor group index $i=\pm,\text{v},\text{s}$,  they allow perturbative expansions of the form  in Mellin-$N$ space
\begin{align}
Z^{i}=1+\sum_n \left(\frac{\alpha_s}{4 \pi}\right)^n z_i^{(n)}\,.
\end{align}
We denote the difference of the splitting functions as $\delta [\Delta P]=\Delta P_{\overline{\text{MS}}}-\Delta P_L$, from DGLAP evolution equations we derive
up to $\text{N}^{3}$LO~\cite{Moch:2014sna}:
\begin{align}
\label{eq:sch-trans1}
\delta[ \Delta P]=&\left(\frac{\alpha_s}{4\pi}\right)^2\left\{
\left[Z_i^{(1)},P_L^{(0)}\right]-\beta_0 Z_i^{(1)}
\right\}
+\left(\frac{\alpha_s}{4\pi}\right)^3
\bigg\{
\left[Z_i^{(1)},P_L^{(1)}\right]+\left[Z_i^{(2)},P_L^{(0)}\right]
\nn\\
+&\left[Z_i^{(1)},P_L^{(0)}\right] Z_i^{(1)}
-\beta_1Z_i^{(1)}+\beta_0\left(
\left(Z_i^{(1)}\right)^2-2 Z_i^{(2)}
\right)
\bigg\}+\cO(\alpha_s^4)\,,
\nn\\
\delta [\Delta P^T]=&\left(\frac{\alpha_s}{4\pi}\right)^2\left\{
\left[P_L^{T,(0)},Z_i^{T,(1)}\right]-\beta_0 Z_i^{T,(1)}
\right\}
+\left(\frac{\alpha_s}{4\pi}\right)^3
\bigg\{
\left[P_L^{T,(1)},Z_i^{T,(1)}\right]+\left[P_L^{T,(0)},Z_i^{T,(2)}\right]
\nn\\
+& Z_i^{T,(1)}\left[P_L^{T,(0)},Z_i^{T,(1)}\right]
-\beta_1Z_i^{T,(1)}+\beta_0\left(
\left(Z_i^{T,(1)}\right)^2-2 Z_i^{T,(2)}
\right)
\bigg\}+\cO(\alpha_s^4)\,,
\end{align}
where   we define the bracket as $[a,b]\equiv a \otimes b-b \otimes a$.
For the non-singlet sectors with i=$\text{ns} (\pm,\text{v})$, we may drop all the brackets and obtain
\begin{align}
\label{eq:sch-trans2}
\delta [\Delta P_{\text{ns}}]=&\left(\frac{\alpha_s}{4\pi}\right)^2\left\{
-\beta_0 z_{\text{ns}}^{(1)}
\right\}
+\left(\frac{\alpha_s}{4\pi}\right)^3
\bigg\{
\beta_0
\left(z_{\text{ns}}^{(1)}\right)^2
-\beta_1z_{\text{ns}}^{(1)}
-2 \beta_0 z_{\text{ns}}^{(2)}
\bigg\}+\cO(\alpha_s^4)\,,
\nn\\
\delta[ \Delta P_{\text{ns}}^T]=&\left(\frac{\alpha_s}{4\pi}\right)^2\left\{
-\beta_0 z_{\text{ns}}^{T(1)}
\right\}
+\left(\frac{\alpha_s}{4\pi}\right)^3
\bigg\{\beta_0
\left(z_{\text{ns}}^{T(1)}\right)^2
-\beta_1z_{\text{ns}}^{T(1)}
-2\beta_0 z_{\text{ns}}^{T(2)}
\bigg\}+\cO(\alpha_s^4)\,.
\end{align}
For the singlet sector with $i=\text{s}$, the transformation factors are $2\times 2$ matrices in Mellin-$N$ space,
but only $q\to q$  entry is non-trivial. Indeed, we have 
\begin{align}
(z_{\text{s}}^{(n)})_{ik}=\delta_{iq}\delta_{kq}
z_{qq}^{(n)}
=
\delta_{iq}\delta_{kq}
\left\{
z^{(n)}_{\text{ns},+}
+z^{(n)}_{\text{ps}}
\right\}
\,,
\end{align}
and 
\begin{align}
\label{eq:Zqq}
Z^{\text{s}}_{ik}=\delta_{ik}+\delta_{iq}\delta_{kq}
\left(
\left(\frac{\alpha_s}{4\pi}\right)
z^{(1)}_{\text{ns},+}
+
\left(\frac{\alpha_s}{4\pi}\right)^2
\left\{
z^{(2)}_{\text{ns},+}
+z^{(2)}_{\text{ps}}
\right\}
+
\left(\frac{\alpha_s}{4\pi}\right)^3
\left\{
z^{(3)}_{\text{ns},+}
+z^{(3)}_{\text{ps}}
\right\}
+\cO(\alpha_s^4)
\right)\,.
\end{align}
Substituting Eq.~(\ref{eq:Zqq}) into Eq.~(\ref{eq:sch-trans1}) , we find for the singlet space-like splitting functions
\begin{align}
\delta[ \Delta P_{qq}^{\text{s}\,(1)}]=&-\beta_0 z_{qq}^{(1)}\,,\quad \delta[ \Delta P_{qg}^{\text{s}\,(1)}]=z_{qq}^{(1)}\otimes \Delta P^{\text{s}\,(0)}_{q g}
\nn\\
\delta[ \Delta P_{gq}^{\text{s}\,(1)}]=&- \Delta P^{\text{s}\,(0)}_{gq}\otimes z_{qq}^{(1)}\,,\quad \delta[ \Delta P_{gg}^{\text{s}\,(1)}]=0\,,
\end{align}
and
\begin{align}
\delta[ \Delta P_{qq}^{\text{s}\,(2)}]=&\beta_0\left[ \left(z_{qq}^{(1)}\right)^2-2z_{qq}^{(2)}\right]-\beta_1 z_{qq}^{(1)}\,,
\nn\\
\delta[ \Delta P_{qg}^{\text{s}\,(2)}]=&z_{qq}^{(2)}\otimes\Delta P^{\text{s}\,(0)}_{qg}+z_{qq}^{(1)}\otimes\Delta P^{\text{s}\,(1)}_{qg,L}\,,
\nn\\
\delta[ \Delta P_{gq}^{\text{s}\,(2)}]=&-
\left(
\Delta P^{\text{s}\,(1)}_{gq,L}-\Delta P^{\text{s}\,(0)}_{gq}\otimes z_{qq}^{(1)}
\right)\otimes z_{qq}^{(1)}
-
\Delta P^{\text{s}\,(0)}_{gq}\otimes z_{qq}^{(2)}\,,
\nn\\
 \delta[ \Delta P_{gg}^{\text{s}\,(2)}]=&0\,.
\end{align}
For the singlet time-like splitting functions, we find
 \begin{align}
\delta[ \Delta P_{qq}^{T,\text{s}\,(1)}]=&-\beta_0 z_{qq}^{T(1)}\,,\quad\delta[ \Delta P_{gq}^{T,\text{s}\,(1)}]=z_{qq}^{T(1)}\otimes \Delta P^{T,\text{s}\,(0)}_{gq}      
\nn\\
 \delta[ \Delta P_{qg}^{T,\text{s}\,(1)}]=&- \Delta P^{T,\text{s}\,(0)}_{qg}\otimes z_{qq}^{T(1)}\,,\quad \delta[ \Delta P_{gg}^{T,\text{s}\,(1)}]=0\,,
\end{align}
and
\begin{align}
\delta[ \Delta P_{qq}^{T,\text{s}\,(2)}]=&\beta_0\left[ \left(z_{qq}^{T(1)}\right)^2-2z_{qq}^{T(2)}\right]-\beta_1 z_{qq}^{T(1)}\,,
\nn\\
\delta[ \Delta P_{gq}^{T,\text{s}\,(2)}]=&
z_{qq}^{T(2)}\otimes\Delta P^{T,\text{s}\,(0)}_{gq}+z_{qq}^{T(1)}\otimes\Delta P^{T,\text{s}\,(1)}_{gq,L}\,,
\nn\\
\delta[ \Delta P_{qg}^{T,\text{s}\,(2)}]=&
-
\left(
\Delta P^{T,\text{s}\,(1)}_{qg,L}-\Delta P^{T,\text{s}\,(0)}_{qg}\otimes z_{qq}^{T(1)}
\right)\otimes z_{qq}^{T(1)}
-
\Delta P^{T,\text{s}\,(0)}_{qg}\otimes z_{qq}^{T(2)}\,,
\nn\\
 \delta[ \Delta P_{gg}^{T,\text{s}\,(2)}]=&0\,.
\end{align}
To extract the scheme transformation factors, 
it is necessary to renormalize physical quantities consistently within both Kreimer’s approach and the \texttt{HVBM} or $\texttt{Larin}^{+}$ schemes to the same perturbative order, 
and define scheme transformations as ratios between them.
In the non-singlet sectors with ns$=\pm$, the two $\gamma_5$-matrices appear in the same trace. 
 One may naively anticommute the two $\gamma_5$-matrices inside the fermion loop to  the same position~\cite{Korner:1991sx,Kreimer:1993bh},
 where they cancel by $\gamma_5^2=1$, leaving  an unpolarized expression.
 The transformation factors are then defined as the ratio between the naive anti-commutative prescription and the \texttt{HVBM} prescription, in Mellin-$N$ space it is
 \begin{align}
 \label{eq:defin-z5-ns}
 Z^{\text{ns},\pm}\equiv\frac{  \Delta\mathcal{I}^{(\mp)}_{\text{naive}}}{  \Delta\mathcal{I}^{(\pm)}_{\texttt{HVBM}}}=1+\sum_{k}\left(\frac{\alpha_s}{4\pi}\right)^k
z^{(k)}_{\text{ns},\pm}\,.
 \end{align}
 With this method we determine
  $Z^{\text{ns}}$ up to three loops, reproducing the known  one- and two-loop results~\cite{Moch:2014sna}.
The first moment of $Z^{\text{ns},+}$ yields
 \begin{align}
\label{Z5}
  Z^{\text{ns},+}|_ {N=1}=&
  1
  - \left(\frac{\alpha_s}{4 \pi}\right)\* 4\, \* C_F 
  +  \left(\frac{\alpha_s}{4 \pi}\right)^2 \* \bigg[ 
            22\, \* C_F
          - \frac{107}{9}\: \* C_F \* C_A
          + \frac{2}{9}\: \* C_F \* N_f
  \bigg]+ \left(\frac{\alpha_s}{4 \pi}\right)^3\* \bigg[
\nonumber \\
  & 
            C_F\* \bigg(
          - \frac{370}{3} 
          + 96\, \* \zeta_3 \bigg)
          + C_F \* C_A \* \bigg( \:
            \frac{5834}{27} 
          - 160\, \* \zeta_3 \bigg)
          + C_F \* C_A \* \bigg(
          - \frac{2147}{27} 
          + 56\, \* \zeta_3 \bigg)
\nonumber \\
          +& C_F \* N_f \* \bigg(
          - \frac{62}{27} 
          - \frac{32}{3}\: \* \zeta_3 \bigg)
          + C_A \* C_F \* N_f \* \bigg( \:
            \frac{356}{81} 
          + \frac{32}{3}\: \* \zeta_3 \bigg)
          + \frac{52}{81}\: \* C_F \* N_f \,
  \bigg]\,,
\end{align}
which is exactly the same finite renormalization constant for the non-singlet axial vector obtained originally by Larin~\cite{Larin:1991tj},
and hence serves as an independent check for our calculation.
 Furthermore, we find at NLO, the Gribov-Lipatov relation breaks down as was originally observed in~\cite{vanNeerven:1998ma,Blumlein:2000wh}
 \begin{align}
 z^{T(2)}_{\text{ns},\pm}(\xi)=-\xi z^{(2)}_{\text{ns},\pm}(1/\xi)+\beta_0 z^{(1)}_{\text{ns},\pm}(\xi)\ln \xi\,.
 \end{align}
For the pure-singlet case, two $\gamma_5$ matrices appear, one in each fermion trace, so that the loops are topologically unconnected. 
 Following the procedure of `reading in' the $\gamma_5$ matrix from the position suggested in Ref.~\cite{Vogelsang:1996im}—namely, by reading in the $\gamma_5$ inside the triangle diagrams (appearing  in $g \to q$ channel at NLO and $q' \to q$ at NNLO)  from the gluon-quark vertex  rather than from the 
projector vertex, and subtracting the collinear poles accordingly,
we reproduce the   NNLO pure-singlet transformation factor $z_{\text{ps}}^{(2)}$~\cite{Matiounine:1998re,Rijken:1997rg} 
as the difference in IR-subtracted coefficient functions  between Kreimer's prescription and the  \texttt{HVBM}  prescription.
The N$^3$LO pure-singlet transformation factor $z_{\text{ps}}^{(3)}$ remains unkown.
 \subsection{The NNLO   helicity-dependent splitting functions}
 \label{sec:nnlo-helicity}
 \subsubsection{Space-like results in ${\overline{\text{MS}}}$}
 The space-like helicity dependent splitting functions were obtained in~\cite{Moch:2014sna,Moch:2015usa,Blumlein:2021enk,Blumlein:2021ryt,Blumlein:2022gpp},
by comparison we find all agreement except for the term with cubic color structure $d_{abc}^2$ for `sea' quark difference $N_f(\Delta P_{qq'}-\Delta P_{q \bar q'})$:
\footnote{Agreement is found at transcendental weight four, whereas discrepancies arise in the weight-three terms.
The author is also grateful to the authors of Ref.~\cite{Behring:2025avs} for pointing out a typo in the $\zeta_2$
 term in the \LaTeX{} output of this formula. Once corrected, the expression is in full agreement with the file \texttt{dPSLarin.m} 
 accompanying the present work and with their independent calculations.}
\begin{dmath}[style={\small},compact]
\Delta P_{d_{abc}^2}^{S,(2)}-\Delta P_{d_{abc}^2}^{\text{Moch:2015}}=
\frac{16 N_f}{3}\frac{d_{abc}^2}{N_c}
\bigg[
-84\zeta_2-18\zeta_3+3H_0+84H_2+36H_3+12H_{-2,0}+
(1-z)\times\left(
174 H_1-12H_{1,0,0}
\right)
+
(1+z)\times\left(
36 H_{-1,0}-48 H_{-1,2} -24 H_{-1,0,0}+48 H_{-1}\zeta_2 -36 H_{0}\zeta_2
\right)
+
z\times\left(
-174 H_0 +90 H_2 +12 H_3 -24 H_{-2,0} +54 H_{0,0} +24 H_{0,0,0} -54 \zeta_2 -72\zeta_3
\right)
\bigg]\,.
\end{dmath}
 It's interesting to investigate the first moment of the helicity splitting functions,
 up to $\text{N}^{3}$LO we found  
 \begin{align}
 \Delta P^{(n)}_{gg}(N=1)=&\beta_n\,,
 \nn\\
 \Delta P_{\text{ns},+}^{(n)}(N=1)=&\Delta P_{qg}^{(n)}(N=1)=0\,,
 \nn\\
\Delta P_{\text{ps}}^{(n)}(N=1)=&-2N_f \Delta P_{gq}^{(n-1)}(N=1)\,.
 \end{align}
 These relations follow from the Adler–Bardeen theorem--ensuring the non-renormalization 
 of the non-singlet axial anomaly in pure QCD~\cite{Adler:1969er}--and from the one-loop exactness of the singlet axial anomaly in QCD~\cite{Larin:1993tq}:
 \begin{align}
 \int_{0}^{1} d\xi \, \xi^{n-1} [ \Delta f_j(\xi) -(-1)^n  \Delta f_{ \bar\jmath} ]=\frac{i^{n-1}}{2 (P^+)^n}
 \langle P|\bar \psi_j \gamma^+\gamma_5(\partial_+)^{n-1} \psi_j | P \rangle_{\text{c}}\,,
\nn\\
 \langle  \psi(x)\slashed\partial\gamma_5 \psi (x)\mathcal{O}(x_1\,,\dots\,,x_n)\rangle = \frac{\alpha_s}{8\pi}
\langle G^a_{\mu\nu}  \widetilde G^{a\mu\nu} (x) \mathcal{O}(x_1\,,\dots\,,x_n) \rangle\,.
  \end{align}

 \subsubsection{Time-like results in ${\overline{\text{MS}}}$}
The NLO time-like helicity-dependent splitting functions were obtained in Refs.~\cite{Stratmann:1996hn,Rijken:1997rg}. Our results extend the precision to NNLO for the first time, reproducing the known NLO expressions. We now present the analytic NNLO results for both  non-singlet and singlet sectors. 
The ‘$\pm$’ non-singlet results coincide with the corresponding unpolarized ones `$\mp$'
\begin{align}
\Delta P_{\text{ns},\pm}^{T,(n)}=P_{\text{ns},\mp}^{T,(n)}\,.
\end{align}
The results for the singlet sectors are
\begin{dmath}[style={\small},compact]
\Delta P_{qq}^{T,\text{s}(0)}=
4 C_F \left[\frac{1}{1-z}\right]_++3 C_F \delta (1-z)+(-2 z-2) C_F\,,
\end{dmath}
\begin{dmath}[style={\small},compact]
\Delta P_{qq}^{T,\text{s}(1)}=
\left[\frac{1}{1-z}\right]_+ \bigg[\left(\frac{268}{9}-8 \zeta _2\right) C_A C_F-\frac{40 C_F N_f}{9}\bigg]
+C_A C_F \bigg[\frac{z^2}{z+1}\left(-8 H_{-1,0}-\frac{446}{9}\right)+\left(-\frac{8 z^2}{(z-1) (z+1)}\right)H_{0,0}+\frac{1}{z+1}\left(\frac{178}{9}-8 H_{-1,0}\right)+\left(-\frac{8}{(z-1) (z+1)}\right)H_{0,0}+\frac{2 z^2}{3 (z-1)}H_0+\left(-\frac{46}{3 (z-1)}\right)H_0+\frac{z \left(8 \zeta _2-\frac{268}{9}\right)}{z+1}\bigg]
+C_F N_f \bigg[z \left(8 H_{0,0}+\frac{440}{9}\right)+8 H_{0,0}+\left(-\frac{80 z^2}{3 (z-1)}\right)H_0+\frac{40}{3 (z-1)}H_0+\frac{16 z}{z-1}H_0-\frac{400}{9}\bigg]
+C_F^2 \bigg[\frac{12 z^3}{(z-1) (z+1)}H_{0,0}+\frac{z^2}{z+1}\left(16 H_{-1,0}+8 \zeta _2+36\right)+\frac{z^2}{z-1}\left(8 H_{0,1}+8 H_{1,0}-20 H_0\right)+\frac{28 z^2}{(z-1) (z+1)}H_{0,0}+\frac{1}{z+1}\left(16 H_{-1,0}+8 \zeta _2-36\right)+\frac{1}{z-1}\left(8 H_{0,1}+8 H_{1,0}+16 H_0\right)+\frac{20}{(z-1) (z+1)}H_{0,0}+\frac{4 z}{(z-1) (z+1)}H_{0,0}+\left(-\frac{8 z}{z-1}\right)H_0\bigg]
+\delta (1-z) \bigg[\left(\frac{44}{3}\zeta _2-12 \zeta _3+\frac{17}{6}\right) C_A C_F+\left(\left(-\frac{8}{3}\right)\zeta _2-\frac{1}{3}\right) C_F N_f+\left(-12 \zeta _2+24 \zeta _3+\frac{3}{2}\right) C_F^2\bigg]\,,
\end{dmath}
\begin{dmath*}[style={\small},compact]
\Delta P_{qq}^{T,\text{s}(2)}=
 C_F N_f^2 \bigg[\left(-\frac{416}{9}\right)H_1+\left(-\frac{152}{9}\right)H_{0,1}+\left(-\frac{32}{3}\right)H_{0,0,0}+\frac{16}{3}H_{0,0,1}+\frac{16}{3}H_{0,1,1}+\frac{40}{3}H_{1,1}+\frac{152}{9}\zeta _2+\frac{64}{3}\zeta _3+\frac{1}{z-1}\left(\frac{8}{3}H_{0,0}+\frac{112}{27}H_0+\frac{16 \zeta _2}{3}H_0\right)+\frac{z}{z-1}\left(\left(-\frac{64}{3}\right)H_{0,0}+\frac{320}{9}H_0\right)+\frac{z^2}{z-1}\left(\left(-\frac{128}{3}\right)H_0+\frac{152}{9}H_{0,0}+\left(-\frac{16 \zeta _2}{3}\right)H_0\right)+40 H_{1,0}+16 H_{0,1,0}+z \left(\left(-\frac{296}{9}\right)H_{0,1}+\left(-\frac{40}{3}\right)H_{1,1}+\left(-\frac{32}{3}\right)H_{0,0,0}+\frac{16}{3}H_{0,0,1}+\frac{16}{3}H_{0,1,1}+\frac{64}{3}\zeta _3+\frac{296}{9}\zeta _2+\frac{416}{9}H_1-40 H_{1,0}+16 H_{0,1,0}+\frac{1952}{27}\right)-\frac{1936}{27}\bigg]
+\left[\frac{1}{1-z}\right]_+ \bigg[C_F \left(\left(-\frac{1072}{9}\right)\zeta _2+\frac{88}{3}\zeta _3+88 \zeta _4+\frac{490}{3}\right) C_A^2+\left(\left(-\frac{112}{3}\right)\zeta _3+\frac{160}{9}\zeta _2-\frac{836}{27}\right) C_F N_f C_A-\frac{16}{27} C_F N_f^2+C_F^2 N_f \left(32 \zeta _3-\frac{110}{3}\right)\bigg]
+\delta (1-z) \bigg[\left(-32 \zeta _3 \zeta _2+18 \zeta _2+68 \zeta _3+144 \zeta _4-240 \zeta _5+\frac{29}{2}\right) C_F^3+\left(\left(-\frac{136}{3}\right)\zeta _3+\frac{20}{3}\zeta _2+\frac{116}{3}\zeta _4-23\right) N_f C_F^2+\left(\left(-\frac{16}{9}\right)\zeta _3+\frac{80}{27}\zeta _2-\frac{17}{9}\right) N_f^2 C_F+C_A^2 \left(\left(-\frac{1552}{9}\right)\zeta _3+\frac{4496}{27}\zeta _2-5 \zeta _4+40 \zeta _5-\frac{1657}{36}\right) C_F+C_A \left(\left(\left(-\frac{494}{3}\right)\zeta _4+\left(-\frac{410}{3}\right)\zeta _2+\frac{844}{3}\zeta _3+16 \zeta _2 \zeta _3+120 \zeta _5+\frac{151}{4}\right) C_F^2+N_f \left(\left(-\frac{1336}{27}\right)\zeta _2+\frac{200}{9}\zeta _3+2 \zeta _4+20\right) C_F\right)\bigg]
\end{dmath*}
\begin{dmath*}[style={\small},compact]
+C_F^3\left[\frac{1}{z-1}\left(96 \zeta _2 H_1-192 \zeta _3 H_1+400 H_1-56 H_{0,1,0}+96 H_{1,0,0}-208 H_{0,0,0,1} -64 H_{0,0,1,1}-48 H_{0,1,0,0}-64 H_{0,1,0,1}-64 H_{0,1,1,0}-128 H_{1,0,-1,0}-128 H_{1,0,0,0}-128 H_{1,0,0,1}-64 H_{1,0,1,0}\right)
+\frac{z}{z-1}\left(-192 \zeta _2 H_1-800 H_1+64 H_{0,1,0}\right)+\frac{z}{z-1}\left(-144 \zeta _2 H_0-72 H_0-12 \zeta _2-192 \zeta _3-324 \zeta _4-352 \zeta _2 H_{0,-1}+144 \zeta _2 H_{0,0}+176 H_{0,0}-96 H_{0,0,0}-64 H_{0,-1,-1,0}+128 H_{0,-1,0,0}-96 H_{0,0,-1,0}-16 H_{0,0,0,0}-64 H_{0,0,1,0}+\frac{1}{z+1}-302\right)+\frac{z^2}{z-1}\left(-16 \zeta _2 H_0+128 \zeta _3 H_0+10 H_0+204 \zeta _2+48 \zeta _3+252 \zeta _4+160 \zeta _2 H_{0,-1}+208 \zeta _2 H_{0,0}+172 H_{0,0}+528 H_{0,0,0}-320 H_{0,-1,-1,0}+192 H_{0,-1,0,0}+288 H_{0,0,-1,0}-592 H_{0,0,0,0}-192 H_{0,0,1,0}+\frac{1}{z+1}-302\right)+\frac{z^2}{z-1}\left(96 \zeta _2 H_1-192 \zeta _3 H_1+400 H_1+88 H_{0,1,0}+96 H_{1,0,0}-240 H_{0,0,0,1}-64 H_{0,0,1,1}-144 H_{0,1,0,0}-64 H_{0,1,0,1}-64 H_{0,1,1,0}-128 H_{1,0,-1,0}-128 H_{1,0,0,0}-128 H_{1,0,0,1}-64 H_{1,0,1,0}\right)+\frac{z^3}{z-1}\left(432 \zeta _2 H_0+114 H_0+12 \zeta _2+336 \zeta _3-28 \zeta _4-352 \zeta _2 H_{0,-1}+48 \zeta _2 H_{0,0}-228 H_{0,0}+240 H_{0,0,0}-64 H_{0,-1,-1,0}-160 H_{0,0,-1,0}-112 H_{0,0,0,0}-128 H_{0,0,1,0}+\frac{1}{z+1}+302\right)+\frac{1}{z+1}\left(-384 H_{-1} \zeta _2+768 H_{-1,-1} \zeta _2-384 H_{-1,0} \zeta _2-96 H_{0,1} \zeta _2-576 H_{-1} \zeta _3-96 H_{-1,0}-12 H_{0,1}+112 H_{-1,0,0}+480 H_{-1,0,1}-64 H_{0,-1,0}+64 H_{0,0,1}-128 H_{-1,-1,0,0}-768 H_{-1,-1,0,1}+128 H_{-1,0,-1,0}
-224 H_{-1,0,0,0}+256 H_{-1,0,0,1}-64 H_{-1,0,1,0}+320 H_{0,-1,0,1}\right)
+\frac{1}{(z-1) (z+1)}\bigg(16 \zeta _2 H_0-64 \zeta _3 H_0-64 H_0-204 \zeta _2-192 \zeta _3-44 \zeta _4+160 \zeta _2 H_{0,-1}+112 \zeta _2 H_{0,0}-224 H_{0,0}-96 H_{0,0,0}-320 H_{0,-1,-1,0}+64 H_{0,-1,0,0}+96 H_{0,0,-1,0}-176 H_{0,0,0,0}-128 H_{0,0,1,0}+302\bigg)
+\frac{z}{z+1}\left(-576 H_{-1} \zeta _2-192 H_{-1,0}-312 H_{0,1}-64 H_{-1,0,0}+768 H_{-1,0,1}-320 H_{0,-1,0}-96 H_{0,0,1}\right)
+\frac{z^2}{z+1}
\bigg(-384 H_{-1} \zeta _2+768 H_{-1,-1} \zeta _2-384 H_{-1,0} \zeta _2-96 H_{0,1} \zeta _2-576 H_{-1} \zeta _3-96 H_{-1,0}-300 H_{0,1}+112 H_{-1,0,0}+480 H_{-1,0,1}-160 H_{0,-1,0}-256 H_{0,0,1}-128 H_{-1,-1,0,0}-768 H_{-1,-1,0,1}+128 H_{-1,0,-1,0}-224 H_{-1,0,0,0}+256 H_{-1,0,0,1}-64 H_{-1,0,1,0}+320 H_{0,-1,0,1}\bigg)
+144 H_{1,0}+192 H_{-1,-1,0}+z \left(192 H_{-1,-1,0}-144 H_{1,0}\right)
\right] 
+C_F^2 N_f \bigg[\frac{668}{3}H_1+\frac{1}{z-1}\left(\left(-\frac{2104}{9}\right)H_{0,1}+\left(-\frac{464}{3}\right)H_{0,1,0}+\left(-\frac{580}{9}\right)H_{1,0}-160 \zeta _2 H_{0,1}\right)+\frac{z}{z-1}\left(\left(-\frac{5048}{9}\right)H_{0,0}
+\left(-\frac{496}{3}\right)H_{0,0,1}+\left(-\frac{80}{3}\right)\zeta _3+\frac{4772}{9}\zeta _2+\frac{7868}{9}H_0+\left(-\frac{112 \zeta _2}{3}\right)H_0-160 H_0 \zeta _3-492 \zeta _4+16 \zeta _2 H_{0,0}-152 H_{0,0,0}+\frac{1}{z+1}-\frac{1727}{3}\right)
+\frac{z}{z-1}\left(\frac{220}{3}H_{0,1}+\frac{280}{3}H_{1,0}+144 H_{0,1,0}\right)
+\frac{z^2}{z-1}\left(\left(-\frac{580}{9}\right)H_{1,0}+\left(-\frac{32}{3}\right)H_{0,1,0}+\frac{1124}{9}H_{0,1}+160 \zeta _2 H_{0,1}\right)
+\frac{z^2}{z-1}\bigg(\left(-\frac{80}{3}\right)\zeta _3+\frac{20}{9}H_0+\frac{56}{3}H_{0,0,0}+\frac{760}{9}H_{0,0}+\frac{1672}{9}\zeta _2-64 H_0 \zeta _2+160 H_0 \zeta _3+492 \zeta _4-16 \zeta _2 H_{0,0}+64 H_{0,0,1}+\frac{1}{z+1}-539\bigg)
+\frac{z^3}{z-1}\bigg(\left(-\frac{7664}{9}\right)H_0+\left(-\frac{5092}{9}\right)\zeta _2+\frac{112}{3}\zeta _3+\frac{368}{3}H_{0,0,1}+\frac{1624}{3}H_{0,0}+\frac{176 \zeta _2}{3}H_0+160 H_0 \zeta _3+492 \zeta _4-16 \zeta _2 H_{0,0}+120 H_{0,0,0}+\frac{1}{z+1}+\frac{1727}{3}\bigg)
\end{dmath*}
\begin{dmath*}[style={\small},compact]
+\left.\frac{1}{z+1}\left(\left(-\frac{512}{9}\right)H_{-1,0}+\left(-\frac{32}{3}\right)H_{-1,0,0}+\left(-\frac{32}{3}\right)H_{0,-1,0}
+\frac{64}{3}H_{-1,0,1}+\left(-\frac{64 \zeta _2}{3}\right)H_{-1}\right)\right.
+\frac{1}{(z-1) (z+1)}\bigg(\left(-\frac{1576}{9}\right)H_{0,0}+\left(-\frac{1352}{9}\right)\zeta _2+\left(-\frac{280}{3}\right)H_{0,0,0}+\left(-\frac{80}{3}\right)\zeta _3+\frac{184}{9}H_0+64 H_0 \zeta _2-160 H_0 \zeta _3-492 \zeta _4+16 \zeta _2 H_{0,0}-64 H_{0,0,1}+539\bigg)
+\left(-\frac{128 z}{3 (z+1)}\right)H_{-1,0}
+\frac{z^2}{z+1}\bigg(\left(-\frac{512}{9}\right)H_{-1,0}+\left(-\frac{32}{3}\right)H_{-1,0,0}+\left(-\frac{32}{3}\right)H_{0,-1,0}+\frac{64}{3}H_{-1,0,1}+\left(-\frac{64 \zeta _2}{3}\right)H_{-1}\bigg)
+400 H_1 \zeta _2-52 H_{1,1}+48 H_{0,1,1}-440 H_{1,0,0}-240 H_{1,0,1}-160 H_{1,1,0}-80 H_{1,1,1}-64 H_{0,0,0,0}-112 H_{0,0,0,1}+80 H_{0,0,1,0}+48 H_{0,0,1,1}-176 H_{0,1,0,0}-96 H_{0,1,0,1}-64 H_{0,1,1,0}+z \bigg(\left(-\frac{668}{3}\right)H_1-400 H_1 \zeta _2+52 H_{1,1}-32 H_{0,1,1}+440 H_{1,0,0}+240 H_{1,0,1}+160 H_{1,1,0}+80 H_{1,1,1}-64 H_{0,0,0,0}-112 H_{0,0,0,1}+80 H_{0,0,1,0}+48 H_{0,0,1,1}-176 H_{0,1,0,0}-96 H_{0,1,0,1}-64 H_{0,1,1,0}-32 H_{0,1,1,1}\bigg)-32 H_{0,1,1,1}\bigg]
+C_A^2 C_F \left[\frac{1}{z-1}\left(\frac{592}{3}H_1+32 H_1 \zeta _2-144 H_1 \zeta _3+16 \zeta _2 H_{1,0}+76 H_{1,0,0}-80 H_{0,-1,-1,0}+36 H_{0,1,0,0}-96 H_{1,0,-1,0}+48 H_{1,0,0,0}-64 H_{1,0,0,1}+64 H_{1,1,0,0}\right)+\frac{z}{z-1}\left(\left(-\frac{1184}{3}\right)H_1-64 H_1 \zeta _2-64 H_{1,0,0}+32 H_{0,-1,-1,0}\right)
+\frac{z}{z-1}\bigg(\left(-\frac{1744}{9}\right)\zeta _2+\left(-\frac{722}{9}\right)H_{0,0}+\left(-\frac{1258}{27}\right)H_0+\frac{160}{3}\zeta _3+\frac{344}{3}H_{0,-1,0}-100 H_0 \zeta _2+7 \zeta _4+60 \zeta _2 H_{0,0}+100 H_{0,0,1}+72 H_{0,-1,0,0}+96 H_{0,-1,0,1}-8 H_{0,0,-1,0}-60 H_{0,0,0,1}+\frac{1}{z+1}+\frac{5333}{27}\bigg)
+\frac{z^2}{z-1}\bigg(\left(-\frac{416}{3}\right)H_{0,0,0}+\left(-\frac{1378}{27}\right)H_0+\left(-\frac{4}{3}\right)H_{0,0,1}+\frac{328}{3}H_{0,-1,0}+\frac{1504}{9}\zeta _2+\frac{332 \zeta _2}{3}H_0-144 H_0 \zeta _3+112 \zeta _3-121 \zeta _4-68 \zeta _2 H_{0,0}-130 H_{0,0}-56 H_{0,-1,0,0}-160 H_{0,-1,0,1}-40 H_{0,0,-1,0}+40 H_{0,0,0,0}+28 H_{0,0,0,1}+\frac{1}{z+1}+\frac{923}{27}\bigg)
+\frac{z^2}{z-1}\bigg(\frac{592}{3}H_1+32 H_1 \zeta _2-144 H_1 \zeta _3+16 \zeta _2 H_{1,0}+76 H_{1,0,0}-80 H_{0,-1,-1,0}+28 H_{0,1,0,0}-96 H_{1,0,-1,0}+48 H_{1,0,0,0}-64 H_{1,0,0,1}+64 H_{1,1,0,0}\bigg)
+\frac{z^3}{z-1}\bigg(\left(-\frac{376}{3}\right)H_{0,-1,0}+\left(-\frac{2266}{27}\right)H_0+\frac{238}{9}H_{0,0}+\frac{224}{3}\zeta _2+188 H_0 \zeta _2+48 \zeta _3-31 \zeta _4+36 \zeta _2 H_{0,0}-188 H_{0,0,1}+88 H_{0,-1,0,0}+96 H_{0,-1,0,1}+8 H_{0,0,-1,0}-36 H_{0,0,0,1}+\frac{1}{z+1}-\frac{5333}{27}\bigg)
+\frac{1}{(z-1) (z+1)}\bigg(\left(-\frac{560}{3}\right)\zeta _3
+\left(-\frac{3442}{27}\right)H_0+\left(-\frac{80}{3}\right)H_{0,0,0}
+\left(-\frac{8}{3}\right)H_{0,-1,0}+\frac{92}{3}H_{0,0,1}+\left(-\frac{100 \zeta _2}{3}\right)H_0
-48 \zeta _2-112 H_0 \zeta _3-63 \zeta _4-28 \zeta _2 H_{0,0}-162 H_{0,0}-40 H_{0,-1,0,0}-160 H_{0,-1,0,1}-24 H_{0,0,-1,0}+24 H_{0,0,0,0}+4 H_{0,0,0,1}-\frac{923}{27}\bigg)
+\frac{z}{z+1}\bigg(\left(-\frac{448}{3}\right)H_{0,1}+\left(-\frac{160}{3}\right)H_{-1,0}-256 H_{-1} \zeta _2-16 \zeta _2 H_{0,1}+96 H_{-1,0,0}+320 H_{-1,0,1}\bigg)
+\frac{z^2}{z+1}\bigg(-\frac{1312}{9}H_{-1,0}
-\frac{224}{3}H_{0,1}
+\frac{56}{3}H_{-1,0,0}+\frac{656}{3}H_{-1,0,1}
-\frac{560 \zeta _2}{3}H_{-1}-192 H_{-1} \zeta _3+256 \zeta _2 H_{-1,-1}
-176 \zeta _2 H_{-1,0}-136 \zeta _2 H_{0,-1}-40 \zeta _2 H_{0,1}-128 H_{-1,-1,0,0}
-256 H_{-1,-1,0,1}+48 H_{-1,0,0,0}+128 H_{-1,0,0,1}\bigg)
+\frac{1}{z+1}\bigg(\left(-\frac{1312}{9}\right)H_{-1,0}+\left(-\frac{224}{3}\right)H_{0,1}+\frac{56}{3}H_{-1,0,0}+\frac{656}{3}H_{-1,0,1}+\left(-\frac{560 \zeta _2}{3}\right)H_{-1}-192 H_{-1} \zeta _3+256 \zeta _2 H_{-1,-1}-176 \zeta _2 H_{-1,0}-120 \zeta _2 H_{0,-1}-40 \zeta _2 H_{0,1}-128 H_{-1,-1,0,0}-256 H_{-1,-1,0,1}+48 H_{-1,0,0,0}+128 H_{-1,0,0,1}\bigg)
+64 z H_{-1,-1,0}+64 H_{-1,-1,0}
\right]
\end{dmath*}
\begin{dmath*}[style={\small},compact]
+C_A C_F^2 \left[\frac{1}{z-1}\bigg(\left(-\frac{2008}{3}\right)H_1+\frac{272}{3}H_{0,1,0}+\frac{2320}{9}H_{1,0}-112 H_1 \zeta _2+384 H_1 \zeta _3-32 \zeta _2 H_{1,0}-112 H_{1,0,0}-80 H_{0,1,0,0}+256 H_{1,0,-1,0}-80 H_{1,0,0,0}+128 H_{1,0,0,1}-128 H_{1,1,0,0}\bigg)
+\frac{z}{z-1}\bigg(\left(-\frac{832}{3}\right)H_{1,0}+\frac{4016}{3}H_1+224 H_1 \zeta _2+128 H_{1,0,0}\bigg)
+\frac{z}{z-1}\bigg(\left(-\frac{1072}{3}\right)H_{0,-1,0}+\left(-\frac{2648}{9}\right)H_{0,1}+\left(-\frac{536}{3}\right)H_{0,0,1}+\left(-\frac{832}{9}\right)H_{0,0}+\frac{56}{3}\zeta _3+\frac{244}{9}H_0+\frac{2296}{9}\zeta _2+\frac{904 \zeta _2}{3}H_0+268 \zeta _4+416 \zeta _2 H_{0,-1}-192 \zeta _2 H_{0,0}+64 \zeta _2 H_{0,1}+140 H_{0,0,0}+128 H_{0,-1,-1,0}-208 H_{0,-1,0,0}-352 H_{0,-1,0,1}+64 H_{0,0,-1,0}+192 H_{0,0,0,1}+\frac{1}{z+1}+\frac{1516}{3}\bigg)
+\frac{z^2}{z-1}\bigg(\left(-\frac{3880}{9}\right)\zeta _2+\left(-\frac{832}{3}\right)\zeta _3+\left(-\frac{416}{3}\right)H_{0,-1,0}+\left(-\frac{350}{9}\right)H_0+\frac{1280}{9}H_{0,0}+\frac{652}{3}H_{0,0,0}+\frac{2296}{9}H_{0,1}-184 H_0 \zeta _2+224 H_0 \zeta _3+28 \zeta _4-352 \zeta _2 H_{0,-1}+32 \zeta _2 H_{0,0}-128 \zeta _2 H_{0,1}+40 H_{0,0,1}+256 H_{0,-1,-1,0}+16 H_{0,-1,0,0}+480 H_{0,-1,0,1}-64 H_{0,0,-1,0}+160 H_{0,0,0,0}-32 H_{0,0,0,1}+32 H_{0,0,1,0}+\frac{1}{z+1}+\frac{1516}{3}\bigg)
+\frac{z^2}{z-1}\bigg(\left(-\frac{2008}{3}\right)H_1+\frac{80}{3}H_{0,1,0}+\frac{2320}{9}H_{1,0}-112 H_1 \zeta _2+384 H_1 \zeta _3-32 \zeta _2 H_{1,0}-112 H_{1,0,0}-16 H_{0,1,0,0}+256 H_{1,0,-1,0}-80 H_{1,0,0,0}+128 H_{1,0,0,1}-128 H_{1,1,0,0}\bigg)
+\frac{z^3}{z-1}\bigg(\left(-\frac{880}{3}\right)\zeta _3+\left(-\frac{1798}{9}\right)H_0+\left(-\frac{152}{9}\right)\zeta _2+\frac{776}{3}H_{0,0}+\frac{992}{3}H_{0,-1,0}+\frac{4792}{9}H_{0,1}+\frac{1672}{3}H_{0,0,1}+\left(-\frac{1688 \zeta _2}{3}\right)H_0-12 \zeta _4+448 \zeta _2 H_{0,-1}-96 \zeta _2 H_{0,0}+96 \zeta _2 H_{0,1}+36 H_{0,0,0}+192 H_{0,-1,-1,0}-176 H_{0,-1,0,0}-352 H_{0,-1,0,1}+64 H_{0,0,-1,0}+96 H_{0,0,0,1}+\frac{1}{z+1}-\frac{1516}{3}\bigg)
+\frac{1}{z+1}\bigg(\left(-\frac{2032}{3}\right)H_{-1,0,1}+\left(-\frac{280}{3}\right)H_{-1,0,0}+\frac{3056}{9}H_{-1,0}+\frac{1696 \zeta _2}{3}H_{-1}+672 H_{-1} \zeta _3-896 \zeta _2 H_{-1,-1}+544 \zeta _2 H_{-1,0}+320 H_{-1,-1,0,0}+896 H_{-1,-1,0,1}-64 H_{-1,0,-1,0}+16 H_{-1,0,0,0}-384 H_{-1,0,0,1}+32 H_{-1,0,1,0}\bigg)
+\frac{1}{(z-1) (z+1)}\bigg(\left(-\frac{80}{3}\right)H_{0,-1,0}+\left(-\frac{152}{9}\right)H_{0,1}+\frac{92}{9}H_0+\frac{436}{3}H_{0,0,0}+\frac{1736}{9}\zeta _2+\frac{1496}{3}\zeta _3+\frac{4504}{9}H_{0,0}+88 H_0 \zeta _2+256 H_0 \zeta _3+268 \zeta _4-320 \zeta _2 H_{0,-1}-160 \zeta _2 H_{0,1}+8 H_{0,0,1}+320 H_{0,-1,-1,0}+48 H_{0,-1,0,0}+480 H_{0,-1,0,1}+32 H_{0,0,0,0}+64 H_{0,0,0,1}+32 H_{0,0,1,0}-\frac{1516}{3}\bigg)
+\frac{z}{z+1}\left(\frac{608}{3}H_{-1,0}+800 H_{-1} \zeta _2-160 H_{-1,0,0}-1024 H_{-1,0,1}\right)+\frac{z^2}{z+1}\left(\left(-\frac{2032}{3}\right)H_{-1,0,1}
+\left(-\frac{280}{3}\right)H_{-1,0,0}+\frac{3056}{9}H_{-1,0}+\frac{1696 \zeta _2}{3}H_{-1}+672 H_{-1} \zeta _3-896 \zeta _2 H_{-1,-1}+544 \zeta _2 H_{-1,0}+320 H_{-1,-1,0,0}+896 H_{-1,-1,0,1}
-64 H_{-1,0,-1,0}+16 H_{-1,0,0,0}-384 H_{-1,0,0,1}+32 H_{-1,0,1,0}\right)-224 z H_{-1,-1,0}-224 H_{-1,-1,0}\right]
+N_f C_A C_F\bigg[-\frac{1960}{9}H_1-\frac{844}{9}H_{0,1}-\frac{88}{3}H_{0,1,1}+\frac{260}{3}H_{1,1}-\frac{356}{z-1}H_{1,0,0}
+\frac{z}{z-1}\bigg(\left(-\frac{544}{3}\right)H_0+\left(-\frac{164}{3}\right)H_{0,0,1}+\frac{472}{3}H_{0,0,0}+\frac{4700}{9}H_{0,0}+\left(-\frac{220 \zeta _2}{3}\right)H_0-248 \zeta _2+160 H_0 \zeta _3-168 \zeta _3+426 \zeta _4+8 \zeta _2 H_{0,0}+\frac{1}{z+1}-\frac{2950}{9}\bigg)
+\frac{696 z}{z-1}H_{1,0,0}
+\left(-\frac{356 z^2}{z-1}\right)H_{1,0,0}
+\frac{z^2}{z-1}\bigg(\left(-\frac{440}{3}\right)H_{0,0,0}+\left(-\frac{1664}{27}\right)H_0+\frac{448}{9}H_{0,0}+\frac{332}{3}\zeta _2+\frac{656}{3}\zeta _3+\frac{548 \zeta _2}{3}H_0-32 H_0 \zeta _3-406 \zeta _4+104 \zeta _2 H_{0,0}-52 H_{0,0,1}+\frac{1}{z+1}-\frac{8014}{27}\bigg)
\end{dmath*}
\begin{dmath}[style={\small},compact]
+\frac{z^3}{z-1}\left(-\frac{1508}{3}H_{0,0}-\frac{472}{3}H_{0,0,0}
+\frac{212}{3}H_{0,0,1}
+\frac{488}{3}\zeta _3
+\frac{6232}{27}H_0+\frac{2392}{9}\zeta _2+\frac{172 \zeta _2}{3}H_0-160 H_0 \zeta _3-426 \zeta _4-8 \zeta _2 H_{0,0}+\frac{1}{z+1}+\frac{2950}{9}\right)
+\frac{1}{z+1}\bigg(-\frac{128}{3}H_{-1,0,1}+\frac{328}{9}H_{-1,0}+\frac{280}{3}H_{0,-1,0}+\frac{568}{3}H_{-1,0,0}+\frac{8 \zeta _2}{3}H_{-1}\bigg)
+\frac{1}{(z-1) (z+1)}\bigg(-\frac{1156}{9}\zeta _2+\frac{16}{3}H_{0,0}+\frac{140}{3}H_{0,0,1}+\frac{1000}{9}H_0-188 H_0 \zeta _2+32 H_0 \zeta _3-192 \zeta _3+406 \zeta _4-104 \zeta _2 H_{0,0}+168 H_{0,0,0}+\frac{8014}{27}\bigg)
+\frac{z}{z+1}\left(\frac{112}{3}H_{-1,0}-16 H_{-1} \zeta _2+368 H_{-1,0,0}-64 H_{-1,0,1}+240 H_{0,-1,0}\right)
+\frac{z^2}{z+1}\bigg(-\frac{128}{3}H_{-1,0,1}+\frac{328}{9}H_{-1,0}+\frac{472}{3}H_{0,-1,0}+\frac{568}{3}H_{-1,0,0}+\frac{8 \zeta _2}{3}H_{-1}\bigg)
-360 H_1 \zeta _2-16 \zeta _2 H_{0,-1}-144 \zeta _2 H_{0,1}+268 H_{1,0}-80 H_{-1,-1,0}-64 H_{0,1,0}+240 H_{1,0,1}+160 H_{1,1,0}+80 H_{1,1,1}-32 H_{0,-1,-1,0}+80 H_{0,-1,0,0}+112 H_{0,0,-1,0}-304 H_{0,0,0,0}+8 H_{0,0,0,1}-144 H_{0,0,1,0}-48 H_{0,0,1,1}+136 H_{0,1,0,0}+96 H_{0,1,0,1}+64 H_{0,1,1,0}+32 H_{0,1,1,1}
+z \bigg(-\frac{2536}{9}H_{0,1}-\frac{260}{3}H_{1,1}+\frac{152}{3}H_{0,1,1}+\frac{1960}{9}H_1+360 H_1 \zeta _2+16 \zeta _2 H_{0,-1}-144 \zeta _2 H_{0,1}-268 H_{1,0}-80 H_{-1,-1,0}+216 H_{0,1,0}-240 H_{1,0,1}-160 H_{1,1,0}-80 H_{1,1,1}+32 H_{0,-1,-1,0}-80 H_{0,-1,0,0}-48 H_{0,0,-1,0}+64 H_{0,0,0,0}+8 H_{0,0,0,1}-144 H_{0,0,1,0}-48 H_{0,0,1,1}+136 H_{0,1,0,0}+96 H_{0,1,0,1}+64 H_{0,1,1,0}+32 H_{0,1,1,1}\bigg)\bigg]
\,.
\end{dmath}

\begin{dmath}[style={\small},compact]
\Delta P_{gq}^{T,\text{s}(0)}=
(4-2 z) C_F
\,,
\end{dmath}

\begin{dmath}[style={\small},compact]
\Delta P_{gq}^{T,\text{s}(1)}=
C_A C_F \left(z \left(8 H_{-1,0}-8 H_{0,0}-8 H_{0,1}+24 H_{1,0}+8 H_{1,1}+44 H_0+16 H_1+8 \zeta _2-16\right)+16 H_{-1,0}-48 H_{0,0}+16 H_{0,1}-48 H_{1,0}-16 H_{1,1}-16 H_0-16 H_1+20\right)+C_F^2 \left(z \left(4 H_{0,0}+16 H_{0,1}-16 H_{1,0}-8 H_{1,1}-18 H_0-16 H_1-28\right)-8 H_{0,0}-32 H_{0,1}+32 H_{1,0}+16 H_{1,1}+32 H_0+16 H_1+38\right)
\,,
\end{dmath}

\begin{dmath*}[style={\small},compact]
\Delta P_{gq}^{T,\text{s}(2)}=
C_F^3 \bigg[\left(-\frac{48}{z}\right)H_{-1,0}+447 H_0+626 H_1+32 H_{-1} \zeta _2+24 H_0 \zeta _2-112 H_1 \zeta _2-424 \zeta _2-400 H_0 \zeta _3+64 H_1 \zeta _3-224 \zeta _3+84 \zeta _4+48 H_{-1,0}+192 \zeta _2 H_{0,-1}-128 \zeta _2 H_{0,0}+174 H_{0,0}+400 \zeta _2 H_{0,1}+40 H_{0,1}+128 \zeta _2 H_{1,0}+304 H_{1,0}+96 \zeta _2 H_{1,1}+240 H_{1,1}+64 H_{-1,-1,0}+64 H_{-1,0,0}-168 H_{0,0,0}-344 H_{0,0,1}+40 H_{0,1,0}-40 H_{0,1,1}+80 H_{1,0,0}-16 H_{1,0,1}+80 H_{1,1,0}+96 H_{1,1,1}-128 H_{-1,-1,0,0}-64 H_{-1,0,-1,0}+32 H_{-1,0,0,0}+384 H_{0,-1,-1,0}-128 H_{0,-1,0,0}-192 H_{0,0,-1,0}+192 H_{0,0,0,0}+128 H_{0,0,0,1}+176 H_{0,0,1,0}+240 H_{0,0,1,1}+80 H_{0,1,0,0}-144 H_{0,1,0,1}-80 H_{0,1,1,0}-208 H_{0,1,1,1}-64 H_{1,0,-1,0}-192 H_{1,0,0,0}-256 H_{1,0,0,1}+32 H_{1,0,1,0}-128 H_{1,0,1,1}-192 H_{1,1,0,0}-96 H_{1,1,0,1}+32 H_{1,1,1,0}+z \left(-60 \zeta _2 H_0-40 \zeta _3 H_0-221 H_0-592 H_1+32 H_{-1} \zeta _2+76 H_1 \zeta _2+44 \zeta _2-32 H_1 \zeta _3+76 \zeta _3-194 \zeta _4+64 H_{-1,0}+64 \zeta _2 H_{0,-1}-32 \zeta _2 H_{0,0}-73 H_{0,0}-232 \zeta _2 H_{0,1}+244 H_{0,1}-64 \zeta _2 H_{1,0}-248 H_{1,0}-48 \zeta _2 H_{1,1}-184 H_{1,1}+64 H_{-1,-1,0}+64 H_{-1,0,0}-144 H_{0,-1,0}+108 H_{0,0,0}+140 H_{0,0,1}+56 H_{0,1,0}+112 H_{0,1,1}+28 H_{1,0,0}+52 H_{1,0,1}-44 H_{1,1,0}-84 H_{1,1,1}-64 H_{-1,-1,0,0}-32 H_{-1,0,-1,0}+16 H_{-1,0,0,0}+128 H_{0,-1,-1,0}-32 H_{0,-1,0,0}-32 H_{0,0,-1,0}-16 H_{0,0,0,0}-64 H_{0,0,0,1}-88 H_{0,0,1,0}-120 H_{0,0,1,1}-8 H_{0,1,0,0}+72 H_{0,1,0,1}+40 H_{0,1,1,0}+104 H_{0,1,1,1}+32 H_{1,0,-1,0}+96 H_{1,0,0,0}+128 H_{1,0,0,1}-16 H_{1,0,1,0}+64 H_{1,0,1,1}+96 H_{1,1,0,0}+48 H_{1,1,0,1}-16 H_{1,1,1,0}-32 H_{1,1,1,1}-\frac{3133}{4}\right)+64 H_{1,1,1,1}+\frac{1569}{2}\bigg]
+N_f  C_F^2 \bigg[\left(-\frac{80}{3}\right)H_{1,0}+\left(-\frac{80}{3}\right)H_{1,0,0}+\left(-\frac{112}{9}\right)H_{1,1}+\left(-\frac{32}{3}\right)H_{0,1,1}+\left(-\frac{16}{3}\right)H_{0,0,0}+\left(-\frac{16}{3}\right)H_{1,0,1}+\left(-\frac{16}{3}\right)H_{1,1,0}+\frac{76}{27}H_1+\frac{16}{3}H_{1,1,1}+\frac{64}{3}H_{0,0,1}+\frac{224}{9}H_{0,1}+\frac{380}{3}H_{0,0}+\frac{4334}{27}H_0-16 H_1 \zeta _2-64 H_{0,0,0,0}+z \left(\left(-\frac{328}{3}\right)H_{0,0,0}+\left(-\frac{2038}{27}\right)H_0+\left(-\frac{32}{3}\right)H_{0,0,1}+\left(-\frac{64}{9}\right)H_{0,1}+\left(-\frac{8}{3}\right)H_{1,1,1}+\frac{8}{3}H_{1,0,1}+\frac{8}{3}H_{1,1,0}+\frac{32}{9}H_{1,1}+\frac{112}{27}H_1+\frac{16}{3}H_{0,1,1}+\frac{40}{3}H_{1,0}+\frac{40}{3}H_{1,0,0}+8 H_1 \zeta _2+154 H_{0,0}+32 H_{0,0,0,0}-\frac{1567}{18}\right)+\frac{767}{9}\bigg]
+C_A^2C_F \bigg[\left(-\frac{7520}{9}\right)\zeta _2+\left(-\frac{7376}{9}\right)H_{1,0}+\left(-\frac{1384}{3}\right)H_{-1,0,0}+\left(-\frac{904}{3}\right)H_{0,-1,0}+\left(-\frac{2248}{9}\right)H_{1,1}+\left(-\frac{2216}{9}\right)H_{-1,0}+\left(-\frac{268}{3}\right)H_{1,0,0}+\frac{8}{3}H_{0,1,1}+\frac{88}{3}H_{1,1,1}+\frac{112}{3}H_{1,1,0}+\frac{932}{3}H_{0,0,1}+\frac{5584}{9}H_{0,0}+\frac{2216}{3}H_{0,0,0}+\frac{7624}{9}H_{0,1}+\frac{28256}{27}H_0+\frac{32362}{27}H_1+\left(-\frac{24}{z}\right)H_{-1,0}+\left(-\frac{1468 \zeta _2}{3}\right)H_0+104 H_{-1} \zeta _2+264 H_1 \zeta _2-16 H_{-1} \zeta _3+480 H_0 \zeta _3
+656 H_1 \zeta _3-712 \zeta _3+1570 \zeta _4-32 \zeta _2 H_{-1,-1}
\end{dmath*}
\begin{dmath*}[style={\small},compact]
+160 \zeta _2 H_{-1,0}-16 \zeta _2 H_{0,-1}-312 \zeta _2 H_{0,0}+496 \zeta _2 H_{0,1}-128 \zeta _2 H_{1,0}+96 \zeta _2 H_{1,1}
+336 H_{-1,-1,0}+64 H_{-1,0,1}+288 H_{0,1,0}-96 H_{1,0,1}-192 H_{-1,-1,-1,0}
+192 H_{-1,-1,0,0}-64 H_{-1,-1,0,1}+352 H_{-1,0,-1,0}-336 H_{-1,0,0,0}+64 H_{-1,0,0,1}-192 H_{-1,0,1,0}-64 H_{-1,0,1,1}+96 H_{0,-1,-1,0}-272 H_{0,-1,0,0}+64 H_{0,-1,0,1}-496 H_{0,0,-1,0}+1264 H_{0,0,0,0}-184 H_{0,0,0,1}+960 H_{0,0,1,0}+320 H_{0,0,1,1}-152 H_{0,1,0,0}-352 H_{0,1,0,1}-224 H_{0,1,1,0}-160 H_{0,1,1,1}-96 H_{1,0,-1,0}+560 H_{1,0,0,0}+160 H_{1,0,0,1}+640 H_{1,0,1,0}+64 H_{1,0,1,1}+160 H_{1,1,0,0}+128 H_{1,1,0,1}+192 H_{1,1,1,0}
+z \left(\left(-\frac{31412}{27}\right)H_1+\left(-\frac{20644}{27}\right)H_0+\left(-\frac{1472}{3}\right)H_{-1,0,0}+\left(-\frac{992}{3}\right)H_{0,-1,0}+\left(-\frac{2752}{9}\right)H_{-1,0}+\left(-\frac{580}{3}\right)H_{0,0,1}+\left(-\frac{484}{3}\right)H_{0,1,1}
+\left(-\frac{328}{3}\right)H_{0,0,0}+\left(-\frac{200}{3}\right)H_{1,1,0}+\left(-\frac{172}{3}\right)H_{1,0,0}+\left(-\frac{44}{3}\right)H_{1,1,1}+\frac{1712}{9}H_{1,1}+\frac{5380}{9}H_{0,1}+\frac{5944}{9}H_{1,0}+\frac{10172}{9}H_{0,0}+104 H_{-1} \zeta _2-156 H_0 \zeta _2-352 H_1 \zeta _2-796 \zeta _2-8 H_{-1} \zeta _3+176 H_0 \zeta _3-328 H_1 \zeta _3-464 \zeta _3+143 \zeta _4-16 \zeta _2 H_{-1,-1}+80 \zeta _2 H_{-1,0}-72 \zeta _2 H_{0,-1}+76 \zeta _2 H_{0,0}+24 \zeta _2 H_{0,1}+64 \zeta _2 H_{1,0}-48 \zeta _2 H_{1,1}+336 H_{-1,-1,0}+64 H_{-1,0,1}-848 H_{0,1,0}+96 H_{1,0,1}-96 H_{-1,-1,-1,0}+96 H_{-1,-1,0,0}-32 H_{-1,-1,0,1}+176 H_{-1,0,-1,0}-168 H_{-1,0,0,0}+32 H_{-1,0,0,1}-96 H_{-1,0,1,0}-32 H_{-1,0,1,1}-80 H_{0,-1,-1,0}+120 H_{0,-1,0,0}+32 H_{0,-1,0,1}+56 H_{0,0,-1,0}-64 H_{0,0,0,0}-76 H_{0,0,0,1}+192 H_{0,0,1,0}+64 H_{0,0,1,1}+116 H_{0,1,0,0}+32 H_{0,1,0,1}+64 H_{0,1,1,0}+32 H_{0,1,1,1}+48 H_{1,0,-1,0}-280 H_{1,0,0,0}-80 H_{1,0,0,1}-320 H_{1,0,1,0}-32 H_{1,0,1,1}-80 H_{1,1,0,0}-64 H_{1,1,0,1}-96 H_{1,1,1,0}-32 H_{1,1,1,1}-\frac{13346}{9}\right)+64 H_{1,1,1,1}+\frac{13840}{9}\bigg]
+ C_A C_F^2 \bigg[\left(-\frac{52072}{27}\right)H_1+\left(-\frac{10868}{9}\right)H_{0,1}+\left(-\frac{1466}{3}\right)H_{0,0}+\left(-\frac{1136}{3}\right)H_{0,0,0}+\left(-\frac{3719}{27}\right)H_0+\left(-\frac{376}{3}\right)H_{1,1,1}+\left(-\frac{368}{3}\right)H_{1,1,0}+\left(-\frac{184}{3}\right)H_{0,0,1}+\left(-\frac{344}{9}\right)H_{1,1}+\left(-\frac{88}{3}\right)H_{1,0}+\frac{56}{3}H_{0,1,1}+\frac{272}{3}H_{1,0,0}+\frac{352}{3}H_{1,0,1}+\frac{1}{z}\left(-72 H_{-1} \zeta _2-24 H_1 \zeta _2+72 H_{-1,0}-48 H_{-1,-1,0}+48 H_{-1,0,0}+48 H_{-1,0,1}\right)-256 H_{-1} \zeta _2+296 H_0 \zeta _2-240 H_1 \zeta _2+716 \zeta _2-176 H_{-1} \zeta _3-752 H_0 \zeta _3-816 H_1 \zeta _3+728 \zeta _3-2144 \zeta _4+288 \zeta _2 H_{-1,-1}-192 \zeta _2 H_{-1,0}+136 H_{-1,0}-32 \zeta _2 H_{0,-1}-80 \zeta _2 H_{0,0}-912 \zeta _2 H_{0,1}+32 \zeta _2 H_{1,0}-192 \zeta _2 H_{1,1}-256 H_{-1,-1,0}+144 H_{-1,0,0}+128 H_{-1,0,1}+96 H_{0,-1,0}-472 H_{0,1,0}+192 H_{-1,-1,-1,0}+64 H_{-1,-1,0,0}-192 H_{-1,-1,0,1}-160 H_{-1,0,-1,0}-16 H_{-1,0,0,0}-64 H_{-1,0,0,1}+128 H_{-1,0,1,0}+64 H_{-1,0,1,1}-192 H_{0,-1,-1,0}-64 H_{0,-1,0,0}-64 H_{0,-1,0,1}+96 H_{0,0,-1,0}+128 H_{0,0,0,0}+752 H_{0,0,0,1}-688 H_{0,0,1,0}-432 H_{0,0,1,1}+416 H_{0,1,0,0}+624 H_{0,1,0,1}+432 H_{0,1,1,0}+368 H_{0,1,1,1}+96 H_{1,0,-1,0}-48 H_{1,0,0,0}+224 H_{1,0,0,1}-544 H_{1,0,1,0}+64 H_{1,0,1,1}+160 H_{1,1,0,0}-32 H_{1,1,0,1}-224 H_{1,1,1,0}-128 H_{1,1,1,1}+z \left(\left(-\frac{6224}{9}\right)H_{0,1}+\left(-\frac{688}{3}\right)H_{0,0,1}+\left(-\frac{452}{3}\right)H_{1,0,1}+\left(-\frac{448}{3}\right)H_{1,0,0}+\frac{56}{3}H_{1,0}+\frac{68}{3}H_{0,1,1}+\frac{376}{9}H_{1,1}+\frac{296}{3}H_{1,1,1}+\frac{340}{3}H_{1,1,0}+\frac{364}{3}H_{0,0,0}+\frac{50168}{27}H_1+\frac{57247}{27}H_0-184 H_{-1} \zeta _2+72 H_0 \zeta _2+380 H_1 \zeta _2+1176 \zeta _2-88 H_{-1} \zeta _3-136 H_0 \zeta _3+408 H_1 \zeta _3+368 \zeta _3-152 \zeta _4+144 \zeta _2 H_{-1,-1}-96 \zeta _2 H_{-1,0}+64 H_{-1,0}-64 \zeta _2 H_{0,-1}-8 \zeta _2 H_{0,0}-889 H_{0,0}+136 \zeta _2 H_{0,1}-16 \zeta _2 H_{1,0}+96 \zeta _2 H_{1,1}-208 H_{-1,-1,0}+24 H_{-1,0,0}+80 H_{-1,0,1}+264 H_{0,-1,0}+528 H_{0,1,0}+96 H_{-1,-1,-1,0}+32 H_{-1,-1,0,0}-96 H_{-1,-1,0,1}-80 H_{-1,0,-1,0}-8 H_{-1,0,0,0}-32 H_{-1,0,0,1}+64 H_{-1,0,1,0}+32 H_{-1,0,1,1}-64 H_{0,-1,-1,0}-16 H_{0,-1,0,0}+32 H_{0,-1,0,1}-16 H_{0,0,-1,0}+32 H_{0,0,0,0}+168 H_{0,0,0,1}-104 H_{0,0,1,0}-8 H_{0,0,1,1}-160 H_{0,1,0,0}-168 H_{0,1,0,1}-168 H_{0,1,1,0}-136 H_{0,1,1,1}-48 H_{1,0,-1,0}+24 H_{1,0,0,0}-112 H_{1,0,0,1}+272 H_{1,0,1,0}-32 H_{1,0,1,1}-80 H_{1,1,0,0}+16 H_{1,1,0,1}+112 H_{1,1,1,0}+64 H_{1,1,1,1}-\frac{2737}{36}\right)+\frac{1577}{18}\bigg]
\end{dmath*}
\begin{dmath}[style={\small},compact, indentstep={0 em}]
+N_f C_A  C_F \bigg[\left(-\frac{728}{27}\right)H_0+\left(-\frac{112}{9}\right)H_{-1,0}+\left(-\frac{32}{3}\right)H_{-1,0,0}+\left(-\frac{32}{3}\right)H_{0,-1,0}+\left(-\frac{32}{3}\right)H_{0,0,1}+\left(-\frac{88}{9}\right)H_{0,1}+\left(-\frac{16}{3}\right)H_{1,1,1}+\frac{8}{9}\zeta _2+\frac{32}{27}H_1+\frac{16}{3}H_{0,1,1}+\frac{32}{3}H_{1,1,0}+\frac{112}{9}H_{1,1}
+\frac{368}{9}H_{1,0}+\frac{160}{3}H_{1,0,0}+\frac{584}{9}H_{0,0}+\frac{256}{3}H_{0,0,0}+\left(-\frac{32 \zeta _2}{3}\right)H_0+32 H_1 \zeta _2-80 \zeta _3+z \left(\left(-\frac{356}{9}\right)H_{0,0}+\left(-\frac{80}{3}\right)H_{1,0,0}+\left(-\frac{208}{9}\right)H_{1,0}+\left(-\frac{518}{27}\right)H_0+\left(-\frac{184}{27}\right)H_1+\left(-\frac{16}{3}\right)H_{-1,0,0}+\left(-\frac{16}{3}\right)H_{0,-1,0}+\left(-\frac{16}{3}\right)H_{1,1,0}+\left(-\frac{32}{9}\right)H_{-1,0}+\left(-\frac{32}{9}\right)H_{1,1}+\left(-\frac{8}{3}\right)H_{0,1,1}+\frac{8}{3}H_{1,1,1}+\frac{32}{9}H_{0,1}+\frac{16}{3}H_{0,0,0}+\frac{16}{3}H_{0,0,1}-16 H_1 \zeta _2-16 \zeta _2+32 \zeta _3+\frac{416}{9}\right)-\frac{526}{9}\bigg]
\,.
\end{dmath}

\begin{dmath}[style={\small},compact]
\Delta P_{qg}^{T,\text{s}(0)}=
(4 z-2) N_f
\,,
\end{dmath}

\begin{dmath}[style={\small},compact]
\Delta P_{qg}^{T,\text{s}(1)}=
C_A N_f \bigg[z \left(-16 H_{-1,0}+16 H_{0,0}-32 H_{0,1}+32 H_{1,0}+16 H_{1,1}+\left(-\frac{40}{3}\right)H_1+\left(-\frac{8}{3}\right)H_0+\frac{740}{9}\right)-8 H_{-1,0}+24 H_{0,0}+16 H_{0,1}-16 H_{1,0}-8 H_{1,1}+\left(-\frac{32}{3}\right)H_0+\left(-\frac{4}{3}\right)H_1-8 \zeta _2-\frac{436}{9}\bigg]
+C_F N_f \bigg[z \left(-8 H_{0,0}+16 H_{0,1}-48 H_{1,0}-16 H_{1,1}+8 H_1-16 \zeta _2-50\right)+4 H_{0,0}-8 H_{0,1}+24 H_{1,0}+8 H_{1,1}+22 H_0+4 H_1+8 \zeta _2+32\bigg]
+N_f^2 \bigg[\left(\left(-\frac{16}{3}\right)H_0+\frac{16}{3}H_1-\frac{32}{9}\right) z+\left(-\frac{8}{3}\right)H_1+\frac{8}{3}H_0-\frac{8}{9}\bigg]\,,
\end{dmath}

\begin{dmath*}[style={\small},compact]
\Delta P_{qg}^{T,\text{s}(2)}=
N_f^3\bigg[\left(-\frac{8}{3}\right)\zeta _2+\left(-\frac{8}{3}\right)H_{0,0}+\left(-\frac{8}{3}\right)H_{1,1}+\left(-\frac{8}{9}\right)H_1+\frac{8}{9}H_0+\frac{8}{3}H_{0,1}+\frac{8}{3}H_{1,0}+z \left(\left(-\frac{16}{3}\right)H_{0,1}+\left(-\frac{16}{3}\right)H_{1,0}+\left(-\frac{32}{9}\right)H_1+\frac{32}{9}H_0+\frac{16}{3}\zeta _2+\frac{16}{3}H_{0,0}+\frac{16}{3}H_{1,1}-\frac{16}{3}\right)+\frac{40}{9}\bigg]
+C_F N_f^2 \bigg[\left(16 \zeta _2+16 H_{-1,0}-16 H_{0,0}\right) z^2+\left(\left(-\frac{128}{3}\right)H_{0,0,1}+\left(-\frac{112}{3}\right)H_{1,1,0}+\left(-\frac{280}{9}\right)H_{0,1}+\left(-\frac{260}{9}\right)H_{0,0}+\left(-\frac{80}{3}\right)H_{1,0,1}+\left(-\frac{80}{3}\right)H_{1,1,1}+\left(-\frac{680}{27}\right)H_1+\frac{280}{9}\zeta _2+\frac{328}{9}H_{1,1}+\frac{128}{3}H_{0,1,1}+\frac{160}{3}H_{0,1,0}+\frac{232}{3}H_{1,0}+\frac{320}{3}H_{0,0,0}+\frac{5090}{27}H_0+\frac{608}{3}\zeta _3+\frac{64 \zeta _2}{3}H_0-16 H_1 \zeta _2-64 H_{-1,0}+64 H_{0,-1,0}+64 H_{1,0,0}-64 H_{0,0,0,0}-\frac{1157}{3}\right) z
\end{dmath*}
\begin{dmath*}[style={\small},compact]
+\left(-\frac{304}{3}\right)\zeta _3+\left(-\frac{848}{9}\right)\zeta _2+\left(-\frac{80}{3}\right)H_{0,1,0}+\left(-\frac{64}{3}\right)H_{0,1,1}+\left(-\frac{56}{3}\right)H_{1,0}+\left(-\frac{32}{9}\right)H_{1,1}+\frac{184}{27}H_1+\frac{40}{3}H_{1,0,1}+\frac{40}{3}H_{1,1,1}+\frac{56}{3}H_{1,1,0}+\frac{64}{3}H_{0,0,1}+\frac{272}{9}H_{0,1}+\frac{898}{9}H_{0,0}+\frac{320}{3}H_{0,0,0}+\frac{3830}{27}H_0+\left(-\frac{128 \zeta _2}{3}\right)H_0+8 H_1 \zeta _2-80 H_{-1,0}-32 H_{0,-1,0}-32 H_{1,0,0}+32 H_{0,0,0,0}+\frac{6587}{18}\bigg] 
+ N_f C_A^2 \bigg[\left(72 H_{-1} \zeta _2-96 H_0 \zeta _2-24 H_1 \zeta _2+24 \zeta _2-120 \zeta _3+24 H_{-1,0}-24 H_{0,0}+48 H_{-1,-1,0}-48 H_{-1,0,0}-48 H_{-1,0,1}-48 H_{0,-1,0}+48 H_{0,0,0}+48 H_{0,0,1}\right) z^2+\left(\left(-\frac{8836}{9}\right)H_{0,1}+\left(-\frac{976}{3}\right)H_{0,0,0}+\left(-\frac{848}{3}\right)H_{-1,-1,0}+\left(-\frac{2348}{9}\right)H_{0,0}+\left(-\frac{592}{3}\right)H_{1,0,1}+\left(-\frac{592}{3}\right)H_{1,1,0}+\left(-\frac{440}{3}\right)H_{1,1,1}+\left(-\frac{3796}{27}\right)H_0+\left(-\frac{404}{3}\right)H_{1,0,0}+\left(-\frac{320}{3}\right)H_{-1,0,1}+\left(-\frac{308}{3}\right)\zeta _2+\left(-\frac{304}{3}\right)H_{-1,0}+\frac{284}{3}H_{0,0,1}+\frac{536}{3}H_{-1,0,0}+\frac{5464}{27}H_1+\frac{1828}{9}H_{1,1}+\frac{2068}{9}H_{1,0}+\frac{824}{3}H_{0,-1,0}+\left(-\frac{104 \zeta _2}{3}\right)H_{-1}+\frac{472 \zeta _2}{3}H_1+60 H_0 \zeta _2+272 H_{-1} \zeta _3+32 H_0 \zeta _3+496 H_1 \zeta _3-184 \zeta _3+170 \zeta _4-416 \zeta _2 H_{-1,-1}+224 \zeta _2 H_{-1,0}+144 \zeta _2 H_{0,-1}-56 \zeta _2 H_{0,0}+336 \zeta _2 H_{0,1}+96 \zeta _2 H_{1,0}-32 \zeta _2 H_{1,1}-24 H_{0,1,0}+192 H_{0,1,1}-192 H_{-1,-1,-1,0}+128 H_{-1,-1,0,0}+320 H_{-1,-1,0,1}+224 H_{-1,0,-1,0}+16 H_{-1,0,0,0}-128 H_{-1,0,1,0}-64 H_{-1,0,1,1}+32 H_{0,-1,-1,0}-208 H_{0,-1,0,0}-128 H_{0,-1,0,1}-112 H_{0,0,-1,0}+128 H_{0,0,0,0}-200 H_{0,0,0,1}+128 H_{0,0,1,0}+352 H_{0,0,1,1}+184 H_{0,1,0,0}-160 H_{0,1,1,1}-32 H_{1,0,-1,0}-240 H_{1,0,0,0}-256 H_{1,0,0,1}-64 H_{1,0,1,1}-320 H_{1,1,0,0}-64 H_{1,1,0,1}-64 H_{1,1,1,0}+64 H_{1,1,1,1}+\frac{5408}{3}\right) z
+\left(-\frac{19762}{27}\right)H_0+\left(-\frac{9620}{27}\right)H_1+\left(-\frac{928}{3}\right)H_{0,0,0}+\left(-\frac{640}{3}\right)H_{-1,-1,0}+\left(-\frac{1502}{9}\right)H_{0,0}+\left(-\frac{352}{3}\right)H_{-1,0,1}+\left(-\frac{566}{9}\right)H_{1,1}+\left(-\frac{434}{9}\right)H_{1,0}+\frac{220}{3}H_{1,1,1}+\frac{232}{3}H_{-1,0}+\frac{304}{3}\zeta _3+\frac{680}{3}H_{1,0,1}+\frac{680}{3}H_{1,1,0}+\frac{776}{3}H_{0,-1,0}+\frac{2390}{9}H_{0,1}+\frac{856}{3}H_{-1,0,0}+\frac{1240}{3}H_{1,0,0}+\frac{1342}{3}\zeta _2+\left(-\frac{224 \zeta _2}{3}\right)H_1+\frac{32 \zeta _2}{3}H_{-1}+\frac{856 \zeta _2}{3}H_0+136 H_{-1} \zeta _3+480 H_0 \zeta _3-248 H_1 \zeta _3+63 \zeta _4-208 \zeta _2 H_{-1,-1}+112 \zeta _2 H_{-1,0}-88 \zeta _2 H_{0,-1}+316 \zeta _2 H_{0,0}-216 \zeta _2 H_{0,1}-48 \zeta _2 H_{1,0}+16 \zeta _2 H_{1,1}-40 H_{0,0,1}+4 H_{0,1,0}-56 H_{0,1,1}-96 H_{-1,-1,-1,0}+64 H_{-1,-1,0,0}+160 H_{-1,-1,0,1}+112 H_{-1,0,-1,0}+8 H_{-1,0,0,0}-64 H_{-1,0,1,0}-32 H_{-1,0,1,1}-176 H_{0,-1,-1,0}+216 H_{0,-1,0,0}+248 H_{0,0,-1,0}-632 H_{0,0,0,0}-388 H_{0,0,0,1}-48 H_{0,0,1,0}-96 H_{0,0,1,1}+204 H_{0,1,0,0}+144 H_{0,1,0,1}+144 H_{0,1,1,0}+128 H_{0,1,1,1}+16 H_{1,0,-1,0}+120 H_{1,0,0,0}+128 H_{1,0,0,1}+32 H_{1,0,1,1}+160 H_{1,1,0,0}+32 H_{1,1,0,1}+32 H_{1,1,1,0}-32 H_{1,1,1,1}-\frac{4670}{3}\bigg]
+N_f C_F^2 \bigg[\left(48 \zeta _2+48 H_{-1,0}-48 H_{0,0}\right) z^2+\left(-80 \zeta _2 H_0+80 \zeta _3 H_0+289 H_0+600 H_1-64 H_{-1} \zeta _2-176 H_1 \zeta _2+612 \zeta _2+992 H_1 \zeta _3-368 \zeta _3+280 \zeta _4-192 \zeta _2 H_{-1,0}+304 H_{-1,0}+128 \zeta _2 H_{0,-1}-80 \zeta _2 H_{0,0}-350 H_{0,0}+208 \zeta _2 H_{0,1}-340 H_{0,1}-224 \zeta _2 H_{1,0}+592 H_{1,0}-32 \zeta _2 H_{1,1}+32 H_{1,1}-128 H_{-1,-1,0}+352 H_{-1,0,0}+56 H_{0,0,0}+16 H_{0,0,1}-16 H_{0,1,0}+32 H_{0,1,1}-64 H_{1,0,0}-48 H_{1,0,1}-144 H_{1,1,0}-48 H_{1,1,1}-384 H_{-1,-1,0,0}-192 H_{-1,0,-1,0}+352 H_{-1,0,0,0}+256 H_{0,-1,-1,0}+64 H_{0,-1,0,0}-192 H_{0,0,-1,0}+32 H_{0,0,0,0}+16 H_{0,0,0,1}+192 H_{0,0,1,0}+128 H_{0,0,1,1}-352 H_{0,1,0,0}-208 H_{0,1,0,1}-240 H_{0,1,1,0}-112 H_{0,1,1,1}-192 H_{1,0,-1,0}+672 H_{1,0,0,0}+160 H_{1,0,0,1}+608 H_{1,0,1,0}+128 H_{1,0,1,1}+32 H_{1,1,0,0}+160 H_{1,1,0,1}+96 H_{1,1,1,0}+64 H_{1,1,1,1}+\frac{3747}{2}\right) z
\end{dmath*}
\begin{dmath}[style={\small},compact]
-1505 H_0-582 H_1-64 H_{-1} \zeta _2+164 H_0 \zeta _2+140 H_1 \zeta _2+394 \zeta _2-152 H_0 \zeta _3-496 H_1 \zeta _3-548 \zeta _3-476 \zeta _4-96 \zeta _2 H_{-1,0}+288 H_{-1,0}+32 \zeta _2 H_{0,-1}+72 \zeta _2 H_{0,0}-947 H_{0,0}-72 \zeta _2 H_{0,1}-242 H_{0,1}+112 \zeta _2 H_{1,0}-390 H_{1,0}+16 \zeta _2 H_{1,1}-34 H_{1,1}-128 H_{-1,-1,0}+352 H_{-1,0,0}+144 H_{0,-1,0}-476 H_{0,0,0}-68 H_{0,0,1}-304 H_{0,1,0}-80 H_{0,1,1}-32 H_{1,0,0}-60 H_{1,0,1}+12 H_{1,1,0}-12 H_{1,1,1}-192 H_{-1,-1,0,0}-96 H_{-1,0,-1,0}+176 H_{-1,0,0,0}+64 H_{0,-1,-1,0}+64 H_{0,-1,0,0}+32 H_{0,0,-1,0}-128 H_{0,0,0,0}-8 H_{0,0,0,1}-96 H_{0,0,1,0}-64 H_{0,0,1,1}+144 H_{0,1,0,0}+104 H_{0,1,0,1}+120 H_{0,1,1,0}+56 H_{0,1,1,1}+96 H_{1,0,-1,0}-336 H_{1,0,0,0}-80 H_{1,0,0,1}-304 H_{1,0,1,0}-64 H_{1,0,1,1}-16 H_{1,1,0,0}-80 H_{1,1,0,1}-48 H_{1,1,1,0}-32 H_{1,1,1,1}-\frac{7379}{4}\bigg]
+C_A N_f^2\bigg[\left(-16 \zeta _2-16 H_{-1,0}+16 H_{0,0}\right) z^2+\left(\left(-\frac{1000}{9}\right)H_{0,0}+\left(-\frac{304}{3}\right)H_{1,0,0}+\left(-\frac{2576}{27}\right)H_0+\left(-\frac{784}{9}\right)H_{1,0}+\left(-\frac{640}{9}\right)H_{1,1}+\left(-\frac{128}{3}\right)H_{0,-1,0}+\left(-\frac{128}{3}\right)H_{0,0,0}+\left(-\frac{64}{3}\right)H_{-1,0,1}+\left(-\frac{32}{3}\right)H_{1,0,1}+\left(-\frac{32}{3}\right)H_{1,1,0}+\frac{16}{3}\zeta _2+\frac{64}{3}H_{-1,0,0}+\frac{80}{3}H_{1,1,1}+\frac{128}{3}H_{-1,-1,0}+\frac{160}{3}H_{-1,0}+\frac{592}{9}H_{0,1}+\frac{208}{3}H_{0,0,1}+\frac{2456}{27}H_1+\left(-\frac{64 \zeta _2}{3}\right)H_1+\frac{128 \zeta _2}{3}H_{-1}-16 H_0 \zeta _2-80 \zeta _3+16 H_{0,1,0}-64 H_{0,1,1}+70\right) z+\left(-\frac{320}{3}\right)H_{0,0,0}+\left(-\frac{748}{9}\right)H_{0,0}+\left(-\frac{902}{27}\right)H_0+\left(-\frac{284}{9}\right)H_{0,1}+\left(-\frac{784}{27}\right)H_1+\left(-\frac{40}{3}\right)H_{1,1,1}+\left(-\frac{32}{3}\right)H_{-1,0,1}+\frac{16}{3}H_{1,0,1}+\frac{16}{3}H_{1,1,0}+\frac{32}{3}H_{-1,0,0}+\frac{64}{3}H_{-1,-1,0}+\frac{64}{3}H_{0,-1,0}+\frac{296}{9}H_{1,1}+\frac{128}{3}H_{-1,0}+\frac{152}{3}H_{1,0,0}+\frac{176}{3}\zeta _3+\frac{196}{3}\zeta _2+\frac{728}{9}H_{1,0}+\frac{8 \zeta _2}{3}H_0+\frac{32 \zeta _2}{3}H_1+\frac{64 \zeta _2}{3}H_{-1}-8 H_{0,0,1}+40 H_{0,1,0}+48 H_{0,1,1}-\frac{988}{9}\bigg]
+C_A C_FN_f \bigg[\left(-72 \zeta _2-72 H_{-1,0}+72 H_{0,0}\right) z^2+\left(\left(-\frac{3484}{3}\right)H_{1,0}+\left(-\frac{25828}{27}\right)H_1+\left(-\frac{25589}{27}\right)H_0+\left(-\frac{7276}{9}\right)\zeta _2+\left(-\frac{2284}{9}\right)H_{1,1}+\left(-\frac{536}{3}\right)H_{0,1,1}+\frac{128}{3}H_{0,0,1}+\frac{272}{3}H_{0,1,0}+\frac{280}{3}H_{0,0,0}+\frac{584}{3}H_{1,1,1}+\frac{784}{3}\zeta _3+\frac{848}{3}H_{1,0,1}+\frac{1168}{3}H_{1,1,0}+\frac{5774}{9}H_{0,0}+\frac{10156}{9}H_{0,1}+\frac{80 \zeta _2}{3}H_0+176 H_{-1} \zeta _2-16 H_1 \zeta _2-80 H_{-1} \zeta _3-112 H_0 \zeta _3-1584 H_1 \zeta _3-44 \zeta _4+160 \zeta _2 H_{-1,-1}+48 H_{-1,0}-128 \zeta _2 H_{0,-1}+64 \zeta _2 H_{0,0}-400 \zeta _2 H_{0,1}+160 \zeta _2 H_{1,0}+64 \zeta _2 H_{1,1}+224 H_{-1,-1,0}-288 H_{-1,0,0}-64 H_{-1,0,1}-176 H_{0,-1,0}+344 H_{1,0,0}+192 H_{-1,-1,-1,0}+128 H_{-1,-1,0,0}-64 H_{-1,-1,0,1}-160 H_{-1,0,-1,0}-48 H_{-1,0,0,0}+192 H_{-1,0,1,0}+64 H_{-1,0,1,1}-256 H_{0,-1,-1,0}+288 H_{0,0,-1,0}-64 H_{0,0,0,0}+128 H_{0,0,0,1}-320 H_{0,0,1,0}-352 H_{0,0,1,1}+272 H_{0,1,0,0}+336 H_{0,1,0,1}+368 H_{0,1,1,0}+272 H_{0,1,1,1}+160 H_{1,0,-1,0}-112 H_{1,0,0,0}+224 H_{1,0,0,1}-480 H_{1,0,1,0}-64 H_{1,0,1,1}+416 H_{1,1,0,0}-96 H_{1,1,0,1}-32 H_{1,1,1,0}-128 H_{1,1,1,1}-\frac{9751}{6}\right) z+\left(-\frac{3016}{9}\right)\zeta _2+\left(-\frac{788}{3}\right)H_{1,1,0}+\left(-\frac{556}{3}\right)H_{1,0,1}+\left(-\frac{220}{3}\right)H_{0,0,1}+\left(-\frac{184}{3}\right)H_{1,1,1}+\frac{640}{9}H_{0,1}+\frac{244}{3}H_{0,0,0}+\frac{1064}{9}H_{1,1}+\frac{400}{3}H_{0,1,1}+\frac{716}{3}H_{0,1,0}+\frac{4607}{9}H_{0,0}+\frac{1904}{3}H_{1,0}+\frac{2872}{3}\zeta _3+\frac{28226}{27}H_1+\frac{31993}{27}H_0+\left(-\frac{364 \zeta _2}{3}\right)H_0+80 H_{-1} \zeta _2+12 H_1 \zeta _2-40 H_{-1} \zeta _3+8 H_0 \zeta _3+792 H_1 \zeta _3+658 \zeta _4+80 \zeta _2 H_{-1,-1}+8 H_{-1,0}-16 \zeta _2 H_{0,-1}-128 \zeta _2 H_{0,0}+296 \zeta _2 H_{0,1}-80 \zeta _2 H_{1,0}-32 \zeta _2 H_{1,1}+128 H_{-1,-1,0}-240 H_{-1,0,0}-16 H_{-1,0,1}+16 H_{0,-1,0}-592 H_{1,0,0}+96 H_{-1,-1,-1,0}+64 H_{-1,-1,0,0}-32 H_{-1,-1,0,1}-80 H_{-1,0,-1,0}-24 H_{-1,0,0,0}+96 H_{-1,0,1,0}+32 H_{-1,0,1,1}-32 H_{0,-1,-1,0}-48 H_{0,-1,0,0}+16 H_{0,0,-1,0}-32 H_{0,0,0,0}+48 H_{0,0,0,1}-80 H_{0,0,1,0}+96 H_{0,0,1,1}-520 H_{0,1,0,0}-312 H_{0,1,0,1}-328 H_{0,1,1,0}-184 H_{0,1,1,1}-80 H_{1,0,-1,0}+56 H_{1,0,0,0}-112 H_{1,0,0,1}+240 H_{1,0,1,0}+32 H_{1,0,1,1}-208 H_{1,1,0,0}+48 H_{1,1,0,1}+16 H_{1,1,1,0}+64 H_{1,1,1,1}+\frac{52313}{36}\bigg]
\,.
\end{dmath}

\begin{dmath}[style={\small},compact]
\Delta P_{gg}^{T,\text{s}(0)}=
4 C_A \left[\frac{1}{1-z}\right]_++\delta (1-z) \left(\frac{11 C_A}{3}-\frac{2 N_f}{3}\right)+\left(-\frac{8 z^2}{z-1}+\frac{12 z}{z-1}-\frac{4}{z-1}\right) C_A
\,,
\end{dmath}

\begin{dmath}[style={\small},compact]
\Delta P_{gg}^{T,\text{s}(1)}=
\left[\frac{1}{1-z}\right]_+ \bigg[\left(\frac{268}{9}-8 \zeta _2\right) C_A^2-\frac{40 C_A N_f}{9}\bigg]
+C_A^2 \bigg[\frac{z^2}{z-1}\left(-64 H_{0,0}+\frac{1}{z+1}-\frac{74}{9}\right)+\frac{z^2}{z-1}\left(32 H_{0,1}+32 H_{1,0}-20 H_0\right)+\frac{32 z^2}{z+1}H_{-1,0}+\frac{1}{z-1}\left(32 H_{0,1}+32 H_{1,0}+\frac{92}{3}H_0-8 \zeta _2+\frac{268}{9}\right)+\frac{z}{z-1}\left(16 H_{0,0}+\frac{1}{z+1}-16 \zeta _2-\frac{74}{9}\right)+\frac{z}{z-1}\left(-48 H_{0,1}-48 H_{1,0}-40 H_0\right)+\frac{32}{z+1}H_{-1,0}+\frac{1}{(z-1) (z+1)}\left(96 H_{0,0}-\frac{194}{9}\right)+\frac{48 z}{z+1}H_{-1,0}+\frac{z^3 \left(32 \zeta _2-\frac{194}{9}\right)}{(z-1) (z+1)}\bigg]
+C_F N_f \bigg[z \left(8 H_{0,0}+4 H_0-28\right)+8 H_{0,0}+28 H_0+28\bigg]
+\delta (1-z) \bigg[-\frac{8 C_A N_f}{3}+\left(12 \zeta _3+\frac{32}{3}\right) C_A^2
-2 C_F N_f\bigg]
+C_A N_f \bigg[\frac{z^2}{z-1}\left(8 H_0+\frac{152}{9}\right)+\frac{1}{z-1}\left(\frac{40}{3}H_0+\frac{152}{9}\right)+\frac{z}{z-1}\left(-16 H_0-\frac{88}{3}\right)-\frac{40}{9 (z-1)}\bigg]
\,,
\end{dmath}

\begin{dmath*}[style={\small},compact]
\Delta P_{gg}^{T,\text{s}(2)}=
 C_A^3
\bigg[\left(\left(-\frac{176}{3}\right)H_{-1,-1,0}+\frac{176}{3}H_{-1,0,1}\right) z^2+\left(336 H_{-1,-1,0}+368 H_{-1,0,1}\right) z+\frac{1}{z-1}\left(\left(-\frac{544}{3}\right)H_{1,0,0}+\left(-\frac{1072}{9}\right)\zeta _2+\left(-\frac{8}{9}\right)H_1+\frac{88}{3}\zeta _3+\frac{2920}{9}H_{1,0}+\frac{416 \zeta _2}{3}H_1+192 H_1 \zeta _3+88 \zeta _4-256 \zeta _2 H_{0,-1}-64 \zeta _2 H_{1,0}-208 H_{0,0,1}-144 H_{0,1,0}-256 H_{0,0,1,1}-608 H_{0,1,0,0}-256 H_{0,1,0,1}-256 H_{0,1,1,0}+128 H_{1,0,-1,0}-640 H_{1,0,0,0}-256 H_{1,0,0,1}-256 H_{1,0,1,0}-256 H_{1,1,0,0}+\frac{490}{3}\right)+\frac{1}{z}\left(\left(-\frac{176}{3}\right)H_{-1,-1,0}+\frac{176}{3}H_{-1,0,1}\right)+\frac{z}{z-1}\left(\left(-\frac{48478}{27}\right)H_0+\left(-\frac{2968}{3}\right)\zeta _3+\left(-\frac{320}{3}\right)H_{0,1}+\left(-\frac{284}{3}\right)H_{0,0}+\left(-\frac{592}{9}\right)\zeta _2+\frac{320}{3}H_{0,-1,0}+\frac{280 \zeta _2}{3}H_0+472 \zeta _4-96 \zeta _2 H_{0,0}+160 \zeta _2 H_{0,1}-112 H_{0,0,0}+64 H_{0,-1,-1,0}-224 H_{0,-1,0,0}-128 H_{0,-1,0,1}-64 H_{0,0,-1,0}+128 H_{0,0,0,0}-160 H_{0,0,0,1}-64 H_{0,0,1,0}+\frac{1}{z+1}+\frac{46102}{9}\right)+\frac{z}{z-1}\left(\left(-\frac{1232}{3}\right)H_{1,0}+\frac{16}{9}H_1-336 H_1 \zeta _2-288 H_1 \zeta _3+416 \zeta _2 H_{0,-1}+96 \zeta _2 H_{1,0}+584 H_{0,0,1}+320 H_{0,1,0}+568 H_{1,0,0}+384 H_{0,0,1,1}+672 H_{0,1,0,0}+384 H_{0,1,0,1}+384 H_{0,1,1,0}-192 H_{1,0,-1,0}+960 H_{1,0,0,0}+384 H_{1,0,0,1}+384 H_{1,0,1,0}+384 H_{1,1,0,0}\right)+\frac{z^2}{z-1}\left(\left(-\frac{3232}{3}\right)H_{0,-1,0}+\left(-\frac{9136}{9}\right)\zeta _2+\left(-\frac{592}{9}\right)H_{0,1}+\frac{16964}{9}H_{0,0}+\frac{6736}{3}H_{0,0,0}+\frac{8990}{3}H_0+\left(-\frac{2024 \zeta _2}{3}\right)H_0-800 H_0 \zeta _3-920 \zeta _3-576 \zeta _4-608 \zeta _2 H_{0,0}-32 \zeta _2 H_{0,1}+320 H_{0,-1,-1,0}-544 H_{0,-1,0,0}-704 H_{0,0,-1,0}+1984 H_{0,0,0,0}+800 H_{0,0,0,1}+512 H_{0,0,1,0}+\frac{1}{z+1}+\frac{46102}{9}\right)
\end{dmath*}

\begin{dmath*}[style={\small},compact]
\left.
+\frac{z^4}{z-1}\bigg(-\frac{176}{3}H_{0,0,0}+\frac{176}{3}H_{0,-1,0}+\frac{440}{3}\zeta _3+\frac{352 \zeta _2}{3}H_0+\frac{1}{z+1}\bigg)
\right.
+\frac{z^2}{z-1}\bigg(-\frac{544}{3}H_{1,0,0}-\frac{8}{9}H_1+\frac{176}{3}H_{0,1,0}+\frac{2920}{9}H_{1,0}+\frac{416 \zeta _2}{3}H_1+192 H_1 \zeta _3-256 \zeta _2 H_{0,-1}-64 \zeta _2 H_{1,0}-112 H_{0,0,1}-256 H_{0,0,1,1}-288 H_{0,1,0,0}-256 H_{0,1,0,1}-256 H_{0,1,1,0}+128 H_{1,0,-1,0}-640 H_{1,0,0,0}-256 H_{1,0,0,1}-256 H_{1,0,1,0}-256 H_{1,1,0,0}\bigg)
+\frac{z^3}{z-1}\bigg(-\frac{176}{3}H_{0,0,1}+\frac{88 \zeta _2}{3}H_1\bigg)
+\frac{z^3}{z-1}\bigg(\frac{1060}{9}H_{0,0}+\frac{3104}{9}H_{0,1}+\frac{2792}{3}\zeta _3+\frac{13402}{9}H_0+\frac{160 \zeta _2}{3}H_0+304 \zeta _2-680 \zeta _4+96 \zeta _2 H_{0,0}-256 \zeta _2 H_{0,1}-48 H_{0,-1,0}+464 H_{0,0,0}+128 H_{0,-1,0,0}+256 H_{0,-1,0,1}-64 H_{0,0,-1,0}-256 H_{0,0,0,0}-96 H_{0,0,0,1}-128 H_{0,0,1,0}+\frac{1}{z+1}-\frac{47572}{9}\bigg)
+\frac{1}{z+1}\bigg(-\frac{6032}{9}H_{-1,0}-288 H_{-1} \zeta _2-384 H_{-1} \zeta _3+512 \zeta _2 H_{-1,-1}-64 \zeta _2 H_{-1,0}-480 H_{-1,0,0}+256 H_{-1,-1,0,0}-512 H_{-1,-1,0,1}+256 H_{-1,0,-1,0}-640 H_{-1,0,0,0}-128 H_{-1,0,1,0}\bigg)
+\frac{1}{(z-1) (z+1)}\bigg(-\frac{89182}{27}H_0-\frac{14612}{9}H_{0,0}-\frac{4448}{3}H_{0,0,0}+\frac{2144}{3}\zeta _3+\frac{2528}{3}H_{0,-1,0}+\frac{9136}{9}\zeta _2+528 H_0 \zeta _2+1088 H_0 \zeta _3+808 \zeta _4+864 \zeta _2 H_{0,0}+304 H_{0,1}-512 H_{0,-1,-1,0}+768 H_{0,-1,0,0}+960 H_{0,0,-1,0}-2496 H_{0,0,0,0}-1184 H_{0,0,0,1}-768 H_{0,0,1,0}-\frac{47572}{9}\bigg)
+\frac{1}{z (z+1)}\left(\frac{176}{3}H_{-1,0,0}-88 H_{-1} \zeta _2\right)
+\frac{z}{z+1}\bigg(-\frac{4736}{3}H_{-1,0}-400 H_{-1} \zeta _2-576 H_{-1} \zeta _3+768 \zeta _2 H_{-1,-1}-96 \zeta _2 H_{-1,0}-1312 H_{-1,0,0}+384 H_{-1,-1,0,0}-768 H_{-1,-1,0,1}+384 H_{-1,0,-1,0}-960 H_{-1,0,0,0}-192 H_{-1,0,1,0}\bigg)
+\frac{z^2}{z+1}\bigg(-\frac{6032}{9}H_{-1,0}-288 H_{-1} \zeta _2-384 H_{-1} \zeta _3+512 \zeta _2 H_{-1,-1}-64 \zeta _2 H_{-1,0}-480 H_{-1,0,0}+256 H_{-1,-1,0,0}-512 H_{-1,-1,0,1}+256 H_{-1,0,-1,0}-640 H_{-1,0,0,0}-128 H_{-1,0,1,0}\bigg)
+\frac{z^3}{z+1}\left(\frac{176}{3}H_{-1,0,0}-88 H_{-1} \zeta _2\right)+\frac{88 \zeta _2}{3 (z-1) z}H_1+336 H_{-1,-1,0}+368 H_{-1,0,1}\bigg]
+N_f C_A^2 \bigg[
\left(-\frac{32}{3}H_{-1,-1,0}+\frac{32}{3}H_{-1,0,1}+\frac{16 \zeta _2}{3}H_1-16 H_{-1} \zeta _2\right) z^2
+32 H_{-1,0,1}+96 H_{0,-1,-1,0}
+\left(-\frac{2174}{9}H_1+56 H_{-1} \zeta _2+88 H_1 \zeta _2-48 \zeta _2 H_{0,-1}+176 H_{-1,-1,0}+32 H_{-1,0,1}-96 H_{0,-1,-1,0}+48 H_{0,-1,0,0}+80 H_{0,0,-1,0}+24 H_{0,0,0,1}+24 H_{0,1,0,0}\right) z
+\frac{2174}{9}H_1-48 H_{0,-1,0,0}-16 H_{0,0,-1,0}
+\frac{1}{z-1}\bigg(-\frac{1844}{9}H_{0,1}-\frac{356}{3}H_{1,0,0}-\frac{928}{9}H_{1,0}-\frac{112}{3}\zeta _3+\frac{160}{9}\zeta _2+48 \zeta _2 H_{0,1}-124 H_{0,0,1}-96 H_{0,1,0}-\frac{836}{27}\bigg)
+16 H_{0,0,0,0}+24 H_{0,0,0,1}+24 H_{0,1,0,0}
+\frac{1}{z}\bigg(-\frac{32}{3}H_{-1,-1,0}+\frac{32}{3}H_{-1,0,1}+\left(-\frac{16 \zeta _2}{3}\right)H_1-16 H_{-1} \zeta _2\bigg)
+\frac{z}{z-1}\bigg(-\frac{1396}{9}H_{0,0}-\frac{368}{3}\zeta _3-\frac{1138}{27}H_0+\frac{376}{3}H_{0,-1,0}+\frac{508}{3}\zeta _2+\left(-\frac{100 \zeta _2}{3}\right)H_0-138 \zeta _4+24 \zeta _2 H_{0,0}-32 H_{0,0,0}+\frac{1}{z+1}+\frac{23740}{27}\bigg)
+56 H_{-1} \zeta _2-88 H_1 \zeta _2+48 \zeta _2 H_{0,-1}+176 H_{-1,-1,0}
+\frac{z}{z-1}\left(\frac{512}{3}H_{0,1}+\frac{512}{3}H_{1,0}+200 H_{0,0,1}+128 H_{0,1,0}+200 H_{1,0,0}\right)
\end{dmath*}
\begin{dmath}[style={\small},compact]
+\frac{z^2}{z-1}\left(-\frac{356}{3}H_{1,0,0}
-\frac{928}{9}H_{1,0}+\left(-\frac{308}{3}\right)H_{0,0,1}+\left(-\frac{224}{3}\right)H_{0,1,0}+\left(-\frac{4}{3}\right)H_{0,1}-48 \zeta _2 H_{0,1}\right)
+\frac{z^2}{z-1}\left(-\frac{832}{3}\zeta _3-\frac{2140}{9}\zeta _2+\frac{1880}{3}H_{0,0}+\frac{6308}{9}H_0-116 H_0 \zeta _2-96 H_0 \zeta _3+78 \zeta _4-40 \zeta _2 H_{0,0}-104 H_{0,-1,0}+392 H_{0,0,0}+\frac{1}{z+1}+\frac{23740}{27}\right)
+\left(-\frac{32 z^3}{3 (z-1)}\right)H_{0,0,1}+\frac{z^3}{z-1}\left(\left(-\frac{1844}{9}\right)\zeta _2+\frac{1246}{9}H_0+\frac{496}{3}\zeta _3+\frac{1780}{9}H_{0,0}+\frac{20 \zeta _2}{3}H_0+138 \zeta _4-24 \zeta _2 H_{0,0}-136 H_{0,-1,0}-32 H_{0,0,0}+\frac{1}{z+1}-\frac{22904}{27}\right)
+\frac{z^4}{z-1}\left(\left(-\frac{32}{3}\right)H_{0,0,0}+\frac{32}{3}H_{0,-1,0}+\frac{80}{3}\zeta _3+\frac{64 \zeta _2}{3}H_0+16 \zeta _2-16 H_{0,0}+\frac{1}{z+1}\right)
+\frac{1}{z+1}\left(\left(-\frac{1864}{9}\right)H_{-1,0}+\left(-\frac{440}{3}\right)H_{-1,0,0}\right)
+\frac{1}{(z-1) (z+1)}\left(\left(-\frac{16324}{27}\right)H_0+\left(-\frac{5432}{9}\right)H_{0,0}+\left(-\frac{1528}{3}\right)H_{0,0,0}+\frac{376}{3}H_{0,-1,0}+\frac{1996}{9}\zeta _2+\frac{880}{3}\zeta _3+100 H_0 \zeta _2+96 H_0 \zeta _3-78 \zeta _4+40 \zeta _2 H_{0,0}-\frac{22904}{27}\right)
+\frac{1}{z (z+1)}\left(\frac{32}{3}H_{-1,0,0}+16 H_{-1,0}\right)
+\frac{z}{z+1}\left(\left(-\frac{1232}{3}\right)H_{-1,0}-272 H_{-1,0,0}\right)
+\frac{z^2}{z+1}\left(\left(-\frac{1864}{9}\right)H_{-1,0}+\left(-\frac{440}{3}\right)H_{-1,0,0}\right)
+\frac{z^3}{z+1}\left(\frac{32}{3}H_{-1,0,0}+16 H_{-1,0}\right)
\bigg] 
+C_A
N_f^2
\bigg[-\frac{82}{9}H_1-\frac{32}{9}H_{0,1}+z \left(-\frac{32}{9}H_{0,1}+\frac{82}{9}H_1\right)+\frac{1}{z-1}\left(\left(-\frac{176}{9}\right)H_{0,0}+\left(-\frac{496}{27}\right)H_0+\left(-\frac{32}{9}\right)\zeta _2+\frac{230}{27}\right)+\frac{z}{z-1}\left(\frac{64}{3}H_{0,0}+30 H_0-\frac{148}{9}\right)+\frac{z^2}{z-1}\left(\left(-\frac{158}{9}\right)H_0+\left(-\frac{80}{9}\right)H_{0,0}+\frac{32}{9}\zeta _2+\frac{230}{27}\right)-\frac{16}{27 (z-1)}\bigg]
+N_f C_F C_A \bigg[\left(\left(-\frac{64}{3}\right)H_{-1,0,0}+\left(-\frac{64}{3}\right)H_{-1,0,1}+\left(-\frac{64}{3}\right)H_{0,-1,0}+\frac{64}{3}H_{-1,-1,0}+\frac{64}{3}H_{0,0,0}+\frac{64}{3}H_{0,0,1}+\left(-\frac{32 \zeta _2}{3}\right)H_1+32 H_{-1} \zeta _2-48 H_{-1,0}+48 H_{0,0}\right) z^2
+\left(\left(-\frac{1756}{9}\right)H_1+\left(-\frac{488}{3}\right)H_{0,1,0}+\left(-\frac{296}{3}\right)H_{0,0,0}+\left(-\frac{152}{3}\right)H_{0,1,1}+\frac{260}{3}H_{1,1}+\frac{2932}{9}H_{0,1}+\frac{1004}{3}H_{1,0}+\frac{1352}{3}H_{-1,0}+\frac{2168}{3}H_{0,0}-176 H_{-1} \zeta _2-640 H_1 \zeta _2+128 \zeta _2 H_{0,-1}-16 \zeta _2 H_{0,0}+288 \zeta _2 H_{0,1}-480 H_{-1,-1,0}+368 H_{-1,0,0}-64 H_{-1,0,1}+272 H_{0,-1,0}+8 H_{0,0,1}+712 H_{1,0,0}+240 H_{1,0,1}+160 H_{1,1,0}+80 H_{1,1,1}+256 H_{0,-1,-1,0}-192 H_{0,-1,0,0}-256 H_{0,0,-1,0}+128 H_{0,0,0,0}-112 H_{0,0,0,1}+80 H_{0,0,1,0}+48 H_{0,0,1,1}-304 H_{0,1,0,0}-96 H_{0,1,0,1}-64 H_{0,1,1,0}-32 H_{0,1,1,1}\right) z
+\left(-\frac{3632}{3}\right)H_{0,0}+\left(-\frac{3056}{3}\right)H_{0,0,0}+\left(-\frac{1004}{3}\right)H_{1,0}+\left(-\frac{260}{3}\right)H_{1,1}+\left(-\frac{128}{3}\right)H_{0,1,0}+\frac{40}{9}H_{0,1}+\frac{88}{3}H_{0,1,1}+\frac{1756}{9}H_1+\frac{1352}{3}H_{-1,0}+\frac{1}{z-1}\left(\left(-\frac{1936}{3}\right)\zeta _3+\left(-\frac{2672}{9}\right)\zeta _2+\frac{51500}{27}H_0+\left(-\frac{416 \zeta _2}{3}\right)H_0-512 H_0 \zeta _3-468 \zeta _4+\frac{88478}{27}\right)
+\frac{1}{z}\left(-\frac{64}{3}H_{-1,0,0}+\left(-\frac{64}{3}\right)H_{-1,0,1}+\frac{64}{3}H_{-1,-1,0}+\frac{32 \zeta _2}{3}H_1+32 H_{-1} \zeta _2-48 H_{-1,0}\right)+\frac{z}{z-1}\left(\left(-\frac{3148}{9}\right)H_0+\frac{1420}{3}\zeta _2+168 H_0 \zeta _2+352 H_0 \zeta _3+1240 \zeta _3+160 \zeta _4-\frac{177946}{27}\right)+\frac{z^2}{z-1}\left(\left(-\frac{41624}{27}\right)H_0+\left(-\frac{1624}{3}\right)\zeta _3+\left(-\frac{1156}{9}\right)\zeta _2+\frac{40 \zeta _2}{3}H_0+160 H_0 \zeta _3+308 \zeta _4+\frac{89468}{27}\right)+\frac{z^3}{z-1}\left(\left(-\frac{160}{3}\right)\zeta _3+\left(-\frac{128 \zeta _2}{3}\right)H_0-48 \zeta _2\right)
\bigg]
\end{dmath}
\begin{dmath}[style={\small},compact, indentstep={0 em}]
-176 H_{-1} \zeta _2+640 H_1 \zeta _2-128 \zeta _2 H_{0,-1}+112 \zeta _2 H_{0,0}+288 \zeta _2 H_{0,1}-480 H_{-1,-1,0}+368 H_{-1,0,0}-64 H_{-1,0,1}+368 H_{0,-1,0}-32 H_{0,0,1}-712 H_{1,0,0}-240 H_{1,0,1}-160 H_{1,1,0}-80 H_{1,1,1}-256 H_{0,-1,-1,0}+192 H_{0,-1,0,0}+128 H_{0,0,-1,0}-448 H_{0,0,0,0}-112 H_{0,0,0,1}+80 H_{0,0,1,0}+48 H_{0,0,1,1}-304 H_{0,1,0,0}-96 H_{0,1,0,1}-64 H_{0,1,1,0}-32 H_{0,1,1,1}
+\delta (1-z) \bigg[\left(\frac{8}{3}\zeta _2+\frac{55}{3}\zeta _4+\frac{536}{3}\zeta _3-16 \zeta _2 \zeta _3-80 \zeta _5+\frac{79}{2}\right) C_A^3+\left(\left(-\frac{80}{3}\right)\zeta _3+\left(-\frac{10}{3}\right)\zeta _4+\left(-\frac{8}{3}\right)\zeta _2-\frac{233}{18}\right) N_f C_A^2+\left(\frac{29 N_f^2}{18}-\frac{241 C_F N_f}{18}\right) C_A+\frac{11}{9} C_F N_f^2+C_F^2 N_f\bigg]
+\left[\frac{1}{1-z}\right]_+ \bigg[\left(\left(-\frac{1072}{9}\right)\zeta _2+\frac{88}{3}\zeta _3+88 \zeta _4+\frac{490}{3}\right) C_A^3+\left(\left(-\frac{112}{3}\right)\zeta _3+\frac{160}{9}\zeta _2-\frac{836}{27}\right) N_f C_A^2+\left(C_F N_f \left(32 \zeta _3-\frac{110}{3}\right)-\frac{16 }{27} N_f^2 \right) C_A\bigg]
+C_F N_f^2 \bigg[\left(-\frac{64}{3}\right)\zeta _3+\left(-\frac{64}{3}\right)H_{0,0,0}+\left(-\frac{40}{3}\right)H_{1,0}+\left(-\frac{40}{3}\right)H_{1,1}+\left(-\frac{328}{27}\right)H_0+\left(-\frac{16}{3}\right)H_{0,1,0}+\left(-\frac{16}{3}\right)H_{0,1,1}+\frac{136}{9}\zeta _2+\frac{224}{9}H_1+\frac{344}{9}H_{0,1}+\frac{16 \zeta _2}{3}H_0-32 H_{0,0}+16 H_{0,0,1}+z \left(\left(-\frac{488}{9}\right)\zeta _2+\left(-\frac{224}{9}\right)H_1+\left(-\frac{64}{3}\right)\zeta _3+\left(-\frac{64}{3}\right)H_{0,0}+\left(-\frac{64}{3}\right)H_{0,0,0}+\left(-\frac{16}{3}\right)H_{0,1,0}+\left(-\frac{16}{3}\right)H_{0,1,1}+\frac{8}{9}H_{0,1}+\frac{40}{3}H_{1,0}+\frac{40}{3}H_{1,1}+\frac{776}{27}H_0+\frac{16 \zeta _2}{3}H_0+16 H_{0,0,1}-\frac{992}{27}\right)+\frac{992}{27}\bigg]
+C_F^2 N_f \bigg[\left(32 \zeta _2+32 H_{-1,0}-32 H_{0,0}\right) z^2+\left(-48 \zeta _2 H_0-160 \zeta _3 H_0+1032 H_0+580 H_1+128 H_{-1} \zeta _2+528 H_1 \zeta _2+580 \zeta _2-48 \zeta _3-380 \zeta _4+80 H_{-1,0}-64 \zeta _2 H_{0,-1}+16 \zeta _2 H_{0,0}-664 H_{0,0}-224 \zeta _2 H_{0,1}-404 H_{0,1}-68 H_{1,0}-52 H_{1,1}+256 H_{-1,-1,0}-128 H_{-1,0,0}-256 H_{0,-1,0}+80 H_{0,0,1}+112 H_{0,1,0}+32 H_{0,1,1}-568 H_{1,0,0}-240 H_{1,0,1}-160 H_{1,1,0}-80 H_{1,1,1}-128 H_{0,-1,-1,0}+64 H_{0,-1,0,0}+128 H_{0,0,-1,0}-32 H_{0,0,0,0}-16 H_{0,0,0,1}-144 H_{0,0,1,0}-48 H_{0,0,1,1}+240 H_{0,1,0,0}+96 H_{0,1,0,1}+64 H_{0,1,1,0}+32 H_{0,1,1,1}-664\right) z+\frac{32}{z}H_{-1,0}+252 H_0-580 H_1+128 H_{-1} \zeta _2+64 H_0 \zeta _2-528 H_1 \zeta _2+152 \zeta _2-288 H_0 \zeta _3-272 \zeta _3-460 \zeta _4+80 H_{-1,0}+64 \zeta _2 H_{0,-1}+16 \zeta _2 H_{0,0}+16 H_{0,0}-224 \zeta _2 H_{0,1}-248 H_{0,1}+68 H_{1,0}+52 H_{1,1}+256 H_{-1,-1,0}-128 H_{-1,0,0}-64 H_{0,0,0}-160 H_{0,0,1}-192 H_{0,1,0}-48 H_{0,1,1}+568 H_{1,0,0}+240 H_{1,0,1}+160 H_{1,1,0}+80 H_{1,1,1}+128 H_{0,-1,-1,0}-64 H_{0,-1,0,0}-32 H_{0,0,0,0}-16 H_{0,0,0,1}-144 H_{0,0,1,0}-48 H_{0,0,1,1}+240 H_{0,1,0,0}+96 H_{0,1,0,1}+64 H_{0,1,1,0}+32 H_{0,1,1,1}+664\bigg]
\,.
\end{dmath}
 \subsection{N$^3$LO coefficient functions and  small-x expansion}
\label{sec:n3lo-coeff-helicity}
The analytic expressions for the  coefficient functions  will be provided in the ancillary files along with the arXiv submission.
In this section we  study the asymptotic behaviors and present their numerical fits\footnote{To determine the 3-loop pure-singlet $\Delta\cI^{(3)}_{qq'}$ in ${\overline{\text{MS}}}$, we also need $z_{\text{ps}}^{(3)}$, which  is currently not known.}.

The coefficient functions develop end-point divergences both in the threshold and high energy limit.
In the $z \to 1$ limit, we have exactly the same results as the unpolarized case~\cite{Luo:2019szz,Luo:2020epw}
\begin{equation}
  \label{eq:largez}
 \lim_{z\to1} \Delta \mathcal{I}^{(2)}_{ij}(z)= \lim_{z\to1}\Delta \mathcal{C}^{(2)}_{ji}(z) = \frac{2 \gamma^R_{1,i}}{(1-z)_+}\delta_{ij} \,, \quad 
  \lim_{z\to1} \Delta \mathcal{I}^{(3)}_{ij}(z)=\lim_{z\to1}\Delta \mathcal{C}^{(3)}_{ji}(z) = \frac{2 \gamma^R_{2,i}}{(1-z)_+} \delta_{ij} \,,
\end{equation}
where $\gamma^R_{1(2)}$ are the two(three)-loop rapidity anomalous dimensions~\cite{Li:2016ctv,Vladimirov:2016dll}. The relation between threshold limit and rapidity anomalous dimension has been anticipated in \cite{Echevarria:2016scs,Lustermans:2016nvk,Billis:2019vxg}.
 We also found that threshold limit exhibits Casimir scaling up to three loops~($n=1,2,3$),
\begin{align}
\lim_{z\to1}\frac{\Delta\mathcal{I}^{(n)}_{gg}(z)}{\Delta\mathcal{I}^{(n)}_{qq}(z)}=\lim_{z\to1}\frac{\Delta\mathcal{C}^{(n)}_{gg}(z)}{\Delta\mathcal{C}^{(n)}_{qq}(z)}=\frac{C_A}{C_F} \,.
\end{align}
It is also instructive to examine the small-x limit and compare with the unpolarized case. While the unpolarized coefficient functions scale as $1/x$, 
the helicity-dependent ones scale as $x^0$.
 A systematic resummation of the associated logarithms is therefore of interest. Significant progress has been made toward NLL small-x resummation~\cite{Catani:1994sq,Ciafaloni:1998iv,Salam:1998tj,Ciafaloni:1999yw,Ciafaloni:2003rd,Ciafaloni:2003kd,Ciafaloni:2003ek,Altarelli:1999vw,Altarelli:2001ji,Altarelli:2003hk,Altarelli:2005ni,Marzani:2007gk,Neill:2023jcd,Kovchegov:2015pbl,Kovchegov:2016weo,Kovchegov:2017jxc,Kovchegov:2017lsr}. For unpolarized TMD FFs, such logarithms were resummed to all orders in $\alpha_s$ up to NNLL in our previous work~\cite{Luo:2020epw},  
 based on the algorithm of Ref.~\cite{Vogt:2011jv}. Below we collect the small-x data for the physical coefficient functions,
we use superscripts `s' to indicate the corresponding results are in the  singlet sector
\begin{align}
\Delta \mathscr{I}^{\text{s} (1)}_{qq} (x)\simeq&
2 C_F
\,,
\end{align}
\begin{align}
\Delta \mathscr{I}^{\text{s}(2)}_{qq} (x)\simeq&
C_A C_F \left(\left(-\frac{2}{3}\right)\ln ^3x+\left(-\frac{11}{6}\right)\ln ^2x+\left(-\frac{31}{9}\right)\ln x-12 \zeta _2+4 \zeta _3+\frac{397}{27}\right)
\nn\\
+&C_F N_f T_F \left(\frac{4}{3}\ln ^3x+\frac{32}{3}\ln ^2x+\frac{344}{9}\ln x-16 \zeta _2+\frac{2452}{27}\right)
\nn\\
+&C_F^2 \left(\ln ^3x+3 \ln ^2x+4 \ln x+18 \zeta _2+20 \zeta _3-52\right)
\,.
\end{align}
\begin{align}
\Delta \mathscr{I}^{\text{s}(1)}_{qg} (x)\simeq&
8 N_f T_F
\,,
\end{align}

\begin{align}
\Delta \mathscr{I}^{\text{s}(2)}_{qg} (x)\simeq&
C_A N_f T_F \left(\frac{4}{3}\ln ^3x+22 \ln ^2x+72 \ln x-32 \zeta _2+8 \zeta _3+140\right)
\nn\\
+&C_F N_f T_F \left(\frac{2}{3}\ln ^3x+13 \ln ^2x+56 \ln x-24 \zeta _2-56 \zeta _3+142\right)
\,,
\end{align}
\begin{align}
\Delta \mathscr{I}^{\text{s}(3)}_{qg} (x)\simeq&
C_A^2 N_f T_F \bigg[\left(\frac{32308}{81}-12 \zeta _2\right) \ln ^3x+\left(\left(-\frac{424}{3}\right)\zeta _2+\left(-\frac{328}{3}\right)\zeta _3+\frac{19276}{9}\right) \ln ^2x
\nn\\
+&\left(\left(-\frac{38}{3}\right)\zeta _4+\frac{376}{3}\zeta _2-1024 \zeta _3+\frac{1376872}{243}\right) \ln x+2 \ln ^5x+\frac{410}{9}\ln ^4x+\frac{11456}{9}\zeta _4
\nn\\
+&\left(\left(-\frac{968}{3}\right)\zeta _3+\frac{170936}{81}\right) \zeta _2-560 \zeta _3+1608 \zeta _5-\frac{16984}{81}\bigg]
\nn\\
+& C_A C_F N_f T_F \bigg[\left(\left(-\frac{124}{9}\right)\zeta _2+\frac{14821}{81}\right) \ln ^3x+\left(\left(-\frac{458}{3}\right)\zeta _2-88 \zeta _3+\frac{59860}{81}\right) \ln ^2x
\nn\\
+&\left(\left(-\frac{1288}{3}\right)\zeta _2+\frac{260}{3}\zeta _4+\frac{1496}{3}\zeta _3+\frac{108581}{243}\right) \ln x+\frac{32}{45}\ln ^5x+\frac{523}{27}\ln ^4x
\nn\\
+&\left(-\frac{3896}{3}\right)\zeta _5+\frac{4438}{9}\zeta _4+\frac{39656}{27}\zeta _3+\zeta _2 \left(-24 \zeta _3-\frac{80504}{81}\right)+\frac{62153}{324}\bigg]
\nn\\
+&C_A N_f^2 T_F^2 \bigg[\left(\frac{64}{3}\zeta _2-\frac{336176}{243}\right) \ln x+\left(-\frac{16}{9}\right)\ln ^4x+\left(-\frac{3320}{81}\right)\ln ^3x
\nn\\
+&\left(-\frac{3116}{9}\right)\ln ^2x+\left(-\frac{304}{9}\right)\zeta _4+\frac{4640}{81}\zeta _2-\frac{163840}{81}\bigg]
\nn\\
+&C_F^2 N_f T_F \bigg[\left(\left(-\frac{110}{9}\right)\zeta _2-\frac{187}{3}\right) \ln ^3x+\left(\left(-\frac{467}{3}\right)\zeta _2+\left(-\frac{344}{3}\right)\zeta _3-\frac{160}{3}\right) \ln ^2x
\nn\\
+&\left(\left(-\frac{2360}{3}\right)\zeta _3+\left(-\frac{1360}{3}\right)\zeta _2+\left(-\frac{472}{3}\right)\zeta _4+\frac{2125}{3}\right) \ln x+\left(-\frac{8}{45}\right)\ln ^5x
\nn\\
-&6 \ln ^4x+\left(-\frac{4904}{3}\right)\zeta _3+\left(\frac{38}{3}\right)\zeta _4+\frac{3632}{3}\zeta _5+\left(\frac{488}{3}\zeta _3-750\right) \zeta _2+\frac{11735}{12}\bigg]
\nn\\
+&C_F N_f^2 T_F^2 \bigg[\left(32 \zeta _2-\frac{127328}{81}\right) \ln ^2x+\left(128 \zeta _2+64 \zeta _3-\frac{1108564}{243}\right) \ln x+\left(-\frac{16}{15}\right)\ln ^5x
\nn\\
+&\left(-\frac{764}{27}\right)\ln ^4x+\left(-\frac{23300}{81}\right)\ln ^3x+\left(-\frac{848}{9}\right)\zeta _4+\frac{2416}{81}\zeta _2+\frac{5984}{27}\zeta _3-\frac{423361}{81}\bigg]
\,.
\end{align}

\begin{align}
\Delta \mathscr{I}^{\text{s}(1)}_{gq} (x)\simeq&
-4 C_F
\,,
\end{align}

\begin{align}
\Delta \mathscr{I}^{\text{s}(2)}_{gq} (x)\simeq&
C_A C_F \left(\left(-\frac{4}{3}\right)\ln ^3x-16 \ln ^2x-50 \ln x+28 \zeta _2+48 \zeta _3-\frac{4454}{27}\right)
\nn\\
+&\frac{544}{27} C_F N_f T_F+C_F^2 \left(\left(-\frac{2}{3}\right)\ln ^3x-10 \ln ^2x-19 \ln x+23\right)
\,,
\end{align}
\begin{align}
\Delta \mathscr{I}^{\text{s}(3)}_{gq} (x)\simeq&
C_A C_F N_f T_F \bigg[\left(\frac{64}{9}\zeta _2+\frac{19232}{81}\right) \ln ^2x+\left(\frac{3904}{27}\zeta _2+\frac{1472}{9}\zeta _3+\frac{110612}{243}\right) \ln x
\nn\\
+&\frac{32}{27}\ln ^4x+\frac{728}{27}\ln ^3x+\left(-\frac{304}{9}\right)\zeta _4+\left(-\frac{3448}{81}\right)\zeta _2+\frac{9296}{27}\zeta _3+\frac{219356}{243}\bigg]
\nn\\
 +&C_A C_F^2 \bigg[\left(\frac{86}{9}\zeta _2-\frac{11407}{81}\right) \ln ^3x+\left(\left(-\frac{56}{3}\right)\zeta _3+\frac{842}{9}\zeta _2-\frac{67001}{162}\right) \ln ^2x
 \nn\\
 +&\left(\left(-\frac{4208}{9}\right)\zeta _3+\frac{556}{3}\zeta _4+\frac{7025}{27}\zeta _2-\frac{34130}{243}\right) \ln x+\left(-\frac{8}{9}\right)\ln ^5x
  +\left(-\frac{523}{27}\right)\ln ^4x
   \nn\\
  +&\left(-\frac{40228}{27}\right)\zeta _3+\left(-\frac{3212}{9}\right)\zeta _4+\left(\left(-\frac{256}{3}\right)\zeta _3-\frac{15107}{81}\right) \zeta _2+856 \zeta _5+\frac{241279}{108}\bigg]
  \nn\\
+&C_A^2 C_F \bigg[\left(\frac{184}{9}\zeta _2-\frac{7852}{27}\right) \ln ^3x+\left(\frac{1486}{9}\zeta _2+\frac{776}{3}\zeta _3-\frac{128950}{81}\right) \ln ^2x
\nn\\
+&\left(\left(-\frac{5342}{27}\right)\zeta _2+\frac{8120}{9}\zeta _3-78 \zeta _4-\frac{811912}{243}\right) \ln x-2 \ln ^5x+\left(-\frac{973}{27}\right)\ln ^4x
\nn\\
+&\left(-\frac{6214}{9}\right)\zeta _4+\frac{22508}{27}\zeta _3+\left(\frac{376}{3}\zeta _3-\frac{29860}{81}\right) \zeta _2-1736 \zeta _5-\frac{209558}{243}\bigg]
\nn\\
+&C_F^2 N_f T_F \bigg[\left(\frac{32}{9}\zeta _2+\frac{68282}{81}\right) \ln ^2x+\left(\left(-\frac{896}{9}\right)\zeta _3+\left(-\frac{2176}{27}\right)\zeta _2+\frac{568504}{243}\right) \ln x
\nn\\
+&\frac{16}{15}\ln ^5x+\frac{608}{27}\ln ^4x+\frac{15428}{81}\ln ^3x+\left(-\frac{14080}{27}\right)\zeta _3+\frac{496}{9}\zeta _4+\frac{10960}{81}\zeta _2+\frac{66509}{27}\bigg]
\nn\\
+&C_F N_f^2 T_F^2\bigg[\left(-\frac{128}{3}\right)\zeta _3-\frac{5696}{243}\bigg]
+C_F^3 \bigg[\left(8 \zeta _2+75\right) \ln ^3x+\left(\frac{200}{3}\zeta _2+72 \zeta _3
+\frac{1129}{6}\right) \ln ^2x
\nn\\
+&
\left(\frac{220}{3}\zeta _2+\frac{1232}{3}\zeta _3-24 \zeta _4+\frac{328}{3}\right) \ln x+\frac{16}{45}\ln ^5x+\frac{28}{3}\ln ^4x+\frac{1184}{3}\zeta _4+\frac{2516}{3}\zeta _3
\nn\\
+&\zeta _2 \left(144 \zeta _3+284\right)-640 \zeta _5-\frac{12365}{12}\bigg]
\,.
\end{align}
\begin{align}
\Delta \mathscr{I}^{\text{s}(1)}_{gg} (x)\simeq&
-8 C_A
\,,
\end{align}
\begin{align}
\Delta \mathscr{I}^{\text{s}(2)}_{gg} (x)\simeq&
C_A N_f T_F \left(\frac{4}{3}\ln ^2x+\frac{136}{9}\ln x+\frac{1882}{27}\right)+C_F N_f T_F \left(\frac{4}{3}\ln ^3x+18 \ln ^2x+84 \ln x+120\right)
\nn\\
+&C_A^2 \left(\left(-\frac{8}{3}\right)\ln ^3x+\left(-\frac{101}{3}\right)\ln ^2x+\left(-\frac{1109}{9}\right)\ln x+56 \zeta _2+96 \zeta _3-\frac{9020}{27}\right)
\,,
\end{align}
\begin{align}
\Delta \mathscr{I}^{\text{s}(3)}_{gg} (x)\simeq&
C_A^2 N_f T_F \bigg[\left(8 \zeta _2+\frac{3416}{81}\right) \ln ^3x+\left(\frac{488}{9}\zeta _2+112 \zeta _3+\frac{10708}{27}\right) \ln ^2x
\nn\\
+&\left(\frac{2992}{27}\zeta _2+\frac{452}{3}\zeta _4+\frac{6160}{9}\zeta _3+\frac{93668}{81}\right) \ln x+\left(-\frac{4}{15}\right)\ln ^5x+\left(-\frac{34}{27}\right)\ln ^4x
\nn\\
+&\frac{586}{9}\zeta _4+\left(\frac{304}{3}\zeta _3-\frac{5158}{81}\right) \zeta _2+\frac{3520}{9} \zeta _3-336 \zeta _5+\frac{1899556}{729}\bigg]
\nn\\
+ &C_A C_F N_f T_F \bigg[\left(\left(-\frac{236}{9}\right)\zeta _2+\frac{52516}{81}\right) \ln ^3x+\left(\left(-\frac{2242}{9}\right)\zeta _2-288 \zeta _3+\frac{287876}{81}\right) \ln ^2x
\nn\\
+&\left(\left(-\frac{14528}{9}\right)\zeta _3+\left(-\frac{20716}{27}\right)\zeta _2+\left(-\frac{1664}{3}\right)\zeta _4+\frac{2760224}{243}\right) \ln x+\frac{16}{5}\ln ^5x+\frac{1886}{27}\ln ^4x
\nn\\
+&\left(-\frac{48208}{27}\right)\zeta _3+\left(-\frac{6224}{9}\right)\zeta _4+\frac{2192}{3}\zeta _5+\left(\left(-\frac{736}{3}\right)\zeta _3-\frac{28640}{81}\right) \zeta _2
+\frac{3080480}{243}\bigg]
\nn\\
+&C_AN_f^2 T_F^2 \bigg[\left(\left(-\frac{128}{27}\right)\zeta _2-\frac{1024}{81}\right) \ln x+\frac{32}{81}\ln ^3x+\frac{224}{81}\ln ^2x+\left(-\frac{2816}{81}\right)\zeta _2
\nn\\
+&\left(-\frac{512}{27}\right)\zeta _3-\frac{75328}{729}\bigg]
+C_A^3 \bigg[\left(\frac{152}{3}\zeta _2-\frac{16990}{27}\right) \ln ^3x+\left(\frac{4055}{9}\zeta _2
+\frac{1376}{3}\zeta _3-\frac{278705}{81}\right) \ln ^2x
\nn\\
+&\left(\frac{1559}{3}\zeta _2+\frac{14896}{9}\zeta _3+24 \zeta _4-\frac{74936}{9}\right) \ln x+\left(-\frac{56}{15}\right)\ln ^5x+\left(-\frac{1948}{27}\right)\ln ^4x
\nn\\
+&\left(-\frac{12338}{9}\right)\zeta _4+\frac{41864}{27}\zeta _3+\left(\frac{896}{3}\zeta _3-\frac{16004}{81}\right) \zeta _2-2848 \zeta _5-\frac{2412238}{729}\bigg]
\nn\\
+&C_F N_f^2 T_F^2 \bigg[\left(\left(-\frac{64}{9}\right)\zeta _2-\frac{2224}{81}\right) \ln ^2x+\left(\left(-\frac{2368}{27}\right)\zeta _2+\left(-\frac{512}{9}\right)\zeta _3-\frac{30400}{243}\right) \ln x
\nn\\
+&\frac{8}{27}\ln ^4x
+\frac{112}{81}\ln ^3x+\left(-\frac{16064}{81}\right)\zeta _2+\left(-\frac{2560}{27}\right)\zeta _3+\left(-\frac{416}{9}\right)\zeta _4-\frac{28672}{243}\bigg]
\nn\\
+&C_F^2 N_f T_F \bigg[\left(\left(-\frac{64}{9}\right)\zeta _2+\frac{1420}{9}\right) \ln ^3x+\left(\left(-\frac{224}{3}\right)\zeta _2+\frac{32}{3}\zeta _3+\frac{2320}{3}\right) \ln ^2x
\nn\\
+&\left(\left(-\frac{736}{3}\right)\zeta _2+\left(-\frac{352}{3}\right)\zeta _3+\frac{784}{3}\zeta _4+\frac{4352}{3}\right) \ln x+\frac{8}{15}\ln ^5x+\frac{130}{9}\ln ^4x
\nn\\
+&\left(-\frac{1088}{3}\right)\zeta _3+\left(-\frac{608}{3}\right)\zeta _5+\frac{296}{3}\zeta _4+\left(\frac{224}{3}\zeta _3-\frac{1840}{3}\right) \zeta _2+\frac{5344}{3}\bigg]
\,.
\end{align}
The small-$z$ expansion of  coefficient functions $\Delta\mathscr{C}_{ij}(z)$ for helicity TMD  FFs   is  provided in the ancillary files.

\section{Conclusion}
\label{sec:conclusion}
We have calculated, for the first time, the complete set of N$^3$LO twist-2 matching coefficients for helicity TMDs in both 
 \texttt{HVBM} and $\overline{\text{MS}}$ schemes. For the latter, the only missing ingredient is the three-loop pure-singlet transformation factor 
$z^{(3)}_{\text{ps}}$, whose numerical impact is expected to be negligible in practice as  $\lim_{x\to 1} z^{(3)}_{\text{ps}}(x)=0$. 
Our computation reproduces the NNLO space-like helicity splitting
 functions (up to a subtlety in the cubic color structure compared to Ref.~\cite{Moch:2015usa}) and yields entirely new NNLO time-like results. 
 With these polarized matching coefficients, 
 we  predict  the spectrum of transverse momentum imbalance to N$^3$LL accuracy  in the current fragmentation region. 
 These developments on the perturbative regime, together with  first-principles  Lattice QCD calculation  in the nonperturbative regime~\cite{Bollweg:2025ecn},  
 provide unprecedented theoretical input for unraveling how quark and gluon spin  contribute to the proton’s longitudinal spin, 
 a central physics goal of the EIC. 
 Looking ahead, extending these results to the small-$x$ regime, 
 with systematic exploration of 
 BFKL dynamics for the resummation of small-$x$ logarithms, 
 promises new insights into spin-dependent Sudakov processes at high-energy limit of QCD.

\acknowledgments
I thank Hua Xing Zhu for encouraging me to look into this problem and Tong-Zhi Yang for collaboration in the early stages of the project.

\appendix

\section{Flavor decomposition}
\label{sec:favor-decom}
In this section, we fix our notations for flavor group decomposition of the coefficient functions, the PDFs (FFs) and  the splitting functions,
our definitions are in consistence with those in~\cite{Vogt:2004ns}.
We begin with the non-singlet sectors,
denoting $q$ to be a quark, $\bar q$  an anti-quark and $q'$  another flavor of quark,
the `valence' quark contributions are  defined as ($f$ can be  identified as partonic PDFs,  the coefficient functions or the splitting functions of  some flavor)
\begin{align}
f^{\text{v}}_{qq}=f_{qq}-f_{qq'}\,,\quad f^{\text{v}}_{q\bar q}=f_{q \bar q}-f_{q \bar q'}\,.
\end{align}
By linear combinations, we define  three independently evolving types of non-singlets
\begin{align}
f^{+}_{\text{ns}}=f^{\text{v}}_{qq}+f^{\text{v}}_{q\bar q}\,,\quad f^{-}_{\text{ns}}=f^{\text{v}}_{qq}-f^{\text{v}}_{q\bar q}\,,
\quad
f^{\text{v}}_{\text{ns}}=f^{-}_{\text{ns}}+N_f(f_{qq'}-f_{q \bar q'})\,,
\end{align}
note that the `sea' quark difference $f_{qq'}-f_{q \bar q'}$ vanishes up to $\alpha_s^2$, and is non-vanishing from the third order.
In particular, we denote the `sea' quark difference as
\begin{align}
\label{eq:sea-diff-defin}
f_{d_{abc}^2}=N_f(f_{qq'}-f_{q \bar q'})\,,
\end{align}
since at 3-loop, they are proportional to the cubic color structure $d_{abc}^2$.
For $i=\pm,\text{v}$, the evolutions of the non-singlets are governed by
\begin{align}
\frac{d}{d \ln \mu^2}f^i_{\text{ns}}=\Delta P^i_{\text{ns}}\otimes f^i_{\text{ns}}(\mu)\,.
\end{align}
For the singlet sector, we define 
\begin{align}
f^{\text{s}}_{qq}=&f^{+}_{\text{ns}}+N_f(f_{qq'}+f_{q \bar q'})\equiv f^{+}_{\text{ns}}+f^{\text{ps}}_{qq}\,,
\nn\\
 f^{\text{s}}_{qg}= &2N_f f_{qg}\,,\quad f^{\text{s}}_{gq}=  f_{gq}\,,\quad  f^{\text{s}}_{gg}=  f_{gg}\,.
\end{align}
It proves useful to use a  matrix notation for the partonic PDFs (FFs)
\begin{align}
{\bf f}^{\text{s}}= \left(
\begin{array}{cc}
f^{\text{s}}_{qq} & f^{\text{s}}_{qg} \\
 f^{\text{s}}_{gq} & f^{\text{s}}_{gg} \\
 \end{array} \right)\,,
\end{align}
and the corresponding  $2\times 2$ evolution system reads 
\begin{align}
\frac{d}{d\ln \mu^2} {\bf f}^{s}(\mu)={\bf P}^{\text{s}}\otimes{\bf f}^{\text{s}}(\mu)\,.
\end{align}
\section{Renormalization Group Consistency}
\label{sec:RG}
From RG invariance of the cross section
\begin{align}
    \frac{\rd }{\rd \ln\mu}
    \left[
    H(\alpha_s(\mu), L_Q)\otimes
    f_1^q(x, \alpha_s(\mu), L_b, L_{\xi_n})\otimes
    D_1^q(z, \alpha_s(\mu), L_b, L_{\xi_{\bar n}})
    \right]\equiv 0
    \,,
\end{align}
we must have 
\begin{align}
\label{RG-relation}
\Gamma_H(\alpha_s(\mu), L_{\xi_n}+L_{\xi_{\bar n}}
)= \Gamma_f(\alpha_s(\mu), L_b, L_{\xi_n})
+\Gamma_D(\alpha_s(\mu), L_b, L_{\xi_{\bar n}}),
\end{align}
where we have used physical TMDs for the factorization formula eq.~(\ref{eq:SIDIS-fac}), and the relation $2 L_Q=L_{\xi_n}+ L_{\xi_{\bar n}}$ for the Collins-Soper scales in eq.~(\ref{eq:CS-scale}).

In RG relation eq.~(\ref{RG-relation}), we can set either $L_{\xi_n}$ or $L_{\xi_{\bar n}}$ to zero, this leads to the following relation for $\Gamma_H$
\begin{align}
    \Gamma_H(\alpha_s(\mu), L_{\xi_n}+L_{\xi_{\bar n}} )= \Gamma_H(\alpha_s(\mu), L_{\xi_n}) + \Gamma_H(\alpha_s(\mu), L_{\xi_{\bar n}})- \Gamma_H(\alpha_s(\mu), 0)\,.
\end{align}
We can perform a redefinition by a shift $\overline \Gamma_H:=\Gamma_H - \Gamma_H(0)$, then additive property follows
\begin{align}
    \overline \Gamma_H (x+y)=\overline \Gamma_H(x) +\overline \Gamma_H(y)\,.
\end{align}
The additive property guarantees that $   \overline \Gamma_H (N x)= N \times   \overline \Gamma_H (x)\,,\forall N \in \mathbb{Z}$. 
We only need to show homogeneity  holds for all $c \in \mathbb{R}$ (or $c \in \mathbb{C}$). This is true as long as $    \overline \Gamma_H (c)=    c\times\overline \Gamma_H (1)
$ holds for all $c \in \mathbb{R}$ (or $c \in \mathbb{C}$). Since $\overline{\Gamma}_H$ is assumed to be a continuous function of its argument and the real (complex) numbers are nothing but the Cauchy sequences of (pairs of) the rationals, it suffice to prove it for the rationals. Since we already have additivity for the integers, it suffice to prove it for unit fractions. Indeed, the following is true
\begin{align}
    \overline \Gamma_H (1/M)=1/M\times M\times\overline \Gamma_H (1/M)=1/M \times \overline \Gamma_H (1)\,,\quad \forall M \in \mathbb{Z}\,,
\end{align}
which complete the proof that $\overline \Gamma_H $ is a linear function. From this we conclude that the anomalous dimensions associated with $\mu$-evolutions is linear in $\ln\mu$ to all-loop orders, e.g.\,,
\begin{align}
\label{eq:RG-example}
\frac{\rd}{\rd \ln\mu^2}
\ln
f_i(x, \alpha_s(\mu), L_b, L_{\xi_i}) 
= 
 - \gamma^i_{\rm cusp}(\alpha_s(\mu))L_{\xi_i} -\gamma^i(\alpha_s(\mu),\dots)\,.
\end{align}
On the other hand, the rapidity evolution with respect to the Collins-Soper scale $\xi_i$ is
\begin{align}
    \Gamma_R^i(L_{\xi_i}, L_b)
    \equiv
    \frac{\rd }{\rd L_{\xi_i}} \ln f_i(x, \alpha_s(\mu), L_b, L_{\xi_i})\,,
\end{align}
and  $\mu$-evolution of the Collins-Soper kernel is controlled by the cusp anomalous dimension
\begin{align}
\frac{\rd}{\rd \ln \mu^2}
    \Gamma_R^i(L_{\xi_i}, L_b)
    =
        \frac{\rd }{\rd L_{\xi_i}}
        \frac{\rd}{\rd \ln \mu^2}
        \ln f_i(x, \alpha_s(\mu), L_b, L_{\xi_i})
        = 
         - \gamma^i_{\rm cusp}(\alpha_s(\mu))\,,
\end{align}
thus the  Collins-Soper kernel takes the form 
\begin{align}
    \Gamma_R^i(L_{\xi_i}, L_b)
    =
-\int^{\mu^2}_{\mu_b^2} \frac{d\bar{\mu}^2}{\bar{\mu}^2} \gamma^i_{\rm cusp} [\alpha_s(\bar{\mu})] + \gamma^i_R[\alpha_s(\mu_b)] \,.
\end{align}
\section{QCD Beta Function}
\label{sec:beta}
The QCD beta function is defined as
\begin{equation}
\frac{d\alpha_s}{d\ln\mu} = \beta(\alpha_s) = -2\alpha_s \sum_{n=0}^\infty \left( \frac{\alpha_s}{4 \pi} \right)^{n+1} \, \beta_n \, ,
\end{equation}
with~\cite{Baikov:2016tgj}
\begin{align}
\beta_0 &= \frac{11}{3} C_A - \frac{4}{3} T_F N_f \, , \nn
\\
\beta_1 &= \frac{34}{3} C_A^2 - \frac{20}{3} C_A T_F N_f - 4 C_F T_F N_f \, ,\nn
\\
\beta_2 &= \left(\frac{158 C_A}{27}+\frac{44 C_F}{9}\right) N_f^2 T_F^2 +\left(-\frac{205 C_A
   C_F}{9}-\frac{1415 C_A^2}{27}+2 C_F^2\right) N_f T_F  +\frac{2857 C_A^3}{54}\,.
\end{align}
\section{Anomalous dimensions}
\label{sec:AD}
For all the anomalous dimensions entering the renormalization group equations of various TMD functions, we define the perturbative expansion in $\alpha_s$ according to
\begin{equation}
\gamma(\alpha_s) = \sum_{n=0}^\infty \left( \frac{\alpha_s}{4 \pi} \right)^{n+1} \, \gamma_n \, ,
\end{equation}
where the coefficients for quark  are given by
\begin{align}
\Gcusp_{0} =& 4 C_F\,, \nn
\\
\Gcusp_{1} =&  \left(\frac{268}{9}-8 
                 \zeta_2\right) C_A C_F -\frac{80 C_F T_F N_f}{9}\,, \nn
\\
\Gcusp_{2} =&\bigg[ \left(\frac{320 \zeta _2}{9}-\frac{224 \zeta _3}{3}-\frac{1672}{27}\right) C_A
   C_F+\left(64 \zeta _3-\frac{220}{3}\right) C_F^2\bigg] N_f T_F  \nn
   \\
 +&\left(-\frac{1072 \zeta
   _2}{9}+\frac{88 \zeta _3}{3}+88 \zeta _4+\frac{490}{3}\right) C_A^2 C_F  -\frac{64}{27} C_F
   N_f^2 T_F^2\,,\nn
     \\
\gamma^S_0 =& 0 \, , \nn
\\
\gamma^S_1 =& \left[ \left( -\frac{404}{27} + \frac{11\zeta_2}{3} + 14\zeta_3 \right) C_A   + \left( \frac{112}{27} - \frac{4 \zeta_2}{3} \right)T_F N_f   \right]  C_F   \,, \nn
\\
\gamma^S_2  =&\left(-\frac{88}{3} \zeta
   _3 \zeta _2+\frac{6325 \zeta _2}{81}+\frac{658 \zeta _3}{3}-88 \zeta
   _4-96 \zeta _5-\frac{136781}{1458}\right) C_A^2 C_F
+\left(\frac{80\zeta _2}{27}-\frac{224 \zeta _3}{27}\right.
\nn\\
+&\left.\frac{4160}{729}\right) C_FN_f^2 T_F^2\nn
    + \left(-\frac{2828 \zeta _2}{81}-\frac{728 \zeta _3}{27}+48 \zeta
   _4+\frac{11842}{729}\right) C_A C_F N_f T_F
   \nn\\
   +&\left(-4 \zeta _2-\frac{304 \zeta _3}{9}-16 \zeta
   _4+\frac{1711}{27}\right) C_F^2 N_f T_F\,.\nn
\nn\\
   \gamma^R_0 = &0 \, , \nn
\\
\gamma^R_1 = & \left[ \left( -\frac{404}{27} + 14\zeta_3 \right) C_A  +  \frac{112}{27} T_F N_f \right] C_F  \,, \nn 
\\
\gamma^R_2 =&\bigg[\left(-\frac{824 \zeta _2}{81}-\frac{904 \zeta _3}{27}+\frac{20 \zeta
   _4}{3}+\frac{62626}{729}\right) C_A N_f T_F 
   +\left(-\frac{88}{3} \zeta _3 \zeta
   _2+\frac{3196 \zeta _2}{81}+\frac{6164 \zeta _3}{27}\right.
   \nn\\
   +&\left.\frac{77 \zeta _4}{3}-96 \zeta_5
   -\frac{297029}{1458}\right)C_A^2 
   + \left(-\frac{304 \zeta _3}{9}-16 \zeta_4+\frac{1711}{27}\right)  C_F N_f T_F 
   +\left(-\frac{64 \zeta_3}{9}\right.
   \nn\\
   -&\left.\frac{3712}{729}\right) N_f^2 T_F^2 \bigg] C_F \,.
\end{align}
Since cusp and  soft and rapidity anomalous dimensions exhibit Casimir scaling,
 the corresponding anomalous dimensions for gluon could be obtained by multiplying from above by $C_A/C_F$.

The beam anomalous dimensions do not exhibits Casimir scaling, thus should be list separately.
The beam anomalous dimensions  for quark are
 \begin{align}
\gamma^B_0 =& 3C_F \, , \nn
\\
\gamma^B_1 =&  \left[  \left( \frac{3}{2} - 12\zeta_2 + 24\zeta_3 \right) C_F + \left( \frac{17}{6} + \frac{44 \zeta_2}{3} - 12\zeta_3 \right)  C_A 
 + \left( -\frac{2}{3} - \frac{16\zeta_2}{3} \right) T_F N_f  \right] C_F  , \nn
\\
\gamma^B_2 = &
 \bigg[ \left(-\frac{2672 \zeta _2}{27}+\frac{400 \zeta _3}{9}+4
   \zeta _4+40\right) C_A C_F+\left(\frac{40 \zeta _2}{3}-\frac{272 \zeta
   _3}{3}+\frac{232 \zeta _4}{3}-46\right) C_F^2\bigg] N_f T_F \nn 
\\   
  +& \left(16 \zeta _3
   \zeta _2-\frac{410 \zeta _2}{3}+\frac{844 \zeta _3}{3}-\frac{494 \zeta
   _4}{3}+120 \zeta _5+\frac{151}{4}\right) C_A C_F^2
   +\left(\frac{320 \zeta _2}{27}-\frac{64 \zeta _3}{9}-\frac{68}{9}\right)
   \nn\\
  \times&  C_F N_f^2T_F^2
  + \left(\frac{4496
   \zeta _2}{27}-\frac{1552 \zeta _3}{9}-5 \zeta _4+40 \zeta _5-\frac{1657}{36}\right) C_A^2 C_F
   +\bigg(-32 \zeta _3 \zeta _2+18 \zeta _2
   \nn\\
   +&68 \zeta _3+144 \zeta_4-240 \zeta _5+\frac{29}{2}\bigg) C_F^3\,.
\end{align}
The beam anomalous dimensions  for gluon are
 \begin{align}
\gamma^B_0 =& \frac{11}{3} C_A - \frac{4}{3} T_F N_f \,, \nn
\\
\gamma^B_1 =& C_A^2 \left( \frac{32}{3}+ 12 \zeta_3 \right) + \left(  -\frac{16}{3} C_A -  4 C_F \right) N_f T_F   \,, \nn
\\
\gamma^B_2 =& C_A^3\left(-80\zeta_5-16\zeta_3\zeta_2+\frac{55}{3}\zeta_4+\frac{536}{3}\zeta_3+\frac{8}{3}\zeta_2+\frac{79}{2}\right)
\nn\\
+&C_A^2 N_f T_F\left(-\frac{20}{3}\zeta_4-\frac{160}{3}\zeta_3-\frac{16}{3}\zeta_2-\frac{233}{9}\right)
+\frac{58}{9}C_A N_f^2 T_F^2
-\frac{241}{9}C_A C_F N_f T_F
\nn\\
+&2 C_F^2 N_f T_F +\frac{44}{9}C_F N_f^2 T_F^2\,.
\end{align}
The cusp anomalous dimension $\Gamma^{\text{cusp}}$ can be found in \cite{Moch:2004pa}. 
 The beam anomalous dimension $\gamma^B$ is related to the soft anomalous dimension $\gamma^S$~\cite{Li:2014afw}
  and  the hard anomalous dimensions $\gamma^H$~\cite{Moch:2005tm,Gehrmann:2010ue,Becher:2009qa} by renormalization group invariance condition $\gamma^B = \gamma^S - \gamma^H$.
 The rapidity anomalous dimension $\gamma^R$ can be found in \cite{Li:2016ctv,Vladimirov:2016dll}. 
 Note that the normalization here differ from those in \cite{Li:2016ctv} by a factor of $1/2$. 
\section{Renormalization Constants}
\label{sec:RC}
The following constants are needed for the renormalization of zero-bin subtracted~\cite{Manohar:2006nz} TMD PDFs through N$^3$LO, see e.g. Ref.~\cite{Luo:2019hmp,Luo:2019bmw}. 
The first three-order corrections to $Z^B $ and $Z^S$ are 
\begin{align}
\label{eqZqZs}
Z^B_1 =& \frac{1}{2\epsilon} \left(2 \gamma^B_0 -\Gamma_0^{\text{cusp}} L_Q \right) \,, \nonumber \\
Z^B_2 =& \frac{1}{8 \epsilon^2} \bigg( ( \Gamma_0^{\text{cusp}} L_Q - 2 \gamma^B_0)^2 + 2 \beta_0 (  \Gamma_0^{\text{cusp}} L_Q - 2 \gamma^B_0)    \bigg)
 + \frac{1}{4\epsilon} \left( 2 \gamma^B_1 - \Gamma_1^{\text{cusp}} L_Q \right) \,, \nonumber \\
Z^B_3  =& \frac{1}{48 \epsilon^3}  \left( 2 \gamma^B_0 -\Gamma^{\text{cusp}}_0 L_Q \right) \biggl( 8 \beta_0^2 + 6 \beta_0 \left( -2 \gamma^B_0 + \Gamma^{\text{cusp}}_0 L_Q \right) 
+ \left( -2 \gamma^B_0 + \Gamma^{\text{cusp}}_0 L_Q \right)^2 \biggl) \nn \\ 
+& \frac{1}{24 \epsilon^2} \biggl( \beta_1 \left(-8 \gamma^B_0 + 4  \Gamma^{\text{cusp}}_0 L_Q \right) + \left(4 \beta_0 - 6 \gamma^B_0 + 3 \Gamma^{\text{cusp}}_0 L_Q \right) \left( -2 \gamma^B_1 + \Gamma^{\text{cusp}}_1 L_Q \right)  \biggl) 
 \nn  \\    
  +& \frac{1}{6 \epsilon} \biggl(  2 \gamma^B_2 -  \Gamma^{\text{cusp}}_2 L_Q  \biggl) 
  \nn\\  
Z^S_1 =& \frac{1}{\epsilon^2} \Gamma^{\text{cusp}}_0  +  \frac{1}{\epsilon} \left( -2 \gamma^S_0 - \Gamma_0^{\text{cusp}} L_\nu \right) \,,\nonumber \\
Z^S_2 =& \frac{1}{2 \epsilon^4} (\Gamma^{\text{cusp}}_0)^2 - \frac{1}{4 \epsilon^3} \bigg(\Gamma^{\text{cusp}}_0 (3 \beta_0 + 8 \gamma^S_0)+4( \Gamma^{\text{cusp}}_0)^2 L_\nu\bigg)   - \frac{1}{2 \epsilon} \left( 2 \gamma^S_1 +  \Gamma^{\text{cusp}}_1 L_\nu \right) \nonumber \\
+& \frac{1}{4 \epsilon^2} \bigg(\Gamma^{\text{cusp}}_1 + 2 ( 2 \gamma^S_0 + \Gamma^{\text{cusp}}_0 L_\nu ) ( \beta_0 + 2 \gamma^S_0 + \Gamma^{\text{cusp}}_0 L_\nu) \bigg) \,, \nn \\
Z^S_3 =& \frac{ 1}{6 \epsilon^6} \left(\Gamma^{\text{cusp}}_0\right)^3  - \frac{1}{4 \epsilon^5} \left(\Gamma^{\text{cusp}}_0 \right)^2 \left(  3 \beta_0 + 4 \gamma^S_0 + 2 \Gamma^{\text{cusp}}_0 L_\nu \right) + \frac{1}{36 \epsilon^4 } \Gamma^{\text{cusp}}_0 \bigg( 22 \beta_0^2 + 45 \beta_0 \left(2 \gamma^S_0 + \Gamma^{\text{cusp}}_0 L_\nu \right)  \nn \\
+& 9 \left( \Gamma^{\text{cusp}}_1 + 2 \left( 2 \gamma^S_0+ \Gamma^{\text{cusp}}_0 L_\nu\right)^2  \right)   \biggl)  + \frac{1}{36 \epsilon^3} \biggl( -16 \beta_1 \Gamma^{\text{cusp}}_0 - 12 \beta_0^2 \left( 2 \gamma^S_0 + \Gamma^{\text{cusp}}_0 L_\nu \right) \nn \\
 -& 2 \beta_0 \left( 5 \Gamma^{\text{cusp}}_1 + 9 \left( 2 \gamma^S_0 + \Gamma^{\text{cusp}}_0 L_\nu \right)^2 \right) - 3 \bigg[ \Gamma^{\text{cusp}}_1 \left(6 \gamma^S_0 + 9 \Gamma^{\text{cusp}}_0 L_\nu \right)  \nn \\
+& 2 \left( 8 \left( \gamma^S_0\right)^3 + 6 \Gamma^{\text{cusp}}_0 \gamma^S_1 + 12 \Gamma^{\text{cusp}}_0  \left(\gamma^S_0\right)^2 L_\nu + 6 \left(\Gamma^{\text{cusp}}_0\right)^2 \gamma^S_0 L_\nu^2 + \left( \Gamma^{\text{cusp}}_0 \right)^3 L_\nu^3 \right) \bigg]    \biggl) \nn \\
 +& \frac{1}{18 \epsilon^2} \biggl(  2 \Gamma^{\text{cusp}}_2 + 3 \left( 2 \beta_1 \left( 2 \gamma^S_0 + \Gamma^{\text{cusp}}_0 L_\nu \right) + \left( 2 \beta_0 + 6 \gamma^S_0 + 3 \Gamma^{\text{cusp}}_0 L_\nu \right) \left( 2 \gamma^S_1 + \Gamma^{\text{cusp}}_1 L_\nu \right) \right)    \biggl) 
 \nn\\
 - & \frac{2 \gamma^S_2 + \Gamma^{\text{cusp}}_2 L_\nu}{3 \epsilon} \,.
\end{align}
Keep in mind that the  anomalous dimensions appeared above depends on the flavor, they should be replaced by the corresponding values in Sec~\ref{sec:AD}.
We also remind the reader that the renormalization constants are formally identical for TMD PDFs and TMD FFs,
the logarithms appeared above should be replaced by their corresponding values in each case,
and we have
\begin{align}
\label{eq:LdefinitionALL}
 L_\perp = \ln \frac{b_T^2 \mu^2}{b_0^2} ,  \quad L_\nu = \ln \frac{\nu^2}{\mu^2} \,,
\end{align}
with $b_0 =2 \, e^{- \gamma_E}$ for both TMD PDFs and TMD FFs.
On the other hand, we have for TMD PDFs
\begin{align}
\label{eq:LdefinitionSLQ}
 L_Q  = 2 \ln \frac{x \, P_+}{\nu} \,,
\end{align}
while for TMD FFs,
\begin{align}
\label{eq:LdefinitionTLQ}
L_Q = 2 \ln \frac{P_+}{ z \, \nu}.
\end{align}

\bibliographystyle{JHEP}
\bibliography{helicity_TMD}

\end{document}